\documentclass[reqno, 12pt]{article}
\usepackage{a4wide,amssymb}
\usepackage{amsthm,amsfonts,amsmath}
\usepackage{enumerate}{}
\begin{document}
\date{}
\newcommand{\version}{}
\newtheorem{theorem}{Theorem}
\newtheorem{lemma}[theorem]{Lemma}
\newtheorem{property}{Property}
\newtheorem{proposition}{Proposition}
\newtheorem{consequence}{Consequence}
\newtheorem{example}{Example}
\newtheorem{definition}{Definition}
\newtheorem{corollary}[theorem]{Corollary}
\renewcommand{\thetheorem}{\thesection.\arabic{theorem}}
\renewcommand{\thelemma}{\thesection.\arabic{lemma}}
\renewcommand{\theproperty}{\thesection.\arabic{property}}
\renewcommand{\theproposition}{\thesection.\arabic{proposition}}
\renewcommand{\theconsequence}{\thesection.\arabic{consequence}}
\renewcommand{\theexample}{\thesection.\arabic{example}}
\renewcommand{\thedefinition}{\thesection.\arabic{definition}}
\renewcommand{\thecorollary}{\thesection.\arabic{corollary}}
\renewcommand{\theequation}{\thesection.\arabic{equation}}
\newcommand{\R}{{\mathord{\mathbb R}}}
\newcommand{\C}{{\mathord{\mathbb C}}}
\newcommand{\Z}{{\mathord{\mathbb Z}}}
\let\th=\vartheta
\let\y=\upsilonpsilon \let\x=\xi \let\z=\zeta
\let\D=\Delta \let\F=\Phi  \let\G=\Gamma  \let\L=\Lambda \let\Th=\Theta
\let\O=\Omega \let\P=\Pi   \let\Ps=\Psi \let\Si=\Sigma \let\X=\Xi
\let\Y=\Upsilon
\let\o=\omega
\let\e=\varepsilon
\let\a=\alpha
\let\pa=\parallel
\let \r=\rho

\title {Stability for Rayleigh-Benard  convective  solutions of the Boltzmann equation.}

\author{L. Arkeryd$^1$, R. Esposito$^2$,  R. Marra$^3$  and A. Nouri$^4$\\  \footnotesize{$^1$ 
Chalmers Institute of Technology, Gothenburg, Sweden.} \\  \footnotesize{$^2$ Dipartimento di
Matematica, Universit\`a di L'Aquila, Coppito, 67100 AQ, Italy}
\\ \footnotesize{$^3$ Dipartimento di Fisica and Unit\`a INFN, Universit\`a di
Roma Tor Vergata, 00133 Roma, Italy.}
\\ \footnotesize{$^4$ LATP,  Universit\'e d'Aix-Marseille I, Marseille, France.}
}

\maketitle

\begin{abstract} We consider the Boltzmann equation for a gas in a horizontal slab, subject to a gravitational force. The boundary conditions are of diffusive type, specifying the wall
temperatures, so that the top temperature is lower than the bottom one (Benard setup). We consider a $2$-dimensional convective stationary solution, which is close for  small Knudsen number to the convective stationary solution of the Oberbeck-Boussinesq  equations, near above the bifurcation point, and  prove its stability under $2$-d small  perturbations, for Rayleigh number above and close to the bifurcation point and for small Knudsen number.
\end{abstract}



\date{10-10-2008}
\section{Introduction}

We study the small Knudsen number solutions to  the  Boltzmann equation in a slab with diffusive boundary conditions in the presence of a gravitational field.

In a previous paper \cite{AEMN} we have proved the existence and stability globally in time for small Knudsen number  of a positive one-dimensional stationary solution to the  Boltzmann equation, which is close to the hydrodynamic laminar solution of the Oberbeck-Boussinesq (O-B) equations. At the hydrodynamical level there is a bifurcation phenomenon: when the vertical  temperature gradient exceeds a certain critical value,  the laminar one-dimensional solution loses stability and various two- or three-dimensional pattern flows appear. In particular, it has been proved the existence of a two-dimensional roll solution of the O-B equations close to the bifurcation point. Its stability under suitable perturbations has also been proved. 

In this paper we construct, by means of perturbative arguments (expansion method), for small Knudsen number, a positive two dimensional solution to the stationary Boltzmann equation, which is close to this roll solution. Moreover, we prove its stability for long times under a suitable class of two dimensional initial perturbations.  These results are true for  values of the  Rayleigh number above and close to the bifurcation value, provided that the force is small enough.  To state our result, we need to introduce some notation.  

Consider a gas in a $2$-dimensional box of height  $2\pi d$ and length $2\pi h $, under the action of a gravitational force $g$ in the direction $z$. The upper and lower walls are kept at temperature $T_+$ and $T_-$ respectively, with $T_+< T_-$, with no-slip conditions, while periodicity is assumed in the horizontal direction. At the kinetic level, the behavior of the gas is given by the following Boltzmann equation with  boundary conditions diffusive in the $z$ direction and periodic in the $x$ direction, written in dimensionless form,
\begin{eqnarray}\label{BoltzEqn}
&&\frac{\partial F}{\partial t}+\frac{1}{\e } \ v_x\frac{\partial F}{\partial x} +\frac{1}{\e } v_z\frac{\partial F}{\partial z} -G\frac{\partial F}{\partial v_z}=\frac{1}{\e ^2}Q(F,F),\nonumber
\\
&&F(0,x,z,v)= F_0(x,z,v),\quad (x,z)\in (-\mu\pi,\mu\pi)\times(-\pi,\pi),\hspace*{0.05in}v\in \mathbb{R}^3,\\
&&F(t,x,\mp\pi,v)\ = M_\mp(v)\int_{w_z \lessgtr 0} |w_z| F(t,x,\mp\pi,w)dw,\ t>0,
\ v_z\gtrless0,\  x\in [-\mu\pi,\mu\pi],\nonumber
\end{eqnarray}
where
$$
F_0\geq 0,\quad M_-=\frac{1}{2\pi }e^{-\frac{v^2}{2}},\quad M_+(v)= \frac{1}{2\pi (1-2\pi\e\lambda)^{2}}e^{-\frac{v^2}{2(1-2\pi\e \lambda)}},
$$
$$\e=\frac{\ell_0}{d},\quad G=\frac{1}{\e}\frac{dg}{ 2T_-},\quad \lambda=\frac{1}{\e}\frac {T_--T_+}{ 2\pi T_-}, \quad \mu=\frac {h} {d},$$
 $$Q(f,g)(z,v,t)= \frac{1}{2}\int_{\mathbb{R}^3}d\/v_*\int_{S_2}d\/\o
B(\o,|v-v_*|)\big\{f'_*g'+f'g'_*-f_*g -g_*f\big\}.$$
Here $h', h'_*,h,h_*$ stand for $h(x,z ,v',t), h(x,z ,v'_*,t),h(x,z ,v,t), h(x,z ,v_*,t)$ respectively, $S_2=\{\o\in\mathbb{R}^3\,|\o^2=1\}$, $B$ is the differential cross section $2B(\o,V)=|V\cdot \o|$ corresponding to hard spheres, and $v$, $v_*$ and $v'$,$v'_*$ are precollisional and postcollisional  velocities or conversely.  The boundary conditions are such that the condition of impermeability of the walls, 
\begin{equation}\int_{\R^3}dv  F v_z dv=0\label{zeroflux},\end{equation}
is satisfied. 
The solution depends on the parameter $\e={2\sqrt 6}{Kn}/{\sqrt{5\pi}}$, where $Kn$ is the Knudsen number given in terms of  $\ell_0$, the mean free path of the gas in equilibrium at temperature $T_-$ and density $\bar \rho$. We have also put  $Ma=\e{\sqrt6}/{\sqrt 5}$, where $Ma$ is the Mach number. With this choice, the Rayleigh number  (\cite{So}) is  $\displaystyle{Ra=\frac{16G (2\pi\lambda)}{\pi  }}\ $, independent of $\e$.  We will fix  through the paper  the parameter $G$ such that $G\le G_0$ with $G_0$ suitably small. Fix $\displaystyle{h=\frac{2\pi d}{\alpha_c}}$ where $\alpha_c$ is the critical wave number for the first bifurcation. The linear analysis of the O-B equations with rigid-rigid boundary conditions, (\ref{O-B}) below, describing the behavior of the fluid at hydrodynamic level, gives a critical value $\alpha_c$ and a corresponding critical Rayleigh number $Ra_c$ \cite{DR}. The parameter $\lambda$ will be chosen in an interval $[\lambda_c, (1+\delta)\lambda_c]$, for $\delta$ small and with $\lambda_c$ determined by the condition that $\displaystyle{Ra_c=\frac{16 G(2\pi\lambda_c)}{\pi  }}$ is the critical value.   With these choices, at the hydrodynamical level, two roll solutions will appear at the bifurcation point,  consisting of one roll, rotating clockwise and anticlockwise respectively. These solutions  are constructed perturbatively for $\delta$ small in a rigorous way  \cite{Iu} and their local non-linear stability has been proved for  small initial perturbations with the same period of the roll solution \cite{Iu}. 

The clockwise solution $h_s$ is then of the form 
 \begin{equation}
 h_s=    h_{\ell}+\delta\  h_{con}+ O(\delta^2),
 \label{hydro}
 \end{equation} where $h_\ell$ is the laminar solution and  $h_{con}$
is the eigenfunction corresponding to the least eigenvalue  
of the linearized Boussinesq   problem  around the laminar solution (see Section 4 for the precise definition).                                     

In this paper, we construct a stationary solution $F_s$ of the Boltzmann equation, which is close for $\e$ small to the  hydrodynamical solution (say, the clockwise one) in the sense that it can be written as a truncated expansion in $\e$,  $F_s=M+\e f_s+ O(\e^2)$ with $\displaystyle{M=\frac{1}{(2\pi)^{3/2}}e^{-\frac{v^2}{2}}}$ and 
$$f_s=M\Big({ \rho_s} + {u_s\cdot v} + T_s\frac{|v|^2- 3}{  2 }\Big)\ , $$
where $\rho_s, u_s, T_s$ are expressed in terms of $h_s$. Moreover, we prove the kinetic non linear stability of $F_s$ under suitable initial perturbations.

 We study the Boltzmann equation for the perturbation $\Phi=M^{-1}(F-F_s)$ with the initial datum   
 \begin{equation}\Phi_0(x,z,v)=\sum_{n=1}^5\e^n \Phi^{(n)}(0,x,z,v)+\e ^5p_5\label{bcf}
\end{equation}
where $\Phi^{(n)}(0,x,z,v)$ is the $n$-th term of the expansion introduced in the next paragraph, computed at time $t=0$, and the $\epsilon $-dependent $p_5$ is arbitrary but for having total mass $\int dvdxdzM(v) p_5(x,z,v)=0$ and satisfying (3.8).

We write also the time dependent solution  in terms of a truncated expansion in $\e$
\begin{equation} \Phi(t,x,z,v)=\sum_{n=1}^5\e^n\Phi^{(n)}(t,x,z,v) +\e R(t,x,z,v), \quad (x,z)\in \Omega_\mu\label{expintro},\end{equation}
where $\Omega_\mu=[-\mu\pi,\mu\pi]\times[-\pi,\pi]$.
 The first term of the expansion in $\e$ is $$\Phi^{(1)}= 
{ \rho^1} + {u^1\cdot v} + 
\theta^1\frac{|v|^2-  3}{  2 }\ ,$$ where  the fields $\rho^1(t,x,z), u^1(t,x,z), \theta^1(t,x,z)$ are solutions of the hydrodynamic equations for the perturbation, with initial datum $(u^1_0,\theta^1_0)$. The initial data are chosen as follows: let $(u^1_0,\theta^1_0)$ be an initial perturbations of the convective solution $(u_s,\theta_s)$  sufficiently small to ensure that the solution $(u(t,x,z),\theta(t,x,z))=(u_s(x,z)+u^1(t,x,z),\theta_s(x,z)+\theta^1(t,x,z))$ of the initial boundary value problem for the O-B (\ref{O-B}) equations 
 exists globally in time
and converges to $(u_s,\theta_s)$ as $t\to +\infty$.

The next orders involve  kinetic corrections in the bulk as well as boundary layer corrections. In fact, at next orders, the standard Hilbert expansion bulk terms do not satisfy the diffusive boundary conditions and boundary layer correction terms are to be included to restore the boundary conditions.  They are computed solving suitable  Milne problems in the presence of a force $\overrightarrow{F}$ and a source term $S$. 
Under suitable assumptions  this problem has $z$-smooth solutions when $\overrightarrow{F}$ is a potential force decaying fast enough at infinity \cite{CME}. We give in Section $3$ a procedure to compute the $\F_n$'s in the time dependent case and show that they have good enough properties as consequence of the smoothness of the hydrodynamic solution. In particular, they inherit the smallness and decay properties of the hydrodynamic solution, such as the  exponential  decay in time. The main difficulty is then the control of the remainder $R$ asymptotically in time.  The equation for the remainder  $R$ is a weakly non-linear  Boltzmann equation with a source $B$, generated by the terms of the expansion. Since it is weakly non-linear, it is enough to get 
good estimates for the associated linear problem, which is  of the form
\begin{eqnarray*}&&\frac{\partial}{\partial t}R+\frac 1 \e (v_x\frac{\partial}{\partial x} R+v_z\frac{\partial}{\partial z} R)-M^{-1}G\frac{\partial}{\partial v_z}(MR)=\frac{1}{\e^2} LR+\frac 1\e J(\F_H,R)+B,\end{eqnarray*}
where $L$ is the linearized Boltzmann operator, defined in Section 2, and $J(\Phi_H,R)$ is a linear operator depending on the perturbed stationary solution and on the  first  terms of the expansion.
The operator $L$ is non positive on $L_2(\R^3, Mdv)$, but has a non trivial null space $\text{\tt Kern}(L)$. The linear operator $J(\Phi _H,R)$ is at least of order $\e$ but it is the main contribution for $R\in \text{\tt Kern}(L)$. The control of the component of $R$ in the orthogonal to $\text{\tt Kern}(L)$ is given by the well known spectral inequality for  $L$  (see e.g. \cite{Ma}),
\begin{equation}-(f,Lf)\ge C((1-P)f,\nu (1-P)f),\label{specl}\end{equation}
where $(f,Lf)$  is the scalar product in $L_2(\R^3, Mdv)$, $P$ is the projector on $\text{\tt Kern}(L)$, $\nu(v)=\int_{\R^3}dv_*\int_{\mathbb S_2}d\omega \frac{1}{2}|(v-v_*)\cdot\omega|M(v_*).$
On the other hand, it is easy to check that 
$$(f,J(\F_H,Pf))\le C\|\nu^{1/2}Pf\|\ \|\nu^{1/2}(1-P)f\|.$$
Using this (and ignoring the boundary contributions) one gets the differential inequality
$$\frac 1 2\frac d{dt}\pa f\pa_{2,2}^2\le C\pa f\pa_{2,2}^2\ +\int_{\Omega_\mu} |(B,f)|,$$
with $\pa\,\cdot\,\pa_{2,2}$ the norm in $L_2(\Omega_\mu\times\R^3, Mdxdzdv)$.
 This produces bounds growing exponentially in time.
 To avoid this we use a spectral inequality for the operator  \newline$L_J(f)= Lf+\e J(\F_H, Pf).$
 The inclusion of the second term in the new operator simplifies all the arguments in the control of the remainder. The price to be paid is that $L_J$ is not self-adjoint and its null space is more complicated. 
This inequality  is an important new ingredient for proving stability results  in the Boltzmann equation framework.  In \cite{AN1} there was no need to use this inequality to get the global stability result because in that paper it was possible to take advantage from the fact  that the constant $C$ in the previous inequality can be taken small, while the parameter controlling the bifurcation can grow beyond the bifurcation value. 
In the Rayleigh-Benard setting, there is not such freedom.

To study the  hydrodynamic part of $R$, $PR$, we use a duality argument which was introduced by N. Maslova \cite{Ma}. 
To illustrate how it works, let us consider the typical equation one has to study  in the stationary case 
$$v\cdot\nabla_x R-\e G\frac{1}{M}\frac{\partial}{\partial v_z}(MR)=\frac{1}{\e} L_JR+q$$
with prescribed incoming data at $z=\pm\pi$, periodic in $x$.
One also considers the dual equation
$$v\cdot\nabla_x \phi-\e G\frac{1}{M}\frac{\partial}{\partial{v_z}}(M\phi)=\frac{1}{\e} L_J^*\phi+h,$$
with vanishing incoming data at $z=\pm\pi$, periodic in $x$.
By taking the inner product of the first by $\phi$ and the second by $R$ and summing, one gets an estimate of $\|h\|$ in terms of $q$ and $\phi$ with suitably small coefficients. For that, we need $G$ small, but we can anyway have large Rayleigh number by increasing $\lambda$.
 Then one chooses $h=PR$ and thus shows that $\|PR\|$ can be bounded in terms of $\|\phi\|$ and $\|q\|$.
The equation for $\phi$ is studied by means of Fourier analysis, which provides a good control on $\phi$ but for the $0$-moment. 
The estimate of the $0$-moment is based on a direct approach,  by means of rather explicit calculations of the first few moments of the solution  
using o.d.e. analysis, for the one-dimensional case.  This allows a reasonably simple analysis when the incoming data are prescribed. Dealing with diffuse reflection requires  more technical steps.
This method has been exploited in \cite{AN1} and  has been extended in \cite{AEMN} to the one-dimensional Benard problem, which  is more difficult to deal with because of the presence of the force and the diffusive boundary conditions. In this paper, we extend the one-dimensional analysis to the two-dimensional case by using that, both in the conductive and convective case,  for  ${\cal R}<{\cal R}_c(1+\delta)$ and $\delta$ small,
the contributions due to the inhomogeneity in the $x$ variable are small and can be included perturbatively. A by-product of this analysis is the extension of the result in \cite{AEMN} to two-dimensional initial perturbations.

Finally, the control of the nonlinearity requires $L^\infty$-estimates (in space), which are more intricate by the presence of the force. We use techniques based on the study of the characteristics, which are no more straight lines because of the presence of the force. This is controlled by looking at the characteristics in the phase space like in \cite{EML} and \cite{AEMN}.

The main result of this paper is summarized in the following theorem.

\goodbreak

 \begin{theorem}
 \label{pro}
 Assume that the gravitational force is such that  $G\le G_0$, with $G_0$ small enough, and $\lambda\in[\lambda_c,(1+\delta)\lambda_c]$. Then, there are  $\delta_0$ and $\e_0$ small enough such that for $\delta\le \delta_0$,    there exists a positive steady solution $F_s$ to the Boltzmann equation, such that for $\e\le \e_0$, 
$$\pa M^{-1}[F_s-(M+\e f_s)] \pa_{2,2}\le c\e^2$$ 
Assume also that the initial perturbation matches the expansion up to order $\e^4$ as detailed in Section $3$ below, and is small as detailed in Section $4$. For such perturbations the stationary solution is stable uniformly in $\e$ for $\e\le \e_0$. 
  \end{theorem}
 Here, stable means that the perturbation vanishes asymptotically in time. This is a consequence of the inequality 
\begin{equation}
\label{stabi}
\int _{0}^{+\infty }dt\int _{\Omega_\mu }dxdz\int _{\mathbb{R}^3} dv|\Phi(t, x,,z,v)|^2 M(v)< \infty \ ,
\end{equation}
 which is proved in Section $4$, and of the regularity of the solution which follows from our construction.

The positivity of the stationary solution is obtained by using the methods in \cite{AN4}. 
We remark that the method presented here  for proving stability strongly relies on the fact that the problem we are dealing with has suitable stability properties at the fluid dynamic level, which we show to be preserved in the kinetic setup by means of a perturbative analysis starting from an Hilbert-type asymptotic expansion plus boundary layer corrections. The preservation of the fluid dynamic stability at kinetic level also occurs in the Taylor-Couette case discussed in   \cite{AN1}, where the bifurcation phenomenon also arises. 

The paper is organized as follows: in Section 2 we construct the stationary solution as a Hilbert asymptotic series in $\e$.
For sake of shortness, we choose not to  give here  the construction of the terms of the expansion, and  refer for that to Section 3, where we show explicitly the analogous construction  in the time-dependent case. We do complete the proof of the existence of the solution in the stationary case  in Section 2, by proving   the main theorem on the remainder.  In Section 4 we deal with the remainder in the time-dependent case and prove the stability result.

\bigskip

\setcounter{section}{1}
\section{Stationary case}

\setcounter{equation}{0}
\setcounter{proposition}{1}

In this section the stationary case is treated.  Here,  (and also in Section $4$), for sake of simplicity, we consider a square box $[-\pi,\pi]^2$ instead of a rectangular box $[-\mu\pi,\mu\pi]\times [-\pi,\pi]$ and hence the $x$-derivative in the equation will have a factor $\mu$ in front. 

We write  $F_s=M( 1+\Phi^\e_s)$ 
in terms of a truncated expansion in the Knudsen number $\e$ plus a rest term:
$$\Phi^\e_s(x,z,v)=\sum_1^5\e^j\Phi_s^{(j)}(x,z,v)+\e R_{s,\e}(x,z,v).$$
The expansion will not be given explicitly here since all the ideas and details are in the previous paper \cite{AEMN}. We only remark  that the construction of the $\Phi_s^{(j)}$'s relies on the solution to a Milne problem with external force given in \cite{CME}.  Moreover, we will give in the next sections more details on the analogous expansion in the time dependent case.  In this section we study the equation for the remainder, beginning with some results on linear existence together with corresponding a priori estimates. The section ends with an existence proof for the nonlinear stationary rest term.

We denote by $H=L^2_M(\R^3)$ the Hilbert space of the measurable functions on $\R^3$ with inner product $(\,\cdot\, , \,\cdot\,)=(\,\cdot\, , \,M\cdot\,)_2$, where $(\,\cdot\, , \,\cdot\,)_2$ is  the standard $L^2$ inner product and $\|\,\cdot\,\pa_2$ the standard $L^2$-norm, while $\|\,\cdot\,\|$ is the norm corresponding to $(\,\cdot\, , \,\cdot\,)$. 

The linearized Boltzmann operator is defined, for any $f$ in a dense subset of $H$ as:
\begin{equation} Lf=2M^{-1}Q(M,Mf)\label{linbolop}.\end{equation}
It is well know that it can be decomposed as
$L=-\nu I +K$, where $I$ is the identity, $K$  a compact operator and  
$\nu(v)$, defined in the Introduction, 
is a smooth function
satisfying the estimates
$\nu_0 (1+|v|)\le \nu(v)\le \nu_1(1+|v|)$
for some positive $\nu_0$ and $\nu_1$. The operator $L$ is a non positive self-adjoint operator with domain ${\cal D}_L=\{f\in H\, |\, \|\nu^{\frac 1 2} f\| <+\infty\}$. 

The functions $\psi_{0}=1, \psi_{1} =\psi_x= v_{x}, \psi_{2}=\psi_y = v_y, \psi_{3}=\psi_z=v_z, \psi _{4} = \frac{1}{\sqrt {6}}(v^2-3)$ form an orthonormal basis for the kernel of ${L}$ in $H$, $\text{\tt Kern}(L)$. For any function in $H$ introduce the orthogonal splitting $f= f_{\parallel }+f_{\perp }:=Pf+(I-P)f$, where $f_\pa$ is called fluid dynamic part and is given by 
\begin{eqnarray*}
&&f_{\parallel }(x,z,v)=\sum_{j=0}^4 f_j\psi_j, \quad f_j:=(f,\psi_j), \quad j=,0,\dots,4,
\end{eqnarray*}
while the non hydrodynamic part $f_\perp$ satisfies
 $(f_\perp,\psi_j)=0, \quad j=,0,\dots,4.$
  $P$ denotes the projector from $H$ on $\text{\tt Kern}(L)$. Note that the range of $L$ is $(I-P)H$. We will use the same symbol for the projections also when dealing with functions depending on $x,z,t$.
We remind that the operator $L$ satisfies the  spectral inequality (\ref{specl}).

For $1\le q\le +\infty$, let $\tilde L^q$ be the Banach space of the measurable functions from $[-\pi,\pi]^2$ in $H$, identified with the space of measurable functions from $[-\pi,\pi]^2\times \R^3$ in $\R$ with norm
$$\pa f\|_{q,2}= \left(\int_{\R^3}d\/vM(v)\left(\int_{[-\pi,\pi]^2}d\/x d\/z|f(x,z,v)|^q\right)^{\frac 2 q}\right)^{\frac 1 2},$$
that is
$$\tilde L^q =  := \{f:[-\pi,\pi]^2\times \R^3\to \R\,|\, \parallel f\parallel_{q,2}<+\infty\}.$$
Moreover $(\,\cdot\, ,\,\cdot\,)_{2,2}$ is the corresponding inner product for $q=2$.
\vskip.2cm

We also need a function space for the boundary functions. We denote by $\R^3_\pm$ the sets $v=(v_x,v_y,v_z)$ such that $v_z\gtrless 0$. We consider the functions on $[-\pi,\pi]\times\{-\pi\}\times \R^3_+\cup [-\pi,\pi]\times\{\pi\}\times \R^3_-$ and define the norm
$$\pa f\pa_{q,2,\sim}=\sup_\pm \left(\int_{\R^3_{\pm}} dv |v_z|M(v) \left(\int_{[-\pi,\pi]}dx|f(x,\mp \pi, v)|^q\right)^{\frac 2 q}\right)^{\frac 1 2}.$$ 
The Banach space $L^+$ is the set of such functions with finite $\pa \, \cdot \,\pa_{2,2,\sim}$ norm. The ingoing and outgoing trace operators $\gamma^\pm$ are defined by 
$$\gamma^\pm f =\begin{cases} f|_{z=-\pi}, & \text{if } v\in \R^3_\pm ,\\f|_{z=\pi}, & \text{if } v\in \R^3_\mp .\end{cases}$$

The function space where the stationary solution will be constructed is the space
$${\cal W}^{q,-}:=\{f: [-\pi,\pi]^2\times \R^3\to \R\,|\, \nu^{\frac 1 2} f\in \tilde L^q, \ \nu^{-\frac 1 2} Df\in \tilde L^q,\ \gamma^+ f\in L^+\}.$$
Note that the norm $\parallel \,\cdot\,\parallel_{2,2,\sim}$ is defined only for incoming velocities. In the sequel, with an abuse of notation we will denote by $\parallel \gamma^- f\parallel_{2,2,\sim}$  the $\parallel \,\cdot\,\parallel_{2,2,\sim}$-norm of $S\gamma^- f$, where $S$ is the reflection of the $z$ component of the velocity.

\medskip

We do not explain here how to construct the  terms of the expansion $\Phi_z^{(j)}$. 
  We simply state a theorem about their  properties. We assume that the Rayleigh number $Ra$ is in $(Ra_c,Ra_c+\delta)$ with $\delta >0$ and sufficiently small and will consider, for sake of definiteness, the clockwise convective solution corresponding to it. 
\setcounter{theorem}{0}
\begin{theorem}
\label{fj0}
The functions $\Phi^{(n)}_s$, $n=1,\dots,5$ and $\psi_{n,\e}$ can be determined so as to satisfy the boundary  conditions
\begin{eqnarray*}
{ \Phi^{(n)}}(x,\mp\pi,v)&&=\frac{M_\mp(v)}{M(v)}\int_{w_z \lessgtr 0} |w_z|M[\Phi^{(n)}(x,\mp\pi,w)-\psi_{n,\e}(x,\mp\pi,w)]dw\\&&
+\psi_{n,\e}(x,\mp\pi,w),
\hspace*{0.05in}t>0, \hspace*{0.05in}v_z\gtrless0,\nonumber
\end{eqnarray*}
and the normalization condition
 $
\int_{\mathbb{R}^3\times[-\pi,\pi]^2}dv\/dx\/dz
\Phi^{(n)}=0,$ 
so that the asymptotic expansion in $\e$ for the stationary problem (\ref{BoltzEqn}), truncated to the order $5$ is given by
$$F^{(exp)}_s(x,z,v)=M(v)\left(1+\sum_{n=1}^5\e^5{ \Phi^{(n)}}(x,z,v)\right).$$
The functions ${ \Phi^{(n)}}$'s satisfy the conditions
$$\parallel \Phi^{(n)}\parallel _{2,2}<\infty,
\quad \parallel \Phi^{(n)}\parallel _{\infty,2}<\infty\ ,
\quad n=1,\dots,5.
$$
Moreover the ${ \Phi^{(n)}}$'s differ from those of the laminar solution  by $O(\delta)$.

\noindent The functions $\psi_{n,\e}$ are such that
$\|\psi_{n,\e}\|_{q,2,\sim}$, $q=2,\infty$ are exponentially small as $\e\to 0$ and $\int_{\R^3}dv v_z M(v) \psi_{n,\e}=0$.
Finally, there exists a stationary solution to (\ref{BoltzEqn}) in the form
$$F_s=F^{(exp)}_s+\e R_{s,\e}.$$
The remainder $R_{s,\e}$, simply denoted by $R$ solves the boundary value problem
\begin{equation}
\mu v_x{\frac\partial{\partial x}} R +v_z{\frac\partial{\partial z}} R -\e G M^{-1}\frac{\partial (MR)
}{\partial v_z}
= {\frac 1\e}{{L}} R + \sum_{i=1}^5\e^{j-1}J(\Phi^{(j)},  R)+
J(R,R) +  A\label{stR} 
\end{equation}
\begin{eqnarray}\label{bcR}
&&\hskip -1cm R(x,\mp\pi,v)=\frac{M_\mp(v)}{M(v)}\int_{w_z\lessgtr 0}|w_z|M(w) \Big(R(x,\mp \pi,w)+\frac{1}{\e}\bar{\psi}_\e(x,\mp\pi,w)\Big)dw-\frac{1}{\e}\bar{\psi}_\e(x,\mp\pi,v),\nonumber\\
&&\hskip 4cm\text{for } v_z\gtrless 0\  \text{ and } x\in [-\pi,\pi]
\end{eqnarray}
where $\frac 1\e\bar\psi_\e=-\sum_{n=1}^5\e^n\psi_{n,\e},$
 $\displaystyle{J(h,g)=\frac{2}{M}Q(M h,M g)}$ 
and $A$ is a smooth function bounded in $\|\,\cdot\,\|_{q,2}$, $q=2,\infty$, such that $\displaystyle{\int_{\R^3} dv M(v) A=0}$.
\end{theorem}
We do not give here the explicit expression of $A$ for which we refer to \cite{AEMN}. It has to be considered as a known term in the rest of this section as well as the $\Phi^{(j)}$'s. The terms of the expansions contribute to $A$ together with their space and velocity derivatives. It suffices to know that it has finite $\|\ \cdot\|_{q,2}$-norm. Note that $M$ and $M_-$ differ just by the normalization and that the $\psi_{n,\e }$'s are known functions exponentially small in $\e$ due to the boundary layer corrections in the expansion. With  these boundary conditions the remainder satisfies the impermeability condition 
\begin{equation}\int_{\R^3}dv v_zM R=0\label{imperm}\end{equation}
at the walls.
\medskip

Now we start the analysis of the equation for the remainder  $R$. 
 The first important result is a spectral property for the operator  $L_J$ below: fix $x,z$ and define, for each $f\in H$
\begin{equation}L_Jf =Lf +\e N P f\ , \end{equation}
with 
$$N=J(q,\,\cdot\,),\quad q=\sum_{n=1}^5\e^{n-1}\Phi^{(n)}_s.$$
Note that in the time-dependent case the form of $q$ will be slightly different but without affecting the argument below.  

The operator $L_J$ is not symmetric in $H$. We denote by $L^*_J$ its adjoint.
To characterize the kernel of $L_J$,  $\text{\tt Kern}(L_J)$, note that the  functions 
 \begin{equation}
 \bar \psi_j=\psi_j-\e L^{-1}N\psi_j,
 \label{bpsi}
 \end{equation}
 are in $\text{\tt Kern}(L_J)$, where, for $\in (I-P)H$,  $L^{-1}f$ denotes the unique solution of $Lg=f$ orthogonal to $\text{\tt Kern}(L)$. 
In fact,
 $L_J\bar\psi_j=L\psi_j-\e N P\psi_j+\e N P\psi_j-\e ^2N P[L^{-1}N P\psi_j]=0$,
 because $\psi_j$ are in $\text{\tt Kern}(L)$ and the last term is zero since
 $P$ kills the terms in the range of $L^{-1}$.

It is easy to check, by using (\ref{specl}),  that, at least for $\e$ sufficiently small, they actually span $\text{\tt Kern}(L_J)$. 
 Indeed, suppose that there is  $g\in \text{\tt Kern}(L_J)$ with $\|g_\perp\| =1$ such that $(g, \bar \psi_j)=0$. This implies $g_\pa= \e \sum_{j=0}^4 (g_\perp, L^{-1}N\psi_j)\psi_j$.  Moreover,  $0=(g,L_Jg)=(g_\perp,Lg_\perp)+\e(g_\perp,Ng_\pa)$. 
Therefore by (\ref{specl}), $C\le -(g_\perp,Lg_\perp)=\e^2\sum_{j=0}^4(g_\perp,N\psi_j)(g_\perp,L^{-1}N\psi_j)\le \alpha\e^2$, for some positive $\alpha$ independent of $\e$. This is in contradiction with $C>0$ for $\e$ sufficiently small. 
  
Let $P_J$ denote the orthogonal  projection onto $\text{\tt Kern}(L_J)$. Since the range of $L_J$ is $(I-P)H$, it follows that $\text{\tt Kern}(L_J^*)=PH=\text{\tt Kern}(L)$.

Consider the space $\mathcal N$ generated by $\{\psi_0,\dots,\psi_4, L^{-1}N\psi_0,\dots,L^{-1}N\psi_4\}$.
 We decompose this  space into $\text{\tt Kern}(L_J)$ and its orthogonal complement in  $\mathcal N$, $L_1$. 
 
Moreover,   $(I-P)H$  can be decomposed in the span of $\{N\psi_1, \dots, N\psi_4\}$ and its orthogonal complement, $\bar L_o$. It is easy to check that
$$
{ H=\text{\tt Kern}(L_J)\oplus L_1\oplus \bar L_o}.
$$
Indeed it is enough to show that $N\psi_\ell, \ell=0,\dots,4$ is a combination of $L^{-1} N\psi_j$'s and  $u\in\bar L_o$: 
 $$N\psi_\ell=\sum_{j=0}^4\alpha_{\ell,j} L^{-1} N\psi_j+ u, \quad (N\psi_k,u)=0, \quad \ell,k=0,\dots,4.$$
 We take the inner product with $N\psi_k$ 
 $$(N\psi_k, N\psi_\ell)=\sum_{j=0}^4\alpha_{\ell,j} (N\psi_j,L^{-1} N\psi_k).$$
 The matrix with elements $(N\psi_j,L^{-1} N\psi_k)$ is non singular, hence the $\alpha_{\ell,j}$ are uniquely determined, and $u$ as a consequence.

\begin{proposition} [Spectral gap property of $L_J$] 
\label{2.2}
There is $\e_0>0$ such that, for $0<\e<\e_0$, there is $c$ independent of $\e$, $(x,z)$ and $\delta$, for which the following inequalities hold:
\begin{eqnarray} 
&&{ -(L_J\varphi,\varphi)\geq c(\nu(I-P_J)\varphi,(I-P_J)\varphi)},\\ &&{ -(L^*_J\varphi,\varphi)\geq c(\nu(I-P)\varphi,(I-P)\varphi)}
\label{spgap}
\end{eqnarray}
\end{proposition}
\noindent\underline{Proof}.  First take $(x,z)$ and the Rayleigh number fixed. It is then enough to consider the set ${\cal H}$ of all functions ${\varphi=(I-P_J)\varphi
= a\varphi_o+b{\varphi_1}} $, where $\varphi_1 \in L_1$ and $ \varphi_o
=(I-P)\varphi_o$ $\in \bar{L}_o$, and with $\nu^{\frac{1}{2}}{\varphi_o}$ and $\varphi_1$ of norm one and ${ a^2+b^2=1}$.
First notice that $(\nu\varphi,\varphi)$ is uniformly bounded in ${\cal H}$.
It remains to show that the left-hand side has a positive bound from below. But $\varphi_1$ can be decomposed as a sum of an element $\varphi_{11}$ in $\text{\tt Kern}(L)$ with norm of order $\e$, and an element ${L}^{-1}N\varphi_{12}$ in the span of $\{{L}^{-1}N\psi_0,...,{L}^{-1}N\psi_4\}$. Then
\begin{eqnarray*}
&& -(L_J\varphi,\varphi)= -\Big(a L\varphi_o+bN\varphi_{12}+b\e N\varphi_{11},a\varphi_o+b\varphi_{11}
 +bL^{-1}N\varphi_{12}\Big)\\ &&=-a^2(L\varphi_o,\varphi_o)
 -b^2(N\varphi_{12}, L^{-1} N\varphi_{12})-\e b^2(N\varphi_{11}, L^{-1} N\varphi_{12})
\\&&\geq C(a^2(\nu\varphi_o,\varphi_o)
+b^2(L^{-1} N\varphi_{12},L^{-1}N\varphi_{12}))
+\e b^2(N\varphi_{11}, L^{-1} N\varphi_{12}).
\end{eqnarray*}
The first equality follows from the fact that $\varphi_{11}$ is in $\text{\tt Kern}(L)$ and is orthogonal to the range of $L$; the second equality is due to the fact that $(\varphi_0,N\varphi_{12})=0=(\varphi_0,N\varphi_{11})$ by the definition of $\bar L_o$; the bound (\ref{specl}) has been used to obtain the inequality. Since the last term is of order $\e ^2$, it follows that for $\e>0$ and small enough
\begin{eqnarray*}
-(L_J\varphi,\varphi)\geq \frac{C}{2}( a^2(\nu \varphi_{\perp},\varphi_{\perp})+b^2({L}^{-1}N\varphi_{12},{L}^{-1}N\varphi_{12}) )\geq c
\end{eqnarray*}
for some $c>0$. The first inequality in the proposition then follows, since the constant depends continuously on $(x,z)$ and the Rayleigh number is in a compact set. The second inequality is obtained by similar arguments. 
$\square$

\bigskip

We only consider Rayleigh numbers in a suitably small neighbourhood of the first bifurcation point, and take $G$ sufficiently small as specified in the sequel.
Constants which, independently of the parameter $\e$, can be made sufficiently small for the purposes of the proofs, will generically be denoted $\eta$. We will first give  estimates in the linear case. The argument is inspired by  the approach in \cite{Ma} and heavily relies on the  study of the space Fourier transform of $R$. 
We use the following definition of Fourier transform:  for any $\xi\in \Z$
$${\cal F}f(\xi)=\frac{1}{2\pi}\int_{[-\pi,\pi]} dx \text{\rm e}^{-i\xi x}f(x).$$

When we want to specify that the Fourier transform is taken with respect to the variable $x$ we write ${\cal F}_x$. In the sequel, if $f$ is a function of $(x,z)\in[-\pi,\pi]^2$, $\hat{f}(\xi_x,\xi_z)=({\cal F}_x{\cal F}_z f)(\xi_x,\xi_z)$. Finally, if $f$ is a function of $x$, $z$ and $v$, we define  $<f>$ as the zero order Fourier coefficient of $f$, that is   
\begin{eqnarray}
<f>:=\frac{1}{(2\pi)^2 }\int _{[-\pi ,\pi ]^2}f(x,z,v)dxdz,\quad a.a. \hspace*{0.05in} v\in \R^3
\end{eqnarray}
and $\tilde{f}: =f\ -<f>$.

An important tool in the analysis is the Green inequality, which we will use extensively in the rest of the paper in various different situations.

Consider the linear boundary value problem
\begin{eqnarray}\label{3.333}
  \mu v_x \frac{\partial f}{\partial x} +v_z \frac{\partial f}{\partial z} -\e G M^{-1}\frac{\partial 
(Mf )}{\partial v_z}= \frac{1}{\e }{L_J}f +g,
\end{eqnarray} 
with $\int Mgdv= 0$ and prescribed incoming data
\begin{equation}f(x,\pm\pi,v)=p(x,\pm\pi,v)\quad v_z\lessgtr0\end{equation}
Due to the presence of the force $G$, the argument requires some care.  We introduce 
$\displaystyle{\kappa(z)=\text {\rm e}^{\e G(z+\pi)}}$.
Then we multiply (2.10) by $2f M\kappa $, integrate over $[-\pi,\pi]^2\times\R^3$, and integrate by parts
to get the Green identity
\begin{equation}
\parallel \kappa^{\frac 1 2} \gamma^-{f} \parallel^2_{2,2,\sim}-\frac 1 \e(\kappa f,L_Jf)_{2,2}=(\kappa g ,f)_{2,2}+\|\kappa^{\frac 1 2}p\|_{2,2,\sim}^2.  
\end{equation}
Apply the spectral inequality (2.7) to obtain, for small $\eta$
\begin{eqnarray}
&&\parallel \kappa^{\frac 1 2} \gamma^-{f} \parallel^2_{2,2,\sim}+\frac{c}{2\e }\parallel \kappa^{\frac 1 2} \nu ^{\frac{1}{2}}(I-P_J){f} \parallel _{2,2}^2 \nonumber\\
&&\leq c( \e    \parallel   \kappa^{\frac 1 2} \nu ^{-\frac{1}{2}}(I-P_J){g}\parallel _{ 2,2}^2+\eta \parallel  \kappa^{\frac 1 2} P_J{f} \parallel _{2,2}^2+\frac{1}{\eta }\parallel   \kappa^{\frac 1 2} P_J{g}\parallel _{2,2}^2+\|\kappa^{\frac 1 2}p\|_{2,2,\sim}^2.
\label{due0}
\end{eqnarray}
Finally, since $1\le \kappa(z)\le \displaystyle{\text {\rm e}^{2\e \pi G}}$, we conclude that, for some constant $C$ it results: 
\begin{eqnarray}
&&\parallel  \gamma^-{f} \parallel^2_{2,2,\sim}+\frac{c}{2\e }\parallel \nu ^{\frac{1}{2}}(I-P_J){f} \parallel _{2,2}^2 \nonumber\\
&&\leq C\Big(\e \parallel \nu ^{-\frac{1}{2}}(I-P_J){g}\parallel _{ 2,2}^2+\eta \parallel P_J{f} \parallel _{2,2}^2+\frac{1}{\eta }\parallel  P_J{g}\parallel _{2,2}^2+\|p\|_{2,2,\sim}^2\Big).
\label{due}
\end{eqnarray}
Inequality (\ref{due}) and its variations will be referred to in the rest of the paper as Green inequality and used extensively, with $L^*_J$ replacing sometimes  $L_J$. In the last case the projector in the r.h.s. has to be replaced by $P_J^*=P$. 

\setcounter{theorem}{2}
\begin{lemma}
\label{2.3}
Let $\varphi(x,z,v)$ be solution to 
\begin{eqnarray}\label{3.2}
  \mu v_x \frac{\partial \varphi}{\partial x} +v_z \frac{\partial \varphi}{\partial z} -\e G M^{-1}\frac{\partial 
(M\varphi)}{\partial v_z}= \frac{1}{\e }{L_J^*}\varphi +g,
\end{eqnarray} 
periodic in $x$ of period $2\pi$, and with zero ingoing boundary values
at $z=-\pi,\pi$.  Then, for some small $\eta$
\begin{eqnarray}
\parallel \nu ^{\frac{1}{2}}(I-P){\varphi} \parallel _{2,2}&&\leq C\Big( {\e} \parallel \nu ^{-\frac{1}{2}}(I-P){g}\parallel _{ 2,2}+\parallel P{g}\parallel _{2,2}+\eta\e\parallel<P\varphi>\parallel_{2}\Big),\label{2.12}\\
\parallel\widetilde{ P\varphi} \parallel _{2,2}&&\leq C\Big( \parallel \nu ^{-\frac{1}{2}}(I-P){g}\parallel _{2,2}+\frac{1}{\e}\parallel P{g}\parallel _{2,2}+\eta\parallel<P\varphi>\parallel_{2}\Big) .\hspace*{0.03in}
\label{2.13}
\end{eqnarray}

\end{lemma}

\hspace{1in}\\
\noindent\underline{Proof of Lemma \ref{2.3}.}\hspace{0.05in} The method from \cite{Ma} (a variant of \cite{Ma} Scn 5.3) can be adapted to the present setting to obtain the existence of a solution to (\ref{3.2}), if one includes the above spectral estimate for $L_{J}^*$, and the new characteristics curves due to the force term.

We write (\ref{3.2}) in Fourier variables. For $\xi\in\Z^2-\{ (0,0)\}$,
\begin{eqnarray}
i(\mu \ \xi_xv_x+\xi_zv_z)\hat{\varphi}= \frac{1}{\e }\widehat{{L}_{J}^*{\varphi}}+\e G M^{-1} \frac{\partial (M \hat{\varphi}) }{\partial v_z} +\hat{g}- (-1)^{\xi_z}|v_z|r.
\label{ft}
\end{eqnarray}
Here 
\begin{equation}
r=\begin{cases}{\cal{F}}_x\varphi(\xi_x,\pi,v)\ & \hbox{for }\  v_z>0,\\ {\cal{F}}_x\varphi(\xi_x,-\pi,v)\ & \hbox{for }\  v_z<0.
\end{cases}
\label{r}
\end{equation} 
Introduce 
$$v^{\mu }= (\mu v_x, v_z),   \quad \Phi = \hat{\varphi },\quad  
 \tilde{Z}= \e ^{-1} \widehat{L_{J}^* {\varphi }}+
\displaystyle{\e G M^{-1} \frac{\partial (M \hat{\varphi}) }{\partial v_z}}
\displaystyle{+\hat{g} -(-1)^{\xi_z}}|v_z|r,$$  $$\displaystyle{ Z= \e ^{-1} \widehat{L_{J}^*{\varphi }}+\hat{g} -(-1)^{\xi_z}}|v_z|r, \quad \displaystyle{ Z'= \e ^{-1} \widehat{L_{J}^* {\varphi }}+\hat{g}},\quad \displaystyle{ \hat{U}= (i\xi \cdot v^{\mu})^{-1}}.$$
Let $\chi $ be the indicatrix function of the set 
$
\{ v\in \R^3\ | \   \mid \xi \cdot v^{\mu}\mid <\alpha |\xi| \} 
$,  
for some positive $\alpha $ to be chosen later.
 Let $\zeta_s(v)= (1+\mid v\mid )^s$. For $\xi \neq (0,0) $
\begin{eqnarray*}
\parallel P(\chi \Phi )\parallel  &&\leq c \sum_{j=0}^4\left|\int_{\R^3}dv \chi(v)\Phi(\xi,v)M(v)\psi_j(v)\right|\\
&&\leq c\parallel \zeta_{-s} \chi\Phi \parallel  \sum_{j=0}^4 \parallel \chi \zeta_s{\psi}_j\parallel   \leq c\sqrt{{\alpha }}\parallel \zeta _{-s}\chi \Phi \parallel.
\end{eqnarray*}
We use this estimate with the following choice of $\alpha$,
$ 
\alpha = \parallel \zeta_{-s}\Phi \parallel ^{-1}\parallel \zeta_{-s}Z'\parallel $.

We also introduce  an indicatrix function $\chi_1$ with $\alpha=\sqrt{\delta_1}$. We fix $\delta_1 $ so that $c\sqrt \delta_1 << 1$. Then we find from the above estimate that the
$P$-part of the right-hand side, $\parallel P( \chi_1\Phi )\parallel $, can be  absorbed by  $\parallel P(\chi_1 \Phi) \parallel $ in the left-hand side. The estimates hold in the same way when  $\chi_1$ is suitably smoothed around  $\sqrt{\delta_1} |\xi|$. For the remaining $\chi^c\chi_1^c\Phi=(1-\chi)(1-\chi _1)\Phi$ we shall use that
$\Phi = -\hat{U}\tilde{Z}$. Then
\begin{eqnarray*}
\parallel P(\chi^c_1\chi^c\Phi )\parallel ^2 &&\leq c\parallel \zeta _{s+2}\chi^c_1\chi^c\hat{U}\parallel ^2 
\parallel \zeta _{-s}Z'\parallel ^2 +\frac{c\parallel  \sqrt{|v_z|} r \parallel ^2}{\delta_1|\xi|^2}
+ \e G  |\Theta| \\
&&\leq \frac{c}{\mid \xi \mid ^2 \mid \alpha \mid }
\parallel \zeta _{-s}Z'\parallel ^2  
+\frac{c\parallel  \sqrt{|v_z|} r\parallel ^2}{\delta_1|\xi|^2}+ \e G|\Theta|,
\end{eqnarray*}
with
$$\Theta=\sum_{j=0}^4\int {\psi }_j\chi^c_1\chi^c \hat{U}\frac{\partial 
(M \hat{\varphi}) }{\partial v_z}dv \left( \int {\psi} _j\chi^c_1 \chi^c(\hat{\varphi}  -\hat{U} Z)M dv\right)^*.$$
We replace $\alpha $ by
 $\parallel \zeta_{-s}\Phi \parallel  ^{-1}\parallel \zeta _{-s}Z'\parallel  $ in the denominator.  That gives
\begin{eqnarray*}
 && \displaystyle{\parallel P\Phi \parallel ^2 \leq c(\parallel \zeta _{-s}\Phi \parallel  \parallel \zeta _{-s}Z'\parallel   +\frac{\parallel \sqrt{|v_z|} r\parallel ^2}{\delta_1|\xi|^2}+\delta_1\parallel \zeta_{-s}(I-P)\Phi \parallel  ^2)}+ \e G|\Theta|.
\end{eqnarray*}
Hence,
\begin{eqnarray*}
\parallel P\Phi \parallel ^2  &&\leq c\Big( (\parallel P\Phi \parallel  +\parallel \zeta _{-s}(I-P)\Phi \parallel   )\parallel \zeta _{-s}Z'\parallel  +\frac{\parallel \sqrt{|v_z|} r\parallel ^2}{\delta_1|\xi|^2}\\ &&+\delta_1\parallel \zeta_{-s}(I-P)\Phi \parallel ^2\Big )+ \e G|\Theta|. 
\end{eqnarray*}
Consequently,
\begin{eqnarray*}
 \parallel P\Phi \parallel ^2 &&\leq c\Big(  \parallel \zeta _{-s}Z'\parallel ^2 +\frac{\parallel \sqrt{|v_z|} r\parallel ^2}{\delta_1|\xi|^2}+\parallel \zeta _{-s}(I-P)\Phi \parallel  \parallel \zeta _{-s}Z'\parallel   \\
&& +\parallel\zeta_{-s}(I-P)\Phi\parallel  ^2 \Big)+ \e G|\Theta|.
\end{eqnarray*}
We next discuss the term $\e G |\Theta|$. The first integral can be bounded by $\e $ times an integral of a product of $M$, $1+|\xi_z |$, a polynomial in $v $, $\mid  \hat \varphi \mid $ and $\hat U$ or $\hat U ^2$. So this factor is bounded by $\e c \parallel \Phi \parallel   $. And so,
\begin{eqnarray*}
\parallel P\Phi \parallel  ^2\leq c\Big( \parallel \zeta _{-s}Z'\parallel  ^2+\frac{\parallel \sqrt{|v_z|} r\parallel ^2}{\delta_1|\xi|^2}+\parallel (I-P)\Phi \parallel  ^2\Big) .\\
\end{eqnarray*}
Therefore for $\xi\neq (0,0)$,
\begin{eqnarray}
\int |P\Phi |^2(\xi ,v)Mdv 
&&\leq c\Big( \frac{1}{\e ^2} \parallel \zeta_{-s}(v)\widehat{{L_J^*}\varphi }( \xi , \cdot )\parallel  ^2
+\parallel (I-P)\Phi (\xi ,\cdot )\parallel  ^2  \nonumber\\
&&+\frac{\parallel  \sqrt{|v_z|} r\parallel ^2}{\delta_1|\xi|^2}+ \parallel 
\nu^{-\frac{1}{2}}\hat{g} (\xi ,\cdot )\parallel  ^2 \Big) \label{3.3}.
\end{eqnarray}
 We remember that  the zero Fourier mode of $\widetilde{P\varphi }$ is zero by definition. Hence, 
 taking $\e$ small enough and summing over all $ 0\neq\xi \in  Z^2 $, implies by Parseval and the spectral estimate on $L_J^*$, that 
\begin{eqnarray}
\int (\widetilde{P\varphi })^2(x,z,v)Mdvdxdz 
\leq c\Big( \frac{1}{\e ^2}\int \nu ((I-P){\varphi} )^2(x ,z,v)Mdvdxdz \nonumber \\
+\int \nu ^{-1}{g}^2(x,z,v)Mdvdxdz+\parallel \gamma^-{\varphi}\parallel^2_{2,2,\sim}\Big).
\label{uno}
\end{eqnarray}
Inequality  (\ref{due}) implies (\ref{2.12}). Replacing (\ref{due}) in (\ref{uno}) gives (\ref{2.13}), and  this concludes the proof of the Lemma \ref{2.3}.  $\square$\\

\noindent{\bf Remark.} The statement of Lemma \ref{2.3} still holds if we replace the operator $L_J^*$ with the operator $L_J$ and the operator $P$ with $P_J$, with some minor modification. The main change is due to the fact that one has to use the basis functions in $\text{\tt Kern}(L_J)$ namely the $\bar \psi_j$'s, instead of the $\psi_j$'s. They depend on $(x,z)$, therefore we fix a point $(x_0,z_0)$ and use the $\bar \psi_j$'s computed at this point and the corresponding projector $P_{J0}$. Then the argument of the proof can be repeated word by word. At the end, we replace $P_{J0}$ with $P_J$ since $P_{J}-P_{J0}=O(\e)$. 

\hspace{1cm}\\
Put $H(R)=\sum_{n=1}^5 \e^{n-1}J(\Phi^{(n)}_s,R)$ and decompose $H$ in accordance with the operator $L_J$. Set $H_1(\,\cdot\,)=H(\,\cdot\,)-{J}(q,P\,\cdot\,)$. We notice that ${H_1}(\cdot)$ is of order zero in $\e$ and only depends on the nonhydrodynamic projection $(I-P)$. 

 To pass from the linear results to the non linear case we will use an iteration procedure that will lead to  Theorem \ref{2.9} below. To separate the difficulties coming from the non linear term and from the boundary conditions, we split the remainder $R$ in two parts, $R_1$ and $R_2$, solutions   of two different equations. In the equation   for $R_1$ the boundary conditions are of given indata type and the nonhydrodynamic known term is included, while  in the equation for $R_2$ the  boundary conditions are of  diffusive type and the  known term is absent. The equation for $R_2$ will be given later (see eq. (\ref{2.17})). We start with a discussion of the equation for $R_1$, 
\begin{eqnarray}
\label{eqnR_1}
&&\mu\ v_x\frac{\partial R_1}{\partial x}+v_z\frac{\partial R_1}{\partial z}-\e GM^{-1}\frac{\partial 
(M  R_1)}{\partial v_z}= \frac{1}{\e }{L}_JR_1+H_1(R_1)+g,\\
&&R_1(x,\mp\pi,v)=-\frac{1}{\e}\bar{\psi}(x,\mp\pi,v),\quad v_z\gtrless0.\nonumber \end{eqnarray}
Here $R_1$ is periodic in $x$ of period $2\pi$, and ${L}_J={L}(\,\cdot\,)+\e {J}(q,P\,\cdot\,)$ has been introduced earlier. 
  An existence proof for this problem can be obtained similarly to that for $\varphi$ above.\\
The nonhydrodynamic part of ${R_1}$ is estimated similarly to the corresponding proof for Lemma \ref{2.3},\begin{eqnarray*}
&&\frac{1}{\e}\parallel \gamma^-{R}_1\parallel^2_{2,2, \sim}+\frac{c}{2\e ^2}\parallel \nu ^{\frac{1}{2}}(I-P_J){R}_1\parallel _{2,2}^2\\
&&\leq C\Big( \parallel \nu ^{-\frac{1}{2}}(I-P_J){g}\parallel _{2,2}^2+\frac{\eta }{2\e }\parallel P_J{R}_1\parallel _{2,2}^2+\frac{1}{2\eta \e }\parallel P_J{g}\parallel _{2,2}^2+\frac{1}{\e ^3}\parallel {\bar{\psi}}\parallel^2_{2,2,\sim}\Big) ,
\end{eqnarray*}
for small $\eta >0$. Here we have used the fact that 
\begin{eqnarray*}
|(R_1,H_1(R_1))_{2,2}|\le C(\parallel \nu ^{\frac{1}{2}}(I-P_J){R}_1\parallel _{2,2}^2+\e ^2\parallel P_J{R}_1\parallel _{2,2}^2).
\end{eqnarray*}
A priori bounds for $P_J R_1 $ will be based on dual techniques involving the problem (\ref{3.2}). Consider first the problem for $R_1$ without the term $H_1(R_1)$. It holds

\begin{lemma}
\label{2.4}
Set $h:=P_JR_1$. Then there is $\delta _0>0$, such that for $0<\delta <\delta _0$,
\begin{equation}
\parallel h \parallel ^2_{2,2} \leq  C ( \parallel \nu^{-\frac{1}{2}}(I-P_J){g} \parallel ^2_{2,2}+\frac{1}{\e ^2}\parallel P_J{g} \parallel ^2_{2,2}+\frac{1}{\e^3}\parallel{\bar{\psi}}\parallel^2_{2,2,\sim}).\nonumber
\end{equation}
\end{lemma} 
\noindent \underline{Proof of Lemma \ref{2.4}.}\hspace*{0.05in}The function $R_1$ is $2\pi$-periodic in $x$, and here solution to
\begin{eqnarray}\label{2.20}
&&\mu \ v_x  \frac{\partial R_1}{\partial x} +v_z  \frac{\partial R_1}{\partial z} -\e  G M^{-1}\frac{\partial 
(M R_1)}{\partial v_z} = \frac{1}{\e }{L}_J  R_1 +g,\label{3.5}\\
&&R_1(x,\mp\pi,v)=-\frac{1}{\e}\bar{\psi}(x,\mp\pi,v),\quad v_z\gtrless0.\nonumber\end{eqnarray}
Let $\varphi $ be a $2\pi$-periodic function in $x$, solution to
\begin{eqnarray*}
\mu\ v_x\frac{\partial \varphi}{\partial x} +v_z \frac{\partial \varphi}{\partial z} -\e  G M^{-1}\frac{\partial 
(M\varphi)}{\partial v_z} = \frac{1}{\e }{L_J^*}\varphi +h,
\end{eqnarray*}
with zero ingoing boundary values  at $z=-\pi,\pi$. 

We consider the equation for $R_1$ multiplied by $M\kappa \varphi $ and the one for $\varphi $ multiplied by $M\kappa R_1$ and add them. After integrating on $[-\pi,\pi]^2\times \R^3$ and integrating by parts,  we obtain,
\begin{eqnarray*}
&&\int dxdz dvM\kappa \Big(\frac{\partial}{\partial z}(v_z{R}_1{\varphi })-\e G\frac{\partial
(R_1\varphi)}{\partial v_z}\Big)=\int dxdzdv{M\kappa }\Big[ \frac{1}{\e }((I-P){\varphi}{L_J}(I-P_J)R_1)\\
&&\qquad\qquad+ \frac{1}{\e }(I-P_J)R_1L_{J^*}(I-P){\varphi} 
+\kappa {g}{\varphi} +\kappa h{R}_1\Big].
\end{eqnarray*}

Using again the bound $1\le\kappa\le \text{\rm e}^{2\pi\e G}$ and the assumption $h=P_JR_1$, this gives 
\begin{eqnarray*}
&&\parallel h\parallel _{2,2}^2\leq 
\frac{K_1}{2}\parallel \gamma ^-R_1\parallel _{2,2,\sim}^2+\frac{1}{2K_1}\parallel \gamma ^-\varphi\parallel _{2,2,\sim}^2+\frac{K_3}{\e }\parallel \nu ^{\frac{1}{2}}(I-P_{J}){R}_1\parallel _{2,2}^2\\
&&+\frac{1}{K_3\e }\parallel \nu ^{\frac{1}{2}}(I-P){\varphi} \parallel _{2,2}^2 
+\frac{K_4}{2} \parallel \nu ^{-\frac{1}{2}}(I-P_{J}){g}\parallel _{2,2}^2+ \frac{1}{2K_4}\parallel \nu ^{\frac{1}{2}}(I-P){\varphi} \parallel _{2,2}^2\\
&&+\frac{\e^2}{2K_4}\parallel \nu^{\frac{1}{2}}\varphi\parallel^2_{2,2}+\frac{K_2}{2}\parallel P_{J}{g}\parallel _{2,2}^2+\frac{1}{2 K_2}\parallel { P\varphi} \parallel _{2,2}^2+\frac{\e^2}{2K_2}\parallel \nu^{\frac{1}{2}}\varphi\parallel^2_{2,2},
\end{eqnarray*}
for arbitrary positive constants $K_j$, $j=1,...,4$. 
It then follows  that
\begin{eqnarray*}
&&\parallel h \parallel _{2,2}^2\leq c[(\frac{K_1}{\e^2}+\frac{K_3}{\e ^2})\parallel{\bar{\psi}} \parallel_{2,2,\sim}^2+(\e K_1+K_4+K_3\e) \parallel \nu^{-\frac{1}{2}}(I-P_J){g} \parallel _{2,2}^2\\ 
&&+(\frac{1}{\e K_1}+\frac{1}{\e ^2 K_2}+\frac{1 }{\e K_3}+\frac{1}{ K_4}+\frac{\eta_1}{\e^2 K_1} )\parallel h\parallel _{2,2}^2+(\frac{K_1}{ \e}+\frac{K_3}{\eta _1} +K_2)\parallel P_J{g}\parallel _{2,2}^2 \\
&&+({\eta_1K_1}+\eta_1 K_3 )\parallel h\parallel ^2_{2,2} +\eta (\frac{\eta_1}{K_1}+\frac{\e}{K_3}+\frac{\e^2}{K_4}+\frac{1}{K_2})\parallel<P\varphi>\parallel_{2}^2.
\end{eqnarray*}
For an estimate of the final $<P\varphi>$-term when $G$ is small and the Rayleigh number for the rolls lies in a neighbourhood of the bifurcation point, we may apply an exact, direct approach based on ordinary differential equations. Namely, $<\varphi>_x(\cdot):= \int \varphi(x,\cdot)dx$ satisfies a $1$-d stationary laminar given indata problem, similar to eq. (3.5) in \cite{AEMN}.  The present case is different from the one in   \cite{AEMN} because there will be new contributions  coming from the terms in the expansion depending on $x$, for example the term $q$ in the ${J}$-term. Since the first order  of the $\e$ expansion is of order $\delta$, the same is true for the higher order (in $\e$)  terms. The new contributions can all be considered as deviations from their $x$-independent values at the bifurcation point, hence are of order $\delta $. We include them in the estimate as $\eta \parallel \varphi \parallel_{2,2}$, $\eta$ small, and obtain  \begin{eqnarray*}
\parallel <P\varphi>\parallel _{2}&&\leq c\parallel<P\varphi>_x\parallel_{2,2}\leq c(\parallel<P\varphi>_x\parallel_{2,2}+\e\parallel \varphi \parallel_{2,2})\\
&&\leq
\frac{c}{\e}\parallel<h>_x\parallel_{2,2}+\eta\parallel \varphi \parallel_{2,2}
\leq \frac{c}{\e}\parallel h \parallel_{2,2}+\eta\parallel\varphi\parallel_{2,2}.
\end{eqnarray*}
For details cf Lemma 3.5 in  \cite{AEMN}.\\
Choosing $\e <<1$, then $K_1$ and $K_3$ (resp.$K_2$) of order $\e ^{-1}$ (resp. $\e ^{-2}$), $\eta _1$ of order $\e$, and using Lemma \ref{2.3}, the inequality of  Lemma \ref{2.4} follows. $\square$\\

\medskip

\noindent{\bf Remark.} From here on,  small factors  $\eta $ in the estimates will  depend also on  the  small $\delta _0$.
\medskip

In the following lemma we get the final estimates for $R_1$.
\begin{lemma}
\label{2.5}
If  $R_1$ is a solution to the system $(\ref{eqnR_1})$, then
\begin{eqnarray*}
\parallel \nu^{\frac{1}{2}}R_1\parallel _{2,2}
&&\leq c\Big( \parallel \nu ^{-\frac{1}{2}}(I-P_J)g\parallel _{2,2}+\frac{1}{\e }\parallel P_Jg\parallel _{2,2}
+\e^{-\frac{3}{2}}\parallel\bar{\psi}\parallel_{2,2,\sim}\Big) , \\
\parallel \nu^{\frac{1}{2}}R_1\parallel _{\infty ,2}
&&\leq c\Big( \frac{1}{{\e} }\parallel \nu ^{-\frac{1}{2}}(I-P_J)g\parallel _{2,2}
+\frac{1}{\e ^2 }\parallel P_J g \parallel _{2,2} 
+\e \parallel \nu^{-\frac{1}{2}}g\parallel_{ \infty, 2}
+\e^{-\frac{5}{2}}\parallel\bar{\psi}\parallel_{2,2,\sim}\Big) .
\end{eqnarray*}
\end{lemma}
\noindent\underline{Proof of Lemma \ref{2.5}.}\hspace*{0.05in} Consider first the solution  to $(\ref{eqnR_1})$ with $H_1= 0$. 
It satisfies
\begin{eqnarray*}
&&\parallel \gamma ^-R_1\parallel _{2,2,\sim}+\e ^{-\frac{1}{2}}\parallel \nu ^{\frac{1}{2}}(I-P_J)R_1\parallel _{2,2}\leq \\
&&c\Big( \frac{1}{\e}\parallel \bar{\psi}\parallel_{2,2,\sim}+\e^{\frac{1}{2}}\parallel \nu^{-\frac{1}{2}}(I-P_J)g\parallel _{2,2}+\eta_1 \parallel P_JR_1\parallel _{2,2}+\frac{1}{\eta_1 }\parallel P_Jg\parallel _{2,2}\Big) ,
\end{eqnarray*}
for any $\eta_1 >0$. By Lemma 2.4 and some additional computations using the solution formula, 
\begin{eqnarray*}
\parallel \nu^{\frac{1}{2}}R_1\parallel _{\infty ,2}\leq c\Big( \frac{1}{\e }\parallel \nu^{\frac{1}{2}} R_1\parallel _{ 2,2}+\e \parallel \nu ^{-\frac{1}{2}}g\parallel _{\infty ,2}+\parallel \gamma ^+R_1\parallel_{2,2,\sim}\Big) .
\end{eqnarray*}
It is easy to see that adding the term $H_1(R_1)$ does not change the above results. That proves the lemma. $\square$

\bigskip

Now we study $R_2$ , the other part of the remainder.
Denote by 
\begin{equation}
f^-(x,\mp\pi,v)= \frac{M_\mp(v)}{M(v)}\int _{w_z\lessgtr 0} \Big(R_1(x,\mp\pi,w)+\frac{1}{\e}\bar{\psi}(x,\mp\pi,w)\Big)|w_z|Mdw,    \quad v_z\gtrless 0, \nonumber
 \end{equation}
the incoming data for $R_2$ which is solution to
\begin{eqnarray}
&&\mu v_x\frac{\partial R_2}{\partial x}+v_z\frac{\partial R_2}{\partial z} -\e G M^{-1}\frac {\partial (M R_2)}{\partial v_z}= \frac{1}{\e }{L_J}R_2+H_1(R_2),  \label{2.17}\\
&&R_{2} (x,\mp\pi,v)=f^-(x,\mp\pi,v)+ \frac{M_\mp(v)}{M(v)}\int _{w_z\lessgtr 0} R_2(x,\mp\pi,w)|w_z|Mdw,    \quad v_z\gtrless 0.\nonumber\end{eqnarray}
Existence and uniqueness for (\ref{2.17}) follow as in the laminar case, cf \cite{AEMN}. The following a priori estimates hold for $R_2$. By Green's formula and the definition of $H_1(R_2)=H_1((I-P)R_2)$, 
\begin{eqnarray*}
\parallel \gamma^-R_2 \parallel_{2,2,\sim}^2 + \frac{c}{\e}\parallel \nu ^{\frac{1}{2}}(I-P_J)R_2 \parallel _{2,2} ^2 \leq \parallel \gamma^+R_2 \parallel_{2,2,\sim} ^2 .
\end{eqnarray*}
Here we face a problem of diffusive boundary conditions. The ingoing flow is given  for $\varphi$ and for $R_1$ but not for $R_2$. The following bound 
is proved as in equation (4.23) in \cite{EML},
\begin{eqnarray*}&\mp\displaystyle{\int_{\mathbb{R}^3}v_zR_2^2(x,\pm\pi,v)}dv\le &c \e\eta\int_{v_z\gtrless0} |v_z| M R_2^2(x,\pm \pi,v)dv\\&&+\frac{1}{\e\eta}\int_{v_z\lessgtr0}|v_z| M (f^-(x,\pm\pi,v))^2dv.
\end{eqnarray*}
The computation in  \cite{EML} has to be adapted to the fact that the boundary conditions for $R_2$ are not purely diffusive but contain the given data $f^-$.   From Green's formula we have also
$$ \frac{c_1}{\e}\parallel \nu ^{\frac{1}{2}}(I-P_J)R_2 \parallel _{2,2} ^2  \leq \int  \big( (v_z, R^2_2(x,-\pi,v))-(v_z, R_2^2(x,\pi,v))\big) dx.
$$
Then, using the previous bound we get for any $\eta>0$,
\begin{equation}
\label{elm}
 \frac{c_1}{\e}\parallel \nu ^{\frac{1}{2}}(I-P_J)R_2 \parallel _{2,2} ^2 \leq  c \e\eta\sum_\pm\int_{v_z\gtrless0} |v_z| MR_2^2(x,\pm \pi,v)dxdv+\frac{1}{\e\eta}\parallel f^-\parallel_{2,2,\sim}^2.
 \end{equation}
We need to estimate the terms involving the outgoing parts of $R_2$ in the r.h.s. in terms of $\|R_2\|_{2,2}$ and $\|f^-\|_{2,2,\sim}$.  This is done separately at $\pm \pi$. We start with $\pi$. Consider the equation for $R_2 $, multiply by $\kappa MR_2$  and integrate in velocity  over the region $v_z\ge q$. Then integrate over space, using a smooth cut-off function $\chi(z)$ which is $0$ in  a small interval close to $-\pi$ and $1$ close to $\pi$. Finally, integrate over $q$, for $q_0<q\le 0$ and $q_0$ small enough.  We get
\begin{eqnarray*}
&&\int_{q_0}^0 dq \int_{v_z\ge q} dv\int_{-\pi}^\pi dx v_z  \kappa(\pi)M R_2^2(x,\pi,v)\le \frac{c_1q_0}{\e^2}\parallel \nu ^{\frac{1}{2}}(I-P_J)R_2 \parallel _{2,2} ^2\\&&
- \int _{q_0}^0 dq \int_{v_z\ge q} dv\int_{[-\pi,\pi]^2} dx dz \kappa(z)\chi'(z) v_z M R_2^2+c\e G\parallel R_2\parallel _{2,2} ^2.
\end{eqnarray*}The term on the l.h.s. equals
$$C|q_0| \parallel\gamma^- R_2\parallel_{2,2\sim}^2+\int_{q_0} ^0dq \int_{q<v_z<0} dv\int_{-\pi}^\pi dx v_z M R_2^2(x,\pi,v).$$
Here in the last term we can replace $R_2$ by the ingoing boundary data so to estimate it as   
$$\left|\int_{q_0}^0 dq \int_{q<v_z<0} dv\int_{-\pi}^\pi dx\big[ M^+\int_{w_z\ge 0} dw\int_{-\pi}^\pi dx w_z M R_2(x,\pi,w)+Mf^-(\pi)\big]^2\right|$$
$$\le c(q_0) \big[\parallel\gamma^- R_2\parallel_{2,2\sim}^2+\parallel f^-\parallel_{2,2,\sim}^2\big],$$
with $c(q_0)=o(|q_0|)$. Adding a similar estimate at $-\pi$ gives
\begin{equation}
\parallel\gamma^- R_2\parallel_{2,2\sim}^2\le  \frac{C}{\e^2}\parallel \nu ^{\frac{1}{2}}(I-P_J)R_2 \parallel _{2,2} ^2+C\parallel f^-\parallel_{2,2,\sim}^2+C\parallel P_JR_2 \parallel _{2,2} ^2.\label{3.7.0}
\end{equation}
Replacing in (\ref{elm}) we get
\begin{equation}
\frac{1}{\e}\parallel \nu ^{\frac{1}{2}}(I-P_J)R _2\parallel _{2,2} ^2 \leq
\e\eta C\parallel P_JR_2 \parallel _{2,2}^2+\frac c{\e\eta}\parallel f^-\parallel_{2,2,\sim}^2.\quad
\label{3.7}
\end{equation}

 We shall next prove an a priori estimate for the hydrodynamic part of $R_2$. First we consider the $1$-d ($x$-independent) case and then include in the argument  the missing terms which, as explained before, will be of order $\delta$. The extension of the $1$-d results to the $2$-d case will be based on perturbative arguments in $\delta$. To take into account these terms of order $\delta$ we add  to the right-hand side an inhomogeneous term $g_1$ with $\int Mg_1dv=0$, which will be of use later on in the proof of  Lemma \ref{2.7}.

\begin{lemma}
\label{2.6}
Let $R_2(z,v)$ be solution of the $1$-d problem and $f^-$ be defined as before,
\begin{eqnarray}
&&v_z\frac{\partial R_2}{\partial z} -\e G M^{-1}\frac {\partial (M R_2)}{\partial v_z}= \frac{1}{\e }{L_J}R_2+H_1(R_2) +g_1,  \label{2.17.1}\\
  &&R_{2} (\mp\pi,v)= \frac{M_\mp(v)}{M(v)}\int _{w_z\lessgtr0} R_2(\mp\pi,w) 
 |w_z|Mdw +f^-(\mp \pi,v),   \quad v_z\gtrless0.\nonumber
\end{eqnarray}
If $g_{10}=\int_{\R^3}dv M g_1=0$, then  it holds that 
\begin{eqnarray*}
\parallel P_JR_2 \parallel^2_{2,2 }\leq \frac{c}{\e^2 }\parallel f^- \parallel^2_{2 \sim }+\parallel \nu ^ {-\frac{1}{2} }  g_1 \parallel^2_{2 }.
\end{eqnarray*}
\end{lemma}
\noindent\underline{Proof of Lemma \ref{2.6}}. 
Here, $L_J$ is   generated by a function $\tilde q$, independent of $x$, defined by   $\tilde q=q+O(\delta)$. In the same way, in the term $H_1$ we retain only terms of order zero in $\delta$, which are independent of $x$, and include the remaining terms in $g_1$. In this way we reduce the equation to a $1$-d equation with a given term.

 Consider equation (\ref{2.17.1})  for  $\hat R_2=\mathcal{F}_z R_2$, the Fourier-transform in $z$  of $R_2$. It satisfies the equation
\begin{equation}
iv_z\xi_z{\hat R_2}-\e GM^{-1}\frac{\partial}{\partial v_z}(M\hat R_2)=\e^{-1}\widehat{L_J  R_2}+\widehat{ H_1( R_2)}-v_z r(-1)^{\xi_z}+\hat g_1
\label{FR_2}
\end{equation}
with $r(v)$ now denoting the difference between ingoing and outgoing boundary values, 
\begin{equation}
r(v)= 
R_{2 }(\pi,v)-R_{2 }(-\pi,v). 
\label{erre}
\end{equation}
 For $\xi _z\neq 0$ we use the method of Lemma \ref{2.3} with $L_J$ instead of its adjoint. Consider  the ingoing boundary values as known, and
follow step by step the proof of Lemma \ref{2.3} with obvious changes. Let $L_{J0} $ be the operator defined in (2.5) with $q$ taken at a fixed $(x_0,z_0)$. Define $P_{J0} $ as the orthogonal projection on $\text{\tt Kern}(L_{J0})$. We reach the analogue of  (\ref{3.3}), \begin{eqnarray}
&&\int |P_{J0}\hat{R}_2 (\xi_z ,v)|^2Mdv
\leq C\Big ( \frac{1}{\e ^2} \parallel \zeta_{-s}(v)\widehat{{L_J}R_2 }( \xi_z , \cdot )\parallel  ^2
+\parallel (I-P_{J0})\hat{R}_2 (\xi _z ,\cdot )\parallel  ^2 \nonumber\\
&&\qquad \qquad +\frac{\parallel P_{J0}(\chi^c_1\chi^c v_{z} r)\parallel ^2}{\delta_1|\xi_z|^2}+\parallel \nu^{-\frac{1}{2}}\hat{g}_1(\xi_z,\cdot)\parallel ^2+ \parallel \nu^{-\frac{1}{2}}\widehat{{H_1}(R_2)}(\xi_z,\cdot)\parallel ^2\Big) .\hspace*{0.4in}
\label{3.8}
\end{eqnarray}
For an estimate of the $r$-term, we express it with the
$\xi _z= 0$ term in the Fourier series for (\ref{FR_2}), 
\begin{eqnarray}\label{r-estimate}
v_z r(v)= \frac{1}{\e }\widehat{{L}_{J}{R}_2}(0,v)+\e G M^{-1} \frac{\partial }{\partial v_z}(M \hat{R}_2(0,v))  +\hat{g_1}(0,v)+ \widehat{{ H_1}(R_2)}(0,v).
\end{eqnarray}
Inserting this into (\ref{3.8}), and summing over 
$ \xi_z\neq 0$, results in
\begin{eqnarray*}
\int (\widetilde{P_{J0}R_2 })^2(z,v)Mdvdz
&&\leq C\Big( \frac{1}{\e ^2}\int \nu ((I-P_{J})R_2)^2(z,v)Mdvdz \nonumber \\
&&+\int \nu ^{-1}g_1^2(x,z,v)Mdvdz+\e^2 \parallel R_2\parallel _{2,2}^2\Big) .
\end{eqnarray*}
We are left with the Fourier component $P_{J0}\hat{R}_2(\xi_z)$ for $\xi_z=0$. Estimate separately the $(I-P)$-component and the $P$-component of $P_{J0}\hat{R}_2(0,\cdot)$. For $(I-P)P_{J0}\hat{R}_2(0,\cdot)$ we obtain
\begin{eqnarray*}
\parallel (I-P)P_{J0}\hat{R}_2(0,\cdot)\parallel\leq C\e(\parallel (I-P_J)R_2 \parallel_{2,2} + \parallel P_J R_2\parallel_{2,2}).
\end{eqnarray*}
The $P$-moments are the more involved and will be discussed each separately. We start from the $v_z$-moment of $\hat{R}_2(0,v)$.  Multiply (\ref{2.17.1}) by $M$ and integrate over $z\in[-\pi,z]$ and $v$. Since $g_{10}=0$, we have 
\begin{eqnarray*}
\int v_zMR_2(z,v)dv&&= \int_{v_z>0}f^-(-\pi,v)v_zMdv
\ .
\end{eqnarray*}

Given two functions  $h(v)$ and  $f((\,\cdot\,), v)$ we use the notation 
$f_{h}(\,\cdot\,):=\int dvh(v) f(\cdot,v)$. In particular, for $h=\psi_j$, $j=0,\dots,4$, we also use the notation $f_j$. We have 
\begin{eqnarray*}
|\hat{R}_{2v_z}(0)|= \left|\int v_zMR_2(z,v)dvdz \right| \leq C\parallel f^-\parallel_{2\sim}.
\end{eqnarray*}
To estimate the moments of $\hat R_2(0,v)$ we use the identity  
\begin{equation}
\hat{R}_2(0,v)= \Delta- \sum_{\xi_z\neq 0}\hat{R}_2(\xi_z,v)(-1)^{\xi_z},
\label{3.91}
\end{equation}
where $\Delta (v) = \pi (R_2(\pi,v)+R_2(-\pi,v))$, which follows from
$$
R_2(\pi,v)+R_2(-\pi,v)=\frac{1}{2\pi}\sum_{\xi_z\in \mathbb{Z}}\hat R_2(\xi_z,v)(e^{i\pi \xi_z}+e^{-i\pi \xi_z}),
$$
by solving for the $\xi_z=0$ coefficient.

We write $\Delta= 2\pi R_2(-\pi,v)+\pi r(v)$  for $v_z>0$  and $\Delta= 2\pi R_2(\pi,v)-\pi r(v)$ for $v_z<0$.
 \medskip
  
 We consider first the $\psi_4$-moment of $\hat R_2(0,v)$, denoted by $\hat R_{24}$. We notice that 
\begin{eqnarray}
\hat{R}_{2v^2_z\bar{A}}(0)=\frac{1}{\sqrt{6}}\hat{R}_{24}(0) \int v^2_z v^2\bar{A}Mdv+\hat{R}^\perp_{2 v^2_ z\bar{A}}(0),
\label{3.10}
\end{eqnarray}
where $\hat R_2^\perp=(1-P)\hat R_2$ and  $\bar{A}$ and $\bar{B}$ are nonhydrodynamic solutions to
\begin{eqnarray}\label{AB}
{L}(v_z\bar{A})= v_z(v^2-5T),\quad {L}(v_xv_z\bar{B})= v_xv_z.
\end{eqnarray}
Hence, we are left with the  control of $\hat{R}_{2v^2_z\bar{A}}(0)$. For that, we use (\ref{3.91}).   A multiplication of (\ref{3.91}) with $Mv_z^2\bar{A}$ followed by a $v$-integration, gives 
\begin{eqnarray*}
\hat{R}_{2v^2_z\bar{A}}(0)=\Delta_{v^2_z\bar{A}}-\sum_{\xi\neq 0}\hat{R}_{2v^2_z\bar{A}}(\xi_z)(-1)^{\xi_z}.
\end{eqnarray*}

Let us first consider the contribution $\Delta_{v^2_z\bar{A}}$. Using the relation $\Delta= 2\pi R_2(\pm\pi,v)\mp \pi r(v)$ for $v_z\lessgtr0$, we  notice that the first part it is computed in terms of the outgoing flow and then can be bounded in terms of $\int dv v_z^2\bar {A} Mf^-$ and  $\int_{v_z\gtrless0} v_z ^2\bar{A}R_2(\pm\pi,v) Mdv$.  Then, we  multiply (\ref{2.17.1}) by $M\chi v_z\bar{A}$ and integrate over $v_z>0$ (similarly at $-\pi$) , to get the bound\begin{eqnarray*}
\mid \int _{v_z\gtrless0} v_z ^2\bar{A}R_2(\pm \pi,v) Mdv\mid^2\leq \eta\parallel P_JR_{2}\parallel_{2,2}^2+c(\frac{1}{\e ^2}\parallel (I-P_J)R_{2}\parallel_{2,2}^2+\parallel  \nu ^ {-\frac{1}{2} } g_{1,\perp }\parallel_{2,2}^2)\ .
\end{eqnarray*}
The second part, namely the  $v_z^2\bar{A}$-moment of the $\pi r$-term, is estimated as before using  (\ref{r-estimate}). 

In order to control $\hat{R}_{2v^2_z\bar{A}}(\xi)$ for $ \xi_z\neq 0$ we take the inner product of  (\ref{FR_2}) with $v_z\bar{A}$ ,
\begin{eqnarray*}
(-1)^{\xi_z}r_{v_z^2\bar{A}}&&+ i\xi_z (v_z^2\bar{A},\hat{R}_2({\xi_z}))+\e G
\int\frac{\partial}{\partial v_z}(v_z\bar{A})M\hat{R}_2(\xi_z)dv\\
&&=\frac{1}{\e}(v_z\bar{A},\widehat{L_JR_2}(\xi_z))+\hat{g}_{1,v_z\bar{A}}(\xi_z)+(v_z\bar{A},\widehat{H_1(R_2)}(\xi_z)).
\end{eqnarray*}
Hence
\begin{eqnarray*}
\hat{R}_{2v_z^2\bar{A}}(\xi_z)&&=\frac{(-1)^{\xi_z}}{\xi_z}ir_{v_z^2\bar{A}}+\frac{i\e G}{\xi_z}\int\frac{\partial}{\partial v_z}(v_z\bar{A})M\hat{R}_2(\xi_z)dv-\frac{i}{\e \xi_z}(v_z\bar{A},\widehat{L_JR_2}(\xi_z))\\
&&-\frac{i}{\xi_z}\hat{g}_{1,v_z\bar{A}}(\xi_z)-\frac{i}{\xi_z}(v_z\bar{A},\widehat{H_1(R_2)}(\xi_z)).
\end{eqnarray*}
Combining this with (\ref{3.91}) and noticing the pairwise cancellation of the $r$-terms with positive and negative $\xi$'s, gives
\begin{eqnarray*}
| \hat{R}_{2v_z^2\bar{A}}(0)|^2\leq c(\frac{1}{\e^2}\parallel\nu^{\frac{1}{2}}(I-P_J)R_2\parallel^2_{2,2}+\parallel\nu^{-\frac{1}{2}}g_{1,\perp}\parallel^2_{2,2}+\eta \parallel P_JR_2\parallel^2_{2,2}),
\end{eqnarray*}
and so  using (\ref{3.7}) and (\ref{3.10})
\begin{eqnarray*}
| \hat{R}_{24}(0)|^2&&\leq c(\frac{1}{\e^2}\parallel\nu^{-\frac{1}{2}}(I-P_J)R_2\parallel^2_{2,2}+\parallel\nu^{-\frac{1}{2}}g_{1,\perp}\parallel^2_{2,2})+\eta \parallel P_JR_2\parallel^2_{2,2}\\&&\leq c(\frac{1}{\e^2}\parallel f^-\parallel _{2,2}^2+\parallel\nu^{-\frac{1}{2}}g_{1,\perp}\parallel^2_{2,2})
+\eta \parallel P_JR_2\parallel^2_{2,2}.
\end{eqnarray*}\\
 The $v_x$- and $v_y$- moments are analogous to each others, so we only discuss the former.
This can be treated similarly to the previous $\hat{R}_{24}(0)$ case but with $\Delta= 2\pi R_2(\mp \pi,v)\pm \pi r(v)$, $v_z\lessgtr 0$. The $2\pi R_2$-term is now ingoing, and its inner product with $v_x$ gives zero. The $\pi r$-term is estimated as before. For the sum in (\ref{3.91}) of the other Fourier coefficients, we notice that 
$\hat{R}_{2v_x}(\xi)=\hat{R}_{2v_z^2v_x}(\xi)-\hat{R}_{2\perp v_z^2v_x}(\xi)$, and we can proceed as before. Since $PP_J\hat R_2$ differs from $P\hat R_2$ by terms of order $\e$ which are already under control, we can summarize the results so far  as
\begin{eqnarray}\label{2.36}
&&\int\left(|\hat{R}_{2\bar{\psi}_1}(0)|^2+|\hat{R}_{2\bar{\psi}_2}(0)|^2+|\hat{R}_{2\bar{\psi}_3}(0)|^2+|\hat{R}_{2\bar{\psi}_4}(0)|^2\right)dz\label{sum}\\
&&\leq c(\frac{1}{\e^2}\parallel f^-\parallel _{2,2}^2 +\parallel\nu^{-\frac{1}{2}}g_{1,\perp }\parallel^2_{2,2})
+\eta \parallel \hat{R}_{2}\parallel^2_{2,2}\nonumber.
\end{eqnarray}
\hspace{1cm}\\

Finally for the $\hat{R}_{20}(0)$-moment, start by considering (\ref{2.17.1}) with the new boundary conditions 
\begin{eqnarray}
&&R_2 (-\pi,v)= f^-(-\pi,v),    \quad v_z>0, \quad\label{nbc}\\
&&R_2(\pi,v)= \frac{M_+(v)}{M(v)}\int_{w_z>0}\big[R_2(\pi,w)- f^-(-\pi,w)\big]w_zM(w)dw   ,\hspace{0.03in}v_z<0.\nonumber
\end{eqnarray}
The new boundary values are constructed in such a way that $\int _{v_z<0}R_2(-\pi,v)v_zMdv=0$ and this  property  will allow the new boundary conditions to be equivalent to the old ones.  
In fact, since $\int_{v_z<0}M_+(v)v_zdv =-1$, 
\begin{eqnarray*}&&\int_{\mathbb{R}^3}v_zR_2(\pi,v)M(v)dv = \int_{v_z<0}v_zR_2(\pi,v)M(v)dv+\int_{v_z>0}v_zR_2(\pi,v)M(v)dv\\
&&= -\int_{w_z>0}w_z[R_2(\pi,w)-f^-(-\pi,w)]M(w)dw+\int_{v_z>0}v_zR_2(\pi,v)M(v)dv\\
&&= \int_{w_z>0}w_zf^-(-\pi,v)M(v)dv.\end{eqnarray*}

Existence and uniqueness for the new  problem are well known. 
 We shall verify that the new problem also satisfies the old boundary conditions of (\ref{2.17.1}) . At $z=-\pi$, for $v_z>0$  the new  ingoing boundary condition for $R_2$ is $R_2(-\pi,v)=f^-(-\pi,v))$, which coincides with the second equation in (\ref{2.17.1}) since $\sqrt{2\pi}\int _{w_z<0} R_2(-\pi,w)|w_z|Mdw=0$. 
At $z=\pi$ the new ingoing boundary condition is 
$$R_2(\pi,v)= M^{-1}M_+\int_{w_z>0}\big[R_2(\pi,w)-f^-(-\pi,w)\big]w_z M(w)dw  ,\quad v_z<0\ .$$ 
This coincides with the old boundary condition at $\pi$, given by  the second equation in (\ref{2.17.1}), that is with 
$$R_2(\pi,v)=f^-(\pi,v)+M^{-1}M_+\\
\int_{w_z>0}R_2(\pi,w)M(w)w_zdw,\quad v_z<0,$$ 
provided that
$$-f^-(\pi,v)=M^{-1}M_+\int_{w_z>0}f^-(-\pi,w)w_z M(w)dw,\quad v_z<0,$$
or, recalling the definition of $f^-$,
$$\int_{w_z>0}\big(R_1(\pi,w)+\frac{\bar \psi(\pi,w)}\e\big)M(w)w_zdw=\int_{w_z<0}\big(R_1(-\pi,w)+\frac{\bar \psi(-\pi,w)}\e\big)M(w)w_zdw.$$
To check this we note that from the assumption that the inhomogeneous term $g$ in the equation for $R_1$ is such that $\int_{\R^3}dv M(v)g(v)=0$, it follows that
$$(v_z, R_1(\pi,v))=(v_z, R_1(-\pi,v)).$$

Then, using the boundary conditions for $R_1$ this becomes
\begin{eqnarray*}
&&\int_{v_z>0}M(v)R_1(\pi,v)v_zdv -\int_{v_z<0}M(v)\frac{\bar\psi}\e(\pi,v)v_zdv\\
&&+\int_{v_z>0}M(v)\frac{\bar\psi}\e(\pi,v)(-\pi,v)v_zdv-\int_{v_z<0}M(v)R_1(-\pi,v)v_zdv=0.\end{eqnarray*}
Using $(\bar\psi(\pm \pi,v), v_z)=0$, the claimed equivalence follows.
 Thus $R_2$ with the new boundary conditions equals the unique solution to (\ref{2.17.1}). 

We write  (\ref{2.17.1}) with $\displaystyle{R_2'=R_2 \kappa(z)}$. The left- hand side becomes
$$\displaystyle{\kappa^{-1}(z)\left(v_z \frac{\partial R_2'}{\partial z} -\e G \frac{\partial 
R_2'}{\partial v_z}\right)}.$$
Multiply  the equation for $R_2'$ by $Mv_z\kappa(z)$  and integrate 
over $ [-\pi,z]\times  \R^{3}$. The l.h.s. gives
\begin{eqnarray*}&&\int R_2'(z,v)Mv_z^2dv -\int R_2'(-\pi,v)Mv_z^2dv\\&&
 -\e G \int _{-\pi}^z dz'\int  R_2'(z',v)Mv_z^2dv +\e G \int _{-\pi}^z dz'\int  R_2'(z',v)Mdv\ .
 \end{eqnarray*}
Since $\displaystyle{\int  dvv_z^2 R_2Mdv = R_{20}+\frac{2 R_{24}}{\sqrt 6}+\int  dvv_z^2 R_2^\perp Mdv}$, by integrating the equation again over $[-\pi,\pi]$ we control  $|\hat R_{20}(0)|$ in terms of known quantities plus terms bounded in the $\|~\cdot~\|_{2,2}$-norm, multiplied by a factor $\e$, and the integral $\int R_2'(-\pi,v)Mv_z^2dv$. The contribution due to the incoming part is given by $f^-$. Therefore, we need  an estimate of the outgoing boundary term $\displaystyle{\int_{v_z<0} R_2(-\pi,v)Mv_z^2 dv}$. To do this, we repeat the steps above  but this time integrate over $\{[-\pi,\pi]\times \R^{3};v_z>0\}$, to get on the l.h.s.
$$\int_{v_z>0} R_2'(\pi,v)Mv_z^2dv -\int_{v_z>0} R_2'(-\pi,v)Mv_z^2dv$$
$$ -\e G \int _{-\pi}^\pi dz'\int_{v_z>0}  R_2'(z',v)Mv_z^2dv +\e G \int _{-\pi}^\pi dz'\int _{v_z>0} R_2'(z',v)Mdv.$$

Since $R'_2$ solves the problem with modified boundary conditions, the incoming part in $-\pi$ can be bounded by  $\|f^-\|_{2,2,\sim}$. Therefore we  get an estimate of $\displaystyle{\int_{v_z>0} R_2(\pi,v)Mv_z^2dv}$  in terms of the norm of $f^-$ and again quantities bounded in the $\|\,\cdot\,\|_{2,2}$-norm \
multiplied by a factor $\e$. To obtain the estimate of $\int_{v_z<0} R_2(-\pi,v)Mv_z^2 dv$, we integrate over $[-\pi,\pi]\times  \R^{3}$. 
 The final  estimate is
 \begin{equation}
|\hat{R}_{20}(0)|\leq c(\frac{1}{\e}\parallel f^-\parallel_{2\sim}+
\frac{1}{\e}\parallel\nu^{\frac{1}{2}}(I-P_J)R_2\parallel_{2,2}+\eta \parallel R_2\parallel_{2,2}+ \parallel \nu ^{-\frac{1}{2}}g_1\parallel_{2,2}).\nonumber
\end{equation}
By combining with the other moment estimates (\ref{sum}),  and using (\ref{3.7}), we obtain  that
\begin{eqnarray}
\parallel P_JR_2\parallel^2_{2,2}\leq \frac{c}{\e^2}\parallel f^-\parallel^2_{2\sim}+ \parallel \nu ^{-\frac{1}{2}}g_1\parallel_{2,2}^2.
\end{eqnarray}
This completes the proof of the lemma. $\square$ 

\bigskip

Based on this $1$-d analysis, it follows in the $2$-d case that 
\begin{lemma}
\label{2.7}
The solution $R_2$ to (\ref{2.17.1}) satisfies
\begin{eqnarray*}
\parallel P_JR_2\parallel^2_{2,2}\leq c\frac{1}{\e^2}\parallel f^-\parallel^2_{2\sim}.
\end{eqnarray*}
\end{lemma}
\noindent\underline{Proof of Lemma \ref{2.7}.} Consider equation (\ref{2.17.1}) for  the Fourier transform   in $x,z$ of $R_2$,  $\hat R_2={\cal F}_x{\cal F}_z R_2$,
\begin{eqnarray}
&&\hskip -1cm i(\mu\xi_x v_x+\xi_z v_z)\hat{R}_2= \frac{1}{\e }\widehat{{L}_{J}{R_2}}+\e G M^{-1} \frac{\partial (M\hat{R_2}) }{\partial v_z} +\widehat{H_1(R_2)}- v_z r(\xi_x,v)(-1)^{\xi_z},\label{3.12}\\ 
&&
r(\xi_x, v)={\cal F}_xR_{2 }(\xi_x,\pi,v)- {\cal F}_xR_{2 }(\xi_x,-\pi,v).
\label{erre1}
\end{eqnarray}

In the case $\xi_x\ne 0$, $\xi_z\neq 0$ we can reach, as in the case of Lemma \ref{2.6}, a bound like (\ref{3.8}). 
If we consider the Fourier components with $\xi$ large,  we see that   in (\ref{3.8}) the $r$-terms are multiplied by a small number  for    $\xi$ large and then can be estimated, by using (\ref{3.7.0}), 
   by $\eta\parallel R_2\parallel_{2,2}$, with a small $\eta$, plus the earlier terms. We notice that in the case $\xi_x=0$,  equation (\ref{3.12}) reduces to (\ref{FR_2}). Hence, we can apply Lemma \ref{2.6} to  $\hat R_2(0,\xi_z,v)=\int R_2(x,z,v)dx$ and, taking into account that $g_1$ is of order $\delta$, get a bound for the Fourier components $P_J\hat{R}_2(0,\xi_z)$, for $\delta$ small.  The remaining components in the  case $\xi_x\ne 0$, $\xi_x,\xi_z$ bounded, can be estimated in the following way. \\
We start from the moment $\hat{R}_{2v_x}$. Notice that $r_{v_z}(\xi_x) =\hat{f}^-_{v_z}(\xi_x)$, where $\hat f^-$ is the function defined as
$$\hat f^-(\xi_x,v)=\begin{cases}({\cal F}_xf^-)(\xi_x,\pi,v)& \text{ for } v_z<0,\\({\cal F}_xf^-)(\xi_x,-\pi,v)& \text{ for } v_z>0.\end{cases}$$
Hence,  by integrating  (\ref{3.12}) we obtain
\begin{eqnarray*}
|\xi_x||\hat{R}_{2v_x}(\xi_x,0)| \leq C|\hat{f}^-_{v_z}(\xi_x)|.
\end{eqnarray*}
Then, we look for a bound for $\hat{R}_{2v_z}(\xi_x,\xi_z)$ when $\xi_z\ne 0$. We use the function $\bar B$ introduced in (\ref{AB}), $\bar B$ being the solution of the equation $L(v_xv_z\bar B)=v_xv_z$.

We have $(v_z^2v_x\bar B,  \hat R_2)\ =\  (v_z^2v_x\bar B, (I-P)\hat R_2)+ (\psi_x, v_xv_z^2 \bar B)(\psi_x ,\hat R_2)$, where $\psi_j$, $j=0,\dots,4$  are as usual the vectors of the orthonormal basis in $\text{\tt Kern}(L)$. We multiply (\ref{3.12}), written for $\xi_z=0$, by $v_z^2\bar{B}-(\psi_x,v_xv_z^2\bar{B})$ and integrate over velocities. We first use the relation so obtained for $\xi_z=0$ and obtain,
\begin{eqnarray*}
|r_{v_z^3\bar{B}}(\xi_x)|&&\leq c(\frac{1}{\e}\parallel \nu^{-\frac{1}{2}}\widehat{L_J(I-P_J)R_2}(\xi_x,0)\parallel_{2}+\e G \parallel \hat{R}_{2}(\xi_x,0)\parallel_{2}+|\hat{f}^-_{v_z}(\xi_x)|\\
&&+\parallel\nu^{-\frac{1}{2}}\widehat{H_1(R_2)}(\xi_x,0)\parallel_{2}+\parallel(I-P)\hat{R}_2(\xi_x,0)\parallel_{2}),
\end{eqnarray*}
having used the fact that $|r_{v_x}|\le c|f^-_{v_x}|$.
We use again that relation for $\xi_z\neq 0$. Since $(v_z^3\bar B,R_2)= (v_z^3\bar B,(I-P)R_2)+ (R_2,\psi_z) (\psi_z,v_z^3\bar B)$, we obtain in this way  an expression for $\xi_z\hat{R}_{2v_z}(\xi_x,\xi_z)$, for $\xi_z\neq 0$, in terms of quantities under control, since with the previous subtraction we have removed from the equation the term $(\psi_x,\hat R_2)$. As a result, for $\xi_z\neq 0$,
\begin{eqnarray*}
|\xi_z|\mid \hat{R}_{2v_z}(\xi_x,\xi_z)\mid &&\leq C\Big(\frac{1}{\e}\parallel \nu^{-\frac{1}{2}}\widehat{L_J(I-P_J){R_2}}(\xi_x,\xi_z)\parallel
\\&&+\parallel\nu^{-\frac{1}{2}}\widehat{H_1(R_2)}(\xi_x,\xi_z)\parallel
+\e \parallel\hat{R}_{2}(\xi_x,\xi_z)\parallel\\
&&+\frac{1}{\e}\parallel \nu^{-\frac{1}{2}}\widehat{L_J(I-P_J)R_2}(\xi_x,0)\parallel+\e  \parallel \hat{R}_{2}(\xi_x,0)\parallel\\
&&+\parallel\nu^{-\frac{1}{2}}\widehat{H_1(R_2)}(\xi_x,0)\parallel
+|\xi_x|\parallel \hat{f}^-(\xi_x)\parallel \Big).
\end{eqnarray*}
Notice that the last term is bounded because $\xi_x$ is bounded in this part of the proof, the terms with $\xi$ large having been estimated before.

In the same way the $v_x$-moment can be controlled when $\xi_z\neq 0$,
\begin{eqnarray*}
&&|\xi_x|\mid \hat{R}_{2v_x}(\xi_x,\xi_z)\mid 
\leq C(\frac{1}{\e}\parallel \nu^{-\frac{1}{2}}\widehat{L_J(I-P_J){R_2}}(\xi_x,\xi_z)\parallel
+\parallel\nu^{-\frac{1}{2}}\widehat{H_1(R_2)}(\xi_x,\xi_z)\parallel
\\&&\qquad \qquad+\e \parallel\hat{R}(\xi_x,\xi_z)\parallel
+\frac{1}{\e}\parallel \nu^{-\frac{1}{2}}\widehat{L_J(I-P_J)R_2}(\xi_x,0)\parallel_{2}+\e \parallel \hat{R}_{2}(\xi_x,0)\parallel\\
&&\qquad \qquad+\parallel\nu^{-\frac{1}{2}}\widehat{H_1(R_2)}(\xi_x,0)\parallel +|f^-_{v_z}(\xi_x)|).
\end{eqnarray*}
For the $v_z$-moment $\hat{R}_{2v_z}(\xi_x,0)$, by using the proof of Lemma \ref{2.6}, 
\begin{eqnarray*}
|\hat{R}_{2v_z}(\xi_x,0)|\leq c(\parallel\hat{f}^-(\xi_x)\parallel+\parallel \mathcal{F}_x{R}_{2v_x}(\xi_x)\parallel).
\end{eqnarray*}
To proceed, we observe that 
\begin{eqnarray*}
 \parallel \mathcal{F}_x{R}_{2v_x}(\xi_x)\parallel&&\leq C\Big(\frac{1}{\e}(\sum_{\xi_z}\parallel \nu^{-\frac{1}{2}}\widehat{L_J(I-P_J){R_2}}(\xi_x,\xi_z)\parallel^2
)^{\frac{1}{2}}\\
&&+(\sum_{\xi_z}\parallel\nu^{-\frac{1}{2}}\widehat{H_1(R_2)}(\xi_x,\xi_z)\parallel^2)^{\frac{1}{2}}
+\e (\sum_{\xi_z} \parallel\hat{R}_{2}(\xi_x,\xi_z)\parallel^2)^{\frac{1}{2}}\\
&&+\frac{1}{\e}\parallel \nu^{-\frac{1}{2}}\widehat{L_J(I-P_J)R_2}(\xi_x,0)\parallel+\e \parallel \hat{R}_{2}(\xi_x,0)\parallel+|\hat{f}^-_{v_z}(\xi_x)|\\
&&+\parallel\nu^{-\frac{1}{2}}\widehat{H_1(R_2)}(\xi_x,0)\parallel \Big).
\end{eqnarray*}
And so,
\begin{eqnarray*}
|\hat{R}_{2v_z}(\xi_x,0)|&&\leq C\Big(\parallel\hat{f}^-(\xi_x)\parallel+\e\parallel R_2 \parallel_{2,2}+\frac{1}{\e}(\sum_{\xi_z}\parallel \nu^{-\frac{1}{2}}\widehat{L_J(I-P_J){R_2}}(\xi_x,\xi_z)\parallel^2
)^{\frac{1}{2}}\\
&&+(\sum_{\xi_z}\parallel\nu^{-\frac{1}{2}}\widehat{H_1(R_2)}(\xi_x,\xi_z)\parallel^2)^{\frac{1}{2}}
+\frac{1}{\e}\parallel \nu^{-\frac{1}{2}}\widehat{L_J(I-P_J)R_2}(\xi_x,0)\parallel\\
&&+\parallel\nu^{-\frac{1}{2}}\widehat{H_1(R_2)}(\xi_x,0)\parallel \Big) .
\end{eqnarray*}
To estimate $\hat{R}_{2v_y}(\xi_x,0)$, we use the method of Lemma \ref{2.6}, that is  the proof of (\ref{sum}). Then, to get the estimate for $\xi_z\neq 0$, we multiply (\ref{3.12}) by $v_yv_z$ and use $v_xv_yv_z\in(\text{\tt Kern}(L))^\perp $ to estimate the term in $r$ as

\begin{eqnarray*}
|r_{v_z^2v_y}(\xi_x)|&&\leq C(\frac{1}{\e}\parallel \nu^{-\frac{1}{2}}\widehat{L_J(I-P_J)R_2}(\xi_x,0)\parallel_{2}+\e \parallel R_{2}(\xi_x,0)\parallel_{2}\\
&&+\parallel\nu^{-\frac{1}{2}}\widehat{H_1(R_2)}(\xi_x,0)\parallel_{2}+\parallel(I-P_J)\hat{R}_2(\xi_x,0)\parallel_{2}).
\end{eqnarray*}
So with $C$ depending on $\xi_x$,
\begin{eqnarray*}
\parallel \hat{R}_{2v_y}(\xi_x,0)\parallel_2&&\leq C(\frac{1}{\e}(\sum_{\xi_z}\parallel \nu^{-\frac{1}{2}}\widehat{L_J(I-P_J){R_2}}(\xi_x,\xi_z)\parallel^2_{2}
)^{\frac{1}{2}}\\
&&+(\sum_{\xi_z}\parallel\nu^{-\frac{1}{2}}\widehat{H_1(R_2)}(\xi_x,\xi_z)\parallel^2_{2})^{\frac{1}{2}}
+\e (\sum_{\xi_z} \parallel R_{2}(\xi_x,\xi_z)\parallel^2_{2})^{\frac{1}{2}}\\
&&+\frac{1}{\e}\parallel \nu^{-\frac{1}{2}}\widehat{L_J(I-P_J)R_2}(\xi_x,0)\parallel_{2}+\e \parallel R_{2}(\xi_x,0)\parallel_{2}\\
&&+\parallel\nu^{-\frac{1}{2}}\widehat{H_1(R_2)}(\xi_x,0)\parallel_{2}
+\parallel(I-P_J)\hat{R}_2(\xi_x,0)\parallel_{2}),
\end{eqnarray*}
and similarly for $\hat{R}_{2v_y}(\xi_x,\xi_z)$ when $\xi_z\neq 0$.\\
For the $\psi_4$-moment we shall use   
\begin{eqnarray}
\hat{R}_{2v^2_z\bar{A}}(\xi)=\frac{1}{\sqrt{6}}\hat{R}_{24}(\xi) \int v^2_z v^2\bar{A}Mdv+\hat{R}_{2\perp v^2_ z\bar{A}}(\xi).
\label{3.13}
\end{eqnarray}
Here in the $2$-d case, 
\begin{eqnarray*}
\xi_z\hat{R}_{2v_z^2\bar{A}}(\xi)&&=-\xi_x\hat{R}_{2v_zv_x\bar{A}}(\xi)+(-1)^{\xi_z}ir_{v_z^2\bar{A}}+{i\e G}
\int\frac{\partial}{\partial v_z}(v_z\bar{A})M\hat{R}_2(\xi)dv\\
&&-\frac{i}{\e }(v_z^2\bar{A},\widehat{L_JR_2}(\xi))
-{i}(v_z\bar{A},\widehat{H_1(R_2)}(\xi)).
\end{eqnarray*}
The additional term in comparison with the $1$-d case, belongs to $R_{2\perp}$, and can be estimated by (\ref{3.7}). An estimate of the boundary term $r_{v_z^2\bar{A}}$ can be obtained by multiplying (\ref{3.12}) with $v_z\bar{A}$ for $\xi_z=0$ and integrating. For $\xi_ z\neq 0$ this gives  
\begin{eqnarray*}
| \hat{R}_{2v_z^2\bar{A}}(\xi_x,\xi_z)|^2\leq C\big(\frac{1}{\e^2}\parallel\nu^{\frac{1}{2}}(I-P_J)R_2\parallel^2_{2,2}+
\e^2\parallel R_2\parallel^2_{2,2}\big)
\end{eqnarray*}
with $C$ depending on $\xi$. Using (\ref{3.13}) the same  estimate holds for $\hat{R}_{24}(\xi_x,\xi_z)$. For $\xi_x\neq 0$ the Fourier component $\hat{R}_{2v_z^2\bar{A}}(\xi_x,0)$ can be expressed by (\ref{3.91}) and (\ref{3.12}), including the pairwise cancellation of the $r(\xi_x)$ terms. Treating the $\Delta $-term as in the $1$-d case, gives
\begin{eqnarray*}
| \hat{R}_{2v_z^2\bar{A}}(\xi_x,0)|^2\leq C\big(\frac{1}{\e^2}\parallel\nu^{\frac{1}{2}}(I-P_J)R_2\parallel^2_{2,2}+
\e^2\parallel R_2\parallel^2_{2,2}+\parallel\hat{f}^-(\xi_x)\parallel^2_{2\sim}\big)
\end{eqnarray*}
with $C$ depending on $\xi_x$.  Again by (\ref{3.13}) the same  estimate holds for $\hat{R}_{24}(\xi_x,0)$.

The $\psi_0$ moments when $\xi_z\neq 0$ may now be obtained by multiplying
(\ref{3.12}) with $v_z$ and integrating. Arguing as above and using the earlier estimate for the $\psi_4$-moment we get    
\begin{eqnarray*}
| \hat{R}_{2 0}(\xi_x,\xi_z)|^2\leq C\big(\frac{1}{\e^2}\parallel\nu^{\frac{1}{2}}(I-P_J)R_2\parallel^2_{2,2}+
\e^2\parallel R_2\parallel^2_{2,2}+\parallel\hat{f}^-(\xi_x)\parallel^2_{2\sim}\big)
\end{eqnarray*}
with $C$ depending on $\xi$. And so there only remains the $\psi_0$ moment for $\hat{R}_2(\xi_x,0)$ when $\xi_x\neq0$. Multiply (\ref{3.12}) for $\xi=(\xi_x,0)$ with $v_x$ and integrate. For the boundary term $r_{v_xv_z}(\xi_x)$, multiply (\ref{3.12}) for $\xi=(\xi_x,1)$ with $v_x$ and integrate. All terms in the upcoming expression for $r_{v_xv_z}(\xi_x)$ are then under control. This gives    
\begin{eqnarray*}
| \hat{R}_{2 0}(\xi_x,0)|^2\leq c(\frac{1}{\e^2}\parallel\nu^{\frac{1}{2}}(I-P_J)R_2\parallel^2_{2,2}+
\e^2\parallel R_2\parallel^2_{2,2}+\parallel\hat{f}^-(\xi_x)\parallel^2_{2\sim}).
\end{eqnarray*}
Combining all the above estimates gives the statement of the lemma,
\begin{eqnarray*}
\parallel P_JR_2\parallel^2_{2,2}\leq \frac{c}{\e^2}\parallel f^-\parallel^2_{2,2,\sim}.\quad \square
\end{eqnarray*}
\hspace{1cm}\\
The step from $L^2$ to $L^{\infty}$ for $R_2$ follows as in the $R_1$-case. These estimates together give
\begin{lemma}
\label{2.8}
A solution to the $R_2$-problem satisfies 
\begin{eqnarray*}
\parallel \nu^{\frac{1}{2}}(I-P_J)R_2\parallel_{2,2}^2&&\leq c\Big(
\e  \parallel \nu^{-\frac{1}{2}}(I-P_J)g\parallel_{2,2}^2+\frac{1}{\e}\parallel P_J g\parallel_{2,2}^2+\frac{1}{\e^2} \parallel \bar{\psi} \parallel_{2,2,\sim}^2\Big) ,\\
\parallel P_J R_2\parallel_{2,2} ^2&&\leq c\Big( \frac{1}{\e}\parallel\nu^{-\frac{1}{2}}(I-P_J)g\parallel_{2,2}^2+\frac{1}{\e^3}\parallel P_J g\parallel_{2,2}^2+\frac{1}{\e^4 }\parallel \bar{\psi} \parallel_{2,2,\sim}^2\Big),\\
\parallel \nu^{\frac{1}{2}}R_2\parallel ^2_{\infty,2}
&&\leq c( \frac{1}{\e^3}\parallel \nu^{-\frac{1}{2}}(I-P_J)g\parallel ^2_{2,2} +\frac {1}{\e ^5}\parallel P_Jg\parallel ^2_{2,2}+\e^2\parallel\nu^{-\frac{1}{2}}g\parallel^2_{\infty,2}\\&&
+\frac{1}{\e^6}\parallel \bar{\psi}\parallel^2_{2,\sim}).
\end{eqnarray*}
\end{lemma}
\bigskip

The previous estimates can be used to prove
\begin{theorem}
\label{2.9}
There exists a solution $R$  in $L_M^{2}( [-\pi,\pi]^2 \times \R^3)$ to the rest term problem
\begin{eqnarray}
&&v^\mu\cdot\nabla R-\e GM^{-1}\frac{\partial (M{R})}{\partial v_z}=\frac{1}{\e }{L}{R}+{J}({R},{R})+H({R})+\e{\alpha },\quad \quad\label{3.14}\\
&& R(x,\mp\pi,v)= \int _{w_z\lessgtr0} ( R(x,\mp\pi,w)+\frac{1}{\e}\bar{\psi}(x,\mp\pi,w))|w_z|M_-dw - \frac{1}{\e}\bar{\psi}(x,\mp\pi,v),
\quad v_z\gtrless0,\nonumber
\end{eqnarray}

\end{theorem}
\noindent\underline{Proof of Theorem \ref{2.9}} \hspace*{.05in} The rest term ${R}$ will be obtained as the limit of the approximating sequence $\{R^n\}$, where $R^0= 0$ and
\begin{eqnarray*}
&&  v^\mu\cdot\nabla R^{n+1}-\e GM^{-1}\frac{\partial(MR^{n+1})}{\partial v_z} 
=\frac{1}{\e }{L}_J R^{n+1}+H_1(R^{n+1})
+{J}(R^{n},R^n)+\e {\alpha },\\
&&R^{n+1}(x,\mp\pi,v)=\frac{M_\mp}{M}\int_{w_z\lessgtr0}(R^{n+1}(x,\mp\pi,w)+\frac{\bar{\psi}}{\e}(x,\mp\pi,w))w_zMdw 
- \frac{\bar{\psi}}{\e}(x,\mp\pi,v),
 v_z\gtrless0.
\end{eqnarray*} 
Here $(I-P_J)g=\e (I-P_J) \alpha$ is of order four, and $P_Jg=\e P_J\alpha$ of order five. In particular, the function $R^1$ is solution to  
\begin{eqnarray*}
&& v^\mu\cdot\nabla R^{1} - \e GM^{-1}\frac{\partial(M R^1)}{\partial v_z}= \frac{1}{\e }L_JR^{1}
+H_1(R^{1})
+\e {\alpha },\\
&&R^{1}(x,\mp\pi,v)= \frac{M_\mp}{M}\int _{w_z\lessgtr0} (R^{1}(x,\mp\pi,w)+\frac{\bar{\psi}}{\e}(x,\mp\pi,w))|w_z|M_-dw 
- \frac{\bar{\psi}}{\e}(x,\mp\pi,v),
  v_z\gtrless0.\\
\end{eqnarray*}
Splitting $R^1$ into two parts $R_1$ and $R_2$, solutions of (\ref{3.5}) and (\ref{2.17}) with $g=\e \alpha$ in (\ref{3.5}), then using the corresponding a priori estimates, Lemma \ref{2.5} and Lemma \ref{2.8}, together with the exponential decrease of $\bar{\psi}$, we obtain, for some constant $c_1$,
\begin{eqnarray*}
\parallel \nu^{\frac{1}{2}}R^1\parallel _{\infty ,2}\leq c_1\e^{\frac{5}{2}},\quad \parallel \nu^{\frac{1}{2}}R^1\parallel _{2,2}\leq c_1 \e^{\frac{7}{2}}.
\end{eqnarray*}
By induction  for $\e$ sufficiently small,
\begin{eqnarray*}
&&\parallel \nu^{\frac{1}{2}} R^j\parallel _{\infty ,2}\leq 2c_1 {\e}^{\frac{5}{2}} ,\quad j\leq n+1, \\
&&\parallel \nu^{\frac{1}{2}}(R^{n+1}-R^n)\parallel _{2,2}\leq c_2{\e^{2}}\parallel\nu^{\frac{1}{2}}( R^n-R^{n-1})\parallel _{2,2},\quad n\geq 1,
\end{eqnarray*}
for some constant $c_2$. Namely,
\begin{eqnarray*}
&&\frac{1}{\e } v^\mu\cdot\nabla(R^{n+2}-R^{n+1})- GM^{-1}\frac{\partial(M (R^{n+2}-R^{n+1}))}{\partial v_z}\\
&&= \frac{1}{\e ^2}L_J(R^{n+2}-R^{n+1})+\frac{1}{\e }H_1(R^{n+2}-R^{n+1})
+\frac{1}{\e }G^{n+1},\\
&&(R^{n+2}-R^{n+1})(x,\mp\pi,v)=\frac{M_\mp}{M}\int _{w_z\lessgtr 0} (R^{n+2}-R^{n+1})(x,\mp\pi,w)|w_z|M_-dw,
\quad v_z\gtrless0.
\end{eqnarray*}
Here, $
G^{n+1}= (I-P)G^{n+1}= {J}(R^{n+1}+R^n,R^{n+1}-R^n).
$ It follows that
\begin{eqnarray*}
\parallel \nu^{\frac{1}{2}}( R^{n+2}-R^{n+1})\parallel _{2,2}&&\leq {c} \e^{-\frac{1}{2}}\parallel \nu^{-\frac{1}{2}}G^{n+1}\parallel _{2,2}\\
&&\leq {c}\e^{-\frac{1}{2}}\Big( \parallel \nu^{\frac{1}{2}} R^{n+1}\parallel _{\infty ,2}+\parallel \nu^{\frac{1}{2}} R^{n}\parallel _{\infty ,2}\Big) \parallel \nu^{\frac{1}{2}}( R^{n+1}-R^{n})\parallel _{2,2}\\
&&\leq c_2\e^2\parallel \nu^{\frac{1}{2}}( R^{n+1}-R^{n})\parallel _{2,2}.
\end{eqnarray*}
Consequently,
\begin{eqnarray*}
\parallel \nu^{\frac{1}{2}}R^{n+2}\parallel _{2,2}\leq \parallel \nu^{\frac{1}{2}}(R^{n+2}-R^{n+1})\parallel _{2,2}+...+\parallel \nu^{\frac{1}{2}}(R^{2}-R^{1})\parallel _{2,2}
+\parallel \nu^{\frac{1}{2}}R^1\parallel _{2,2}
\leq 2c_1\e ^{\frac{7}{2}} ,
\end{eqnarray*}
for $\e $ small enough. Similarly $\parallel R^{n+2} \parallel _{\infty,2}\leq 2c_1 {\e}^{\frac{5}{2}}$. In particular $\{R^n\}$ is a Cauchy sequence in $L_M^{2}( [-\pi,\pi]^2 \times \R^3)$. The existence of a solution ${R}$ to (\ref{3.14}) follows. $\square$ \\

From here Theorem \ref{fj0} follows, and as a consequence the first part of Theorem \ref{pro}.

\bigskip

\section {Stability: the expansion.}
\setcounter{section}{3}
\setcounter{theorem}{0}
\setcounter{equation}{0}

In the previous section we have constructed a stationary solution $F_s$  of the Boltzmann equation close to the clockwise roll hydrodynamic solution $h_s$. In  the next two sections we study the behavior in time of a small perturbation of  $F_s$ by writing the perturbation as a truncated $\e$-expansion and in particular in this section we show the decay to zero in time of the first terms of the expansion. This result relies crucially on the hydrodynamical stability under small perturbations of the hydrodynamic roll solution $h_s$. Hence, before starting the construction of the Boltzmann solution, let us recall some known hydrodynamic results. The Oberbeck-Boussinesq (O-B) equations with periodic and rigid boundary conditions (see \cite{Jo}), describing the hydrodynamic behavior of the fluid in the present setup in dimensionless form, are:
\begin{eqnarray}\label{O-B}
& &\partial_t u+ u\cdot \nabla u=\hat \eta \triangle u 
-\nabla  p -  e_zG \theta ,\\
& &{\frac 5 2}(\partial_t \theta+u\cdot \nabla 
\theta  +\lambda u_z)= {\frac5 2}\hat k \Delta
\theta,\nonumber \\
&&\hbox {div}\ u=0.\nonumber
\end{eqnarray}
where $e_z$ is the unit vector in he positive $z$-direction, $u\in \R^2$ and $\theta\in \R$ are the velocity field and the deviation from the linear temperature profile respectively, $\hat \eta $ and $\hat k$ are the dimensionless kinematic viscosity and  conductivity respectively. 
The initial conditions are
$$u(x,z, 0)=u_0(x,z),\quad \hbox{div}\  u_0=0,\quad \theta(x, z, 0)=\theta_0(
x, z)$$ 
for any $x, z\in \Omega_\mu=(-\mu\pi,\mu\pi)\times (-\pi,\pi)$.
The boundary conditions for this problem are
\begin{eqnarray} 
&&u(x,-\pi,t)=u(x,\pi,t)=\theta(x,-\pi,t)=\theta(x,\pi,t)=0, \quad x\in [-\pi,\pi], \quad t>0.
\end{eqnarray} 
Here the notations are as in the Introduction. For a proof of the existence of a global in time solution for small initial data see for example  \cite{FMT}.  
The laminar solution is  the trivial stationary solution $u=\theta=0$. It is the unique solution for $Ra\le Ra_c$ and is asymptotically stable for  $Ra < Ra_c$ \cite{Iu}, \cite {FMT}. After $Ra_c$ a pair of new stationary solutions appear. In \cite {Iu} and \cite{Iu1} it is proved that there exists  $\delta_0$ such that for any $0<\delta \le \delta_0$, there are two stationary roll solutions $(u_s^{\mp},\theta^{\mp}_s)$ corresponding to the Rayleigh number $Ra= Ra_c(1+\delta)$, of the form
 \begin{eqnarray}   
&&\displaystyle{u_s^{\mp}(x,z)=\mp  \delta\  C_0\  \phi + O(\delta^2)}\label{hydro1}\\ 
&& \nonumber\\ &&\displaystyle{\theta_s^{\mp}(x,z)=\mp \delta\ C_0 \  \tau + O(\delta^2)}.
\nonumber
\end{eqnarray}
Here $C_0$ is a positive constant, the couple $(\phi,\tau)$ is the eigenfunction corresponding to the smallest eigenvalue $d_0$ of the linearized problem, namely the solution of  
\begin{eqnarray*}
&&\hat\eta\triangle \phi -\nabla p=e_zG\tau,\quad \hbox {div}\  \phi=0,\quad  \hat k\triangle \tau=d_0\phi_z,\\
&& (\phi,\tau)(x,-\pi)=(\phi,\tau)(x,\pi)=0,\quad (\phi,\tau)(x,z,t)=(\phi,\tau)(x+\mu \pi,z,t). \end{eqnarray*}
In \cite{Iu2} both  solutions are proved to be stable for small perturbations (see also \cite{MW}). All the previous results are stated in the Sobolev spaces $H_2$, but, by general theorems on PDE of  parabolic type \cite{La} (or by the method in \cite{Gh}), the regularity can be improved to higher Sobolev spaces $H_k$. Hence we can state the  stability theorem in a form suited to our purposes. Let $(u_s,\theta_s)\in ( H_k)^3$, $k$ large enough, be the clockwise solution of (\ref{hydro1}).
\begin{theorem}
\label{hydrostab}
Let $(u,\theta)$ be the periodic solution of the following equation
\begin{eqnarray}
\label{hydrostabeq}
&& \partial_t u+ u_s\cdot \nabla u+u\cdot \nabla u_s+u\cdot \nabla u=\hat\eta \triangle u 
-\nabla  p -  e_z G \theta \nonumber\\
&&{\frac 5 2}(\partial_t \theta+u_s\cdot \nabla 
\theta  +u \cdot \nabla \theta_s+\lambda u_z)= {\frac 5 2}\hat k \Delta
\theta\nonumber\\
&&\text{\rm div}\ u=0,\nonumber\\
&&u(x,z, 0)=u_0(x,z),\quad \theta(x, z, 0)=\theta_0(x,z) \quad (x, z)\in [-\mu\pi,\mu\pi]\times [-\pi,\pi]\nonumber\\ 
&&u(x,-\pi,t)=u(x,\pi,t)=\theta(x,-\pi,t)=\theta(x,\pi,t)=0, \quad x\in [-\pi,\pi], \quad t>0.\nonumber
\end{eqnarray}
If  $(u_0,\theta_0)\in (H_k)^3$, $k$ sufficiently large, and $\pa u_0\pa_{H_k}+\pa \theta_0\pa_{H_k}<n_0$, for $n_0$ small enough, then, $(u,\theta)(x,z,t)$ is in $(H_k)^3$ and 
 $\displaystyle{\lim_{t\to\infty}(u,\theta)(t,x,z)=0}$ exponentially in time in $(H_{k'})^3$, for any $k'<k$.
\end{theorem}

Notice that the convective solution $h_{conv}=(u_s, T_s)$ in the Introduction,  (\ref{hydro}) , is related to $u_s,\theta_s$ in (\ref{hydro1}) through the shift $T_s=\theta_s+\lambda (z+\pi)$.
\vskip.2cm
In the previous section, we have constructed a positive stationary solution of the Boltzmann equation (\ref{BoltzEqn}) as  $F_s=M(1+\Phi^\e_s)$. Here, we want to study the evolution of positive perturbations of  $F_s$. The perturbation $\Phi^\e$,  defined through $F=M(1+\Phi^\e_s+\Phi^\e)$, with $F$ a solution to (\ref{BoltzEqn}), has to solve the initial boundary value problem
\begin{eqnarray}
&&\frac{\partial {\Phi^\e }}{\partial t}+\frac{1}{\e } v^\mu\cdot\nabla
{\Phi^\e} -\frac{G}{M}\frac{\partial (M 
\Phi^\e)}{\partial v_z}= \frac{1}{\e ^2}\Big(
{L}{\Phi^\e}+{J}(\Phi^\e,\Phi^\e)+{J}(\Phi^\e_s,\Phi^\e)\Big),\label{teq}\\&&
{\Phi^\e}(0,x,z,v)= \zeta_0(x,z,v),\quad (x,z)\in (-\pi,\pi)^2,\hspace*{0.05in}v\in \mathbb{R}^3,\nonumber\\
&&{\Phi^\e }(t,x,\pm\pi,v)=\frac{M_\pm}{M}\int_{w_z \gtrless 0} |w_z|M{\Phi^\e}(t,x,\pm\pi,w)dw,
 \hspace*{0.05in}v_z\lessgtr 0, \hspace*{0.05in}t>0, \hspace*{0.05in}x\in[-\pi,\pi].\nonumber
\end{eqnarray}
The following initial perturbations $F(0,x,z,v)-F_s:=\Phi^\e(0,x,z,v)=\zeta_0(x,z,v)$ are considered,
\begin{equation}\zeta_0(x,z,v)=\sum_{n=1}^5\e^n \Phi^{(n)}(0,x,z,v)+\e ^5p_5,\label{bcfn}
\end{equation}
where 
\begin{equation}
 F(0, \,\cdot\,,\,\cdot\,,\,\cdot\,) \geq 0,\quad \parallel p_5 \parallel_{\infty ,2} := \sup_{\e>0}\left(\int \sup_{(x,z)\in[-\pi,\pi]^2}
p^2_5(x,z,v) Mdv\right)^{\frac{1}{2}}<c,\label{3.8.0}
\end{equation}
for some constant $c$.  The nonhydrodynamic part of the functions $\Phi^{(n)}(0,x,z,v)$ is  determined by the expansion as  explained below together with some terms  of the hydrodynamic part. We will denote by $I_i^{(n)}(t,x,z)$ the coefficients of the functions $\psi_i$ in the hydrodynamic part of $\Phi^{(n)}(t,x,z,v)$. The functions  $I_i^{(1)}(t,x,z)$ will be determined by the solution   $(u,\theta)$ in Theorem \ref{hydrostab}.   Finally, we  require
 $$\int _{ [-\pi,\pi] ^2\times\mathbb{R}^3} \zeta_0(x,z,v)M\psi_0(v) dxdz dv=0.$$
 Since the Boltzmann equation conserves the total  mass, it follows that $\Phi ^{\e }$ will satisfy  
 \begin{eqnarray*}
 \int _{[-\pi,\pi]^2\times\mathbb{R}^3}  M\Phi ^{\e }(x,z,v)\psi_0(v) dv dxdz=0, \quad t>0.
\end{eqnarray*}
We write an $\e$-expansion for $\Phi^\e$ in the form
$$\Phi^\e(t,x,z,v)=\sum_{n=1}^5 \Phi^{(n)}(t,x, z,v)\e^n + \e  R(t,x,z,v).$$
 For the proof of stability we need to show that $\Phi^{(n)}(t,x,z,v)$ converge to zero, when time tends to infinity in a suitable norm. To this end we will construct explicitly the first terms of the expansions. The behavior of the higher order terms will then be evident from this analysis. This construction is by now standard and contained in many papers. We give here a sketch of the argument for sake of completeness, and follow closely the analysis in \cite{AEMN}. 

In the following we use the notation $\langle h\rangle=\int_{\R^3}dv h(v)$.
 By plugging the expansion  into  (\ref{teq}), as a first condition, $\Phi^{(1)}$ has to be  a combination of the
collision invariants $\psi_i, i=0,\dots,4$
$$\Phi^{(1)}= 
\Big({ \rho^1} + {u^1\cdot v} + 
\theta^1\frac {|v|^2-  3} {2} \Big)\ ,$$
so that $\rho^1\equiv I^{(1)}_0, \quad u^1_x\equiv I^{(1)}_1$, $0\equiv I^{(1)}_2$, $u^1_z\equiv I^{(1)}_3$, $\theta^1\equiv \frac{2}{\sqrt6} I^{(1)}_4$.
We require that $u^1, \theta^1$ satisfy the  initial and boundary conditions in Theorem \ref{hydrostab} and, as a  consequence, do not need boundary layer correction
to the first order in $\e$. Indeed, in $z=-\pi$ the solution is already of the right type.  On the other hand, 
 $M+\e \Phi^{(1)}$, when evaluated for $z=\pi$, cannot satisfy the boundary conditions, but
differs from it by terms of order $\e^2$, which will appear in the corrections of higher order.  Hence, for $n>1$ the higher order corrections are decomposed into a bulk term $B^{(n)}$ and two boundary layer terms $b^{(n)}_\pm$. 

To determine the functions $\rho^1$, $u^1$ and $\theta^1$ which give $\Phi^{(1)} (= B^{(1)})$, we consider the equation obtained by equating the terms of next order. Note from the previous sections that the stationary solution can also be expanded in $\e$, and denote by $\Phi_s ^{(n)}$ the terms of this expansion. The equation which we get at  next order, by ignoring  boundary layer corrections,  is
\begin{equation}
 v_x
\frac{\partial}{\partial x}{\Phi^{(1)}}+v_z
\frac{\partial}{\partial z}{\Phi^{(1)}} = 
{ L}{B^{(2)}}+J({\Phi^{(1)} }, \Phi^{(1)})+{J}(\Phi^{(1)},\Phi_s ^{(1)})
\label{3.12.1}
\end{equation}
It can be seen as an equation in $B^{(2)}$,  whose solvability conditions give the usual incompressibility  condition and the Boussinesq condition
\begin{equation}
\hbox{div\ } u=0, \phantom{...}\quad\quad
\nabla( \theta^1+ \rho^1)=0.
\label{3.23}
\end{equation}
  The Boussinesq condition fixes $\rho^1=-\theta^1$, up to a constant. To determine $\theta^1$ and  $u^1$ we look at the solvability condition at next order in $\e$. Indeed, once
(\ref{3.23}) is satisfied, we can deduce from
(\ref{3.12.1}) the following expression for $B^{(2)}$, where ${ L}^{-1}$ 
denotes the inverse of the restriction of ${ L}$ to the orthogonal of its null
space,
\begin{equation}
B^{(2)}={ L} ^{-1}\Big[v\cdot\nabla \Phi^{(1)} -J(\Phi^{(1)},\Phi^{(1)})-J(\Phi^{(1)},\Phi_s ^{(1)})\Big] +\sum_{i=0}^ 4\psi_i\  I^{(2)}_i(t, z)\ .
\label{3.24}
\end{equation}
The coefficients $I^{(2)}_i$ are undetermined at this point and will be  partly fixed by the solvability condition for the equation at next order in $\e$ and the rest of them in some later step,
\begin{equation}
\frac{\partial}{\partial t}B^{(1)}+v\cdot\nabla
B^{(2)}-\frac{1}{M}G \frac{\partial}{\partial v_z}(MB^{(1)})= 
{ L}{B^{(3)}}+J({\Phi^{(2)} }, \Phi^{(1)})+{J}(\Phi^{(2)},\Phi_s ^{(1)})+{J}(\Phi^{(1)},\Phi_s ^{(2)}).
\end{equation}
The solvability conditions for this equation,  
\begin{equation}
( \psi_i,[\frac{\partial}{\partial t} B^{(1)}  + v\cdot\nabla B^{(2)}-
\frac{1}{M} G\frac{\partial}{\partial{v_z}} (MB^{(1)})])=0,\quad
i=0,\cdots,4,
\label{3.25}
\end{equation}
produce the equations for $u^1$
and $\theta^1$.  Let us fix  $i=1, 2, 3$ in (\ref{3.25}). Then 
the first term gives the time derivative of $ u^1$. The third  one
reduces to $0$ for $i=1,2$, and to $- G\/ \rho^1$ for $i=3$ after integrating by parts. The term
$ \langle v\otimes v B^{(2)}\rangle$  
gives rise to dissipative transport terms and a term   which can be interpreted as the
second order correction to the pressure $P_2$. The term  ${J}(\Phi^{(1)},\Phi_s ^{(1)})$ in (\ref{3.24}) produces the linear transport terms, depending on the stationary flow. The result is
$$\frac{\partial}{\partial t} u^1 +u^1\cdot \nabla u^1+u^1\cdot \nabla u^1_s+u^1_s\cdot \nabla u^1=\hat\eta \triangle u^1_z-\nabla P_2 + e_z G \rho^1\ .$$
Using the Boussinesq condition we replace the term $G\rho^1$ by $-G\theta^1+const$. The constant can be absorbed in the pressure term that we rename $p$. 

{ \bf Remark.} There are constants (one coming from the Boussinesq condition, another from the pressure condition) at any order which will be determined in the end  by the total mass condition. Since we are asking
that the total mass  of the perturbation is  zero we can put to zero all the constants.

 To get the equation for the temperature, one has to 
 look at (\ref{3.25}) for $i=4$. It is actually more convenient to replace $\psi_4$ with the equivalent $\tilde\psi_4=\frac{1}{2}(v^2-5)$. We have
$$( \tilde\psi_4, f_1)=\frac{5}{2}\theta^1,\quad {G}( \tilde\psi_4,\frac{\partial}{\partial v_z} f_1)=- u^1_z G,\quad 
( v\tilde\psi_4, B^{(2)})
=- \frac{5}{2}\hat k\nabla \theta^1+\frac{5}{2}u^1\theta^1+u^1_s \theta^1+\theta^1_s u^1\ .$$
Putting all the terms together, we get
$$\frac{5}{2}\left[\frac{\partial}{\partial t}  \theta^1+u^1\nabla \theta^1+u^1_s\nabla \theta^1+u^1\nabla \theta^1_s\right ]-Gu^1_z=\frac{5}{2}\hat k\triangle \theta^1\ .$$
This equation has to be solved with boundary conditions $\theta^1(\pm 1,t)=0$ for $t>0$, and an initial condition $\theta^1_0$, which is completely arbitrary.

{ \bf Remark.}  In the previous equation there is a term $-Gu^1_z$ which does not appear in the usual O-B equations. This term can be absorbed by changing the boundary conditions. 
Here $\theta^1_s,u^1_s$ are the  hydrodynamic terms of first order in $\e$ in the expansion of $\Phi_s$, and hence they coincide with $u_s,T_s$ in (\ref{hydro}). The boundary conditions for $T_s$ are: $T_s(x,-\pi,t)=~0, T_s(x,\pi,t)=2\lambda \pi$. The shifted temperature $\tilde T_s=T_s -Gz$ will satisfy  the usual Boussinesq equation, in which the term $G u_z$ is missing and a different boundary condition, $T_s(x,\pi,t)=(2\lambda- G) \pi$.
This aspect was discussed in \cite{EML}. It was pointed out that starting from the compressible Navier-Stokes equation or the Boltzmann equation in the scaling we are considering, one obtains a set of equations 
which differ from the usual O-B ones  for this shift in the boundary condition for the temperature. By scaling the variables, this amounts to the usual O-B equations  in dimensionless form, with a new Rayleigh number given by $Ra(1-G)$. We conclude that nothing changes in our analysis.

To summarize what we got so far, $(u^1, \theta^1)$ has to satisfy the O-B   equations in Theorem \ref{hydrostab}. By fixing the initial conditions so that the assumptions of the Theorem are satisfied, we get that  $(u^1, \theta^1)$
vanishes exponentially in time, with its spatial derivatives. Since $\theta^1$ differs from $\rho^1$ by a constant, which can be taken as zero,   we may conclude that for $q\in [1,+\infty]$, $\parallel \Phi^{(1)}\parallel_{q,2}$ is finite and converges to zero exponentially in time.

 The second order term in the expansion, $\Phi^{(2)}$, is not yet completely determined.
 Equation (\ref{3.25}) with $i=0$ gives
 $\displaystyle{\frac{\partial}{\partial t}  \rho^1=\hbox {div}\ {I}^{(2)}}$, fixing 
 $\hbox {div}\ {I}^{(2)}$. 
Moreover, a combination  of 
$I^{(2)}_0$ and $I^{(2)}_4$ contributes to the pressure $p$ which is
determined by the previous equations, so that these parameters are not
independent. 

The nonhydrodynamic part of $B^{(2)}$ is a linear function of   
the derivatives of $\rho^1, \theta^1$ which are in general
different from zero at the boundaries. Therefore the non hydrodynamical part of $B^{(2)}$ is completely fixed (even at time zero) and  violates the boundary conditions. We need to introduce $b_{\pm}^{(2)}$ to
restore the boundary conditions by compensating the non hydrodynamical part of $B^{(2)}$ which is not Maxwellian. We explain how to find the correction  $b^{(2)}_-$. The correction $b^{(2)}_+$ is found in a similar way. Here $L^-=2M_-^{-1}Q(M_-,M_-\,\cdot\,)$. We choose 
$b^{(2)}_-$
by solving, for any $t>0$, the Milne problem for $z^->0$,
\begin{equation}v_z{\frac\partial {\partial z^-}} h -\e^2 \frac{1}{M}G^-{\frac \partial{\partial v_z}}(Mh)={L}^- h, \phantom{...}\langle v_zh\rangle:=\int_{\R^3}dv 
v_z  h =0,\label{Milne}\end{equation}
where $z^-=\e^{-1}(z+\pi)$ is defined  as  the rescaled $z$ variable near the bottom plate, and $G^-$ is a smooth force rapidly decaying to zero far from the bottom plate. Indeed, the gravity force has been decomposed in three parts, a   force constant in the bulk and two boundary parts $G^\pm$ (see \cite{EML}, \cite{ELM2} for details). We impose the boundary condition at $z^-=0$ in such a way that the incoming flux of $h$ at $z=-\pi$, $v_z>0$,  is  given
by $(I-P) B^{(2)}(-1,v;t)$. The results in \cite{CME} tell
us that as $z^-\to +\infty$ the solution approaches a function 
$q^{(2)}_-(v,t)$ in  $\hbox{\tt Kern}\ {L}^-$. 
Note that in $q^{(2)}_-$ there is no term proportional to $\psi_3$ because of the vanishing mass flux condition in the direction of the $z$ axis $\langle
v_z  h \rangle=0$.
Thus we set $b^{(2)}_-(x,z_-,v,t)= h(x,z_-,v,t)-q^{(2)}_-(x,v,t)$, which will go
to zero at infinity exponentially in $z^-$. This produces a term $b^{(2)}_-(x,2\pi\e^{-1},v,t)=\psi_{2,\e}(x,\pi,v,t)$, exponentially small in $\e^{-1}$ on the opposite boundary. Scaling again to the variable $z$, the resulting term in the expansion is thus $\Phi^{(2)}=B^{(2)}+b_+^{(2)}+b_-^{(2)}$, and is such that in $z=-\pi$, for example, it has zero non hydrodynamic part, while the hydrodynamic part  is 
$$\Phi^{(2)}(x,-\pi,v; t)=\sum_{i=0}^4I_i^{(2)}(x,-\pi; t)\psi_i(v)+ 
b^{(2)}_+( x,2\e^{-1},v, t) -q^{(2)}_-, \phantom{..} v_z>0, \phantom{..}
t> 0.$$ 
We are not yet done since $M\Phi^{(2)}(x,-\pi,v)$ is not Maxwellian for $v_z>0$, (as it should, in order to satisfy the boundary conditions) because of the presence of terms proportional to $\psi_i$, $i=1,2,4$ in $q^{(2)}_-$ and $b^{(2)}_+( t,x,2\e^{-1})$. The latter is not important because it is a correction exponentially small in $\e$ and will be compensated  in the remainder. The former will be compensated by the coefficients $I_i^{(2)}$, $i\ne 0,3,$ that can be chosen arbitrarily on the boundaries. Moreover, 
 we have to choose $I^{(2)}_3=0$ on the boundaries, because $\langle v_zq^{(2)}_-\rangle=0$.  We are left with
$$\Phi^{(2)}(t,x,\pm \pi, {v_z \lessgtr 0} )=\alpha^{\pm}_2 M_{\pm} + 
\psi_{2,\e}(x,\pm\pi,v,t),\phantom{...}\alpha^{\pm}_2 =I^{(2)}_0(\pm 1)- 
\langle q_{\pm}^{(2)}\rangle ,$$
where $\psi_{2,\e}$ are terms exponentially small in $\e$. 
Finally,  we impose the impermeability condition $\langle v_z\Phi^{(2)}\rangle=0$ by choosing
 $$\a_2^\pm=\frac{M_\pm}{M}\int_{v_z\gtrless0}v_zM[\Phi^{(2)}(t,x,\pm \pi, v )-\psi_{2,\e}(t,x,\pm \pi, v )]dv,\quad {v_z \lessgtr 0}, t>0 .$$

The coefficients $I^{(2)}_i, i=1,2,4$ of  the hydrodynamical part
of $B^{(2)}$  are determined by the compatibility condition for the equation at  next order in $\e$,
$$\left(\psi_i,\big[ \frac{\partial}{\partial t} B^{(2)} +v\cdot\nabla B^{(3)} +
G\frac{\partial}{\partial v_z}B^{(2)}\big]\right)=0,$$
where 
$$
B^{(3)}={ L} ^{-1}\Big[\frac{\partial}{\partial t} \Phi^{(1)} +v\cdot\nabla B^{(2)} + \frac{1}{M}G\frac{\partial}{\partial v_z}
(M\Phi^{(1)})-J(\Phi^{(1)},B^{(2)})$$
$$-J(\Phi_s^{(1)},B^{(2)})-J(\Phi^{(1)},B_s^{(2)})\Big]
 +\sum_{i=0}^ 4\psi_i\  I^{(3)}_i,$$
together  with the boundary conditions
$I^{(2)}_i=(q_-^{(2)})_{ i}, \  i=1, 2, 4$.  Then
$I^{(2)}_0$ is found up to a constant that is chosen so that the total
mass associated to $\Phi^{(2)}$ vanishes.
Proceeding as in the determination of the Boussinesq equation, we find now a set of
three linear time-dependent nonhomogeneous  Stokes equations 
for $I^{(2)}_i$,
\begin{eqnarray*}
&&\partial_t\rho^2+\hbox{div}\ u^2+\hbox{div}\ (\rho^1u^1)+\hbox{div}\ (u^3)=N_0\\
&&\partial_t u^2 =u^2\cdot\nabla u^1+u^1\cdot\nabla u^2=\hat\eta\triangle u^2- \nabla P^3+G\rho^2+\nabla \hbox{div}\ u^2+N\\
&&\partial_t \theta^2+\frac{2}{3}[\hbox{div}\ u^3+ (\rho^1+\theta^1)\hbox{div}\ u^2]+\rho^1[\partial_t \theta^1+\frac{2}{3}\hbox{div}\ u^2]=\hat k\frac{2}{3}[\triangle \theta^2+(\nabla u_1)^2]+N_4,\\
\end{eqnarray*}
where $N_0, N_4$ depend on the third order spatial derivatives of $\rho^1, \theta^1$ and $N$ depend on the third order spatial derivatives of  $u^1$. We remember that $P^2$ is determined by $p$ which has been found at the previous step by solving the O-B for $u^1,\theta^1$. On the other hand, $P^2=\rho^2+\theta^2+\rho^1\theta^1$ allows us to eliminate $\rho^2$ from the previous equations. Replacing $\hbox{div}\ (u^3)$ as given from the first equation in the last one, and using the condition on $P^2$, we get a set of two coupled equations for $\theta^2, u^2$ with a constraint on $\hbox{div} u^2$. $P^3$ plays the role of Lagrangian multiplier for this constraint. The nonhomogeneous term is controlled by the results at the previous step and hence is known to decay to zero exponentially in time in the right norms.
Then, general theorems for the Stokes equation assures the existence of a solution for 
 the  chosen boundary  conditions, vanishing exponentially  in time.

Once $B^{(2)}$ is completely determined, the last equation gives the non-hydrodyna\-mical part of
$B^{(3)}$. As before, we introduce the terms $b_\pm^{(3)}$ to compensate $(I-P)B^{(3)}$
on the boundaries $z=\pm \pi$. The term $b_\pm^{(3)}$ is found as a solution of a Milne
problem with a source term, which depends on the previous boundary corrections $b_\pm^{(2)}$ and $\Phi^{(1)}$. The procedure can be continued to any order. 

We notice that  $(I-P)\Phi^{(n)}$ at time zero are not arbitrary, since they depend on $\Phi^{(n-1)}$ and its derivatives. We can instead assign at time zero $I^{(n)}_i, i=1,2,4$. Notice that the rest term $R$ at time zero is of order $\e^4$. 
By using the results in \cite{CME} and  the exponential decay in time of  
$\Phi^{(n)}$ we can state the following theorem

\begin{theorem}
\label{fjt}
Assume that at time zero, for some  suitably large $k$,
$$\parallel M \partial^kI_i^{(n)}(0,x,z)\parallel_{L^2}< \infty,\quad i=1,2,4,\quad n=1,\dots 5\ ,$$
where $\partial^k$ denotes any space derivative of order $k$.
Then, it is possible to determine the functions $\Phi^{(n)}$, $n=2,\dots,5$ in the asymptotic
expansion  satisfying the boundary  conditions
\begin{eqnarray*}&&
{ \Phi^{(n)}}(t,x,\mp \pi,v)=\frac{M_\pm(v)}{M(v)}\int_{w_z \lessgtr 0} |w_z|M\big[\Phi^{(n)}(t,x,\mp\pi,w)-\psi_{n,\e}(t,x,\mp\pi,w)\big]dw\\
&&+\psi_{n,\e}(t,x,\mp \pi,v),
\hspace*{0.05in}t>0, \hspace*{0.05in}v_z\gtrless 0, t>0,\nonumber\\&&
 \end{eqnarray*}
the normalization condition $\displaystyle{
\int_{\mathbb{R}^3\times\Omega_\mu}M
\Phi^{(n)}dv\/dzdx =0,\quad t\in \mathbb{R}^+,}$ 
and
$$\parallel \Phi^{(n)}\parallel _{2,2,2}<\infty,\quad \parallel \Phi^{(n)}\parallel _{\infty ,\infty,2}<\infty\ .
$$
Here,
$$\parallel f\parallel _{2,2,2}= \Big( \int _0^\infty\int_{\Omega_\mu}\int _{\mathbb{R}^3}|f(s,x,z,v)|^2M(v)dsdxdzdv\Big) ^{\frac{1}{2}},$$
$$
\parallel f\parallel _{\infty ,\infty ,2}= \sup _{t>0} \Big( \int _{\mathbb{R}^3} \sup
_{(x,z)\in\Omega_\mu} |f(t,x,z,v)|^2M(v)dv\Big) ^{\frac{1}{2}}\ .$$
\end{theorem}

\bigskip

\section{Stability: the remainder}
\setcounter{section}{4}
\setcounter{theorem}{0}
\setcounter{equation}{0}

We remember that $\Phi^\e$, the solution to (\ref{teq}), is written  as $\Phi^\e=\sum_{i=1}^5 \e ^i \Phi^{(i)} +\e R$. In the previous section we have constructed the terms $\Phi^{(i)}$ and shown that they decay to zero in suitable norms. In this section we construct the rest term $R$,  solution of
\begin{eqnarray}
&&\frac{\partial R}{\partial t}+\frac{1}{\e }\mu\ v_x\frac{\partial R}{\partial x}+\frac{1}{\e }v_z\frac{\partial R}{\partial z}-GM^{-1}\frac{\partial (M R)}{\partial v_z}=\frac{1}{\e ^2} L R+\frac{1}{\e } J(R,R)+\frac{1}{\e }H(R)+ A,\label{4.1}\\
&&R(0,x,z,v)= R_0(x,z,v)=\e ^4 p_5(x,z,v),\quad \quad\nonumber\\
&&R(t,x,\mp\pi,v)= \frac{M_\mp}{M}\int _{w_z\lessgtr 0} (R(t,x,\mp\pi,w)+\frac{\bar{\psi}}{\e}(t,x,\mp\pi,w))|w_z|Mdw - \frac{\bar{\psi}}{\e}(t,x,\mp\pi,v),\nonumber\\
&&\qquad \qquad \qquad x\in[-\pi,\pi], \quad t>0,\hspace*{0.03in}v_z>0,\nonumber
\end{eqnarray}
Here $\bar{\psi}(t,x,\pm\pi,v)=\sum_n\e ^n \psi_{n,\e}(t,x,\pm\pi,v)$  is the Knudsen part of the asymptotic expansion from $(t,x,\mp\pi,v) $, exponentially small when evaluated at $(t,x,\pm\pi,v)$. ${A }$ contains all the terms fully coming from the asymptotic expansion, 
\begin{eqnarray*}
H(R)=\frac{1}{\e}J( R,\bar{\Phi}+{\Phi_s}),
\end{eqnarray*}
where $\bar{\Phi}=\sum_1^5\Phi^{(j)}\e^j$, and $\int p_5Mdxdzdv=0$. We shall require that the initial value of $\Phi ^\e $ is close to zero, and in the sequel introduce smallness assumptions.

 The following norms will be used,
\begin{eqnarray*}
\parallel R\parallel _{2t,2,2}&&= \Big( \int _0^t\int_{-\pi}^{\pi}\int_{-\pi}^{\pi}\int _{ \R^3}R^2(s,x,z,v)M(v)dsdxdzdv\Big) ^{\frac{1}{2}},\\
\parallel R\parallel _{\infty ,2,2}&&= \sup _{t>0}\Big( \int_{-\pi}^{\pi}\int_{-\pi}^{\pi}\int _{ \R ^3}R^2(t,x,z,v)M(v)dxdzdv\Big) ^{\frac{1}{2}},\\
\parallel R\parallel _{\infty ,\infty ,2}&&= \sup _{t>0} \Big( \int _{ \R^3} \sup _{-\pi<x,z<\pi} R^2(t,x,z,v)M(v)dv\Big) ^{\frac{1}{2}},
\end{eqnarray*}
\begin{eqnarray*}
\parallel f \parallel _{2t,2,\sim}&&=\Big( \int _0^t\int_{-\pi}^{\pi}\int _{v_z>0} v_z M(v) \mid f(s,x,-\pi,v)\mid ^{2}dvdxds\Big)^{\frac{1}{2}}+\\
&&\Big( \int_0^t \int_{-\pi}^{\pi}\int _{v_z<0} \mid v_z \mid M(v) \mid f(s,x,\pi,v)\mid ^{2}dvdxds\Big)^{\frac{1}{2}} ,\\
\parallel f \parallel _{\infty,2,\sim}&&=\Big( \sup _{t>0}\int_{-\pi}^{\pi}\int _{v_z>0} v_z M(v) \mid f(t,x,-\pi,v)\mid ^{2}dxdv\Big)^{\frac{1}{2}}+\\
&&\Big( \sup _{t>0}\int_{-\pi}^{\pi}\int _{v_z<0} \mid v_z \mid M(v) \mid f(t,x,\pi,v)\mid ^{2}dxdv\Big)^{\frac{1}{2}} .
\end{eqnarray*}

\setcounter{theorem}{0}

We will prove the existence of $\Phi ^{\e }$ and the stability result (\ref{stabi}).
 We follow closely the approach in Section 2, starting from
 dual, space-periodic solutions
to a linear problem (in the rescaled time variable $\bar{\tau}=\e^{-1} t$) discussed in the following lemma. We use the notations introduced in Section 2,  ${L}_J=L(\,\cdot\,)+\e J(q,P\,\cdot\,)$, but here the function $q$ has the expression  $q=\e^{-1}(\bar{\Phi}+{\Phi}_s)$ which is also time dependent.
\medskip

\begin{lemma}
\label{l4.1}
Let $\varphi(\bar{\tau },x,z,v)$ be solution to 
\begin{eqnarray}
  \frac{\partial \varphi }{\partial \bar{\tau }}+\mu\ v_x \frac{\partial \varphi}{\partial x} +v_z \frac{\partial \varphi}{\partial z} -\e G M^{-1}\frac{\partial 
(M\varphi)}{\partial v_z}= \frac{1}{\e }{L_J^*}\varphi +g, \label{4.3}
\end{eqnarray} 
periodic in $x$ of period $2\pi$, with zero initial and ingoing boundary values
at $z=-\pi,\pi$, and $g$ $x$-periodic of period $2\pi$. 
Set $\tilde \varphi =\varphi-<\varphi>=\varphi-(2\pi)^{-2 } \int \varphi dxdz$. 

Then, if $\e\le\e_0$, $\delta\le\delta_0$, for $\e_0,\delta_0$ small enough, there exists $\eta$  small such that,
\begin{eqnarray*}
\parallel {\varphi} \parallel _{\infty ,2,2}\leq c\Big( \e^{\frac{1}{2}} \parallel \nu ^{-\frac{1}{2}}(I-P){g}\parallel _{2,2,2}+\e ^{-\frac{1}{2}}\parallel Pg\parallel _{2 ,2,2}+\eta\e^{\frac{1}{2}}\parallel<P\varphi>\parallel_{2,2}\Big) ,\\
\parallel \nu ^{\frac{1}{2}}(I-P){\varphi} \parallel _{2 ,2,2}\leq c\Big( {\e} \parallel \nu ^{-\frac{1}{2}}(I-P)g\parallel _{2 ,2,2}+\parallel Pg\parallel _{2,2,2}+\eta\e\parallel<P\varphi>\parallel_{2,2}\Big),\\
\parallel\widetilde{ P\varphi} \parallel _{2 ,2,2}\leq c\Big( \parallel \nu ^{-\frac{1}{2}}(I-P)g)\parallel _{2 ,2,2}+\e ^{-1}\parallel Pg\parallel _{2,2,2}+\eta\parallel<P\varphi>\parallel_{2,2}\Big) .
\end{eqnarray*}
\end{lemma}

\noindent\underline{Proof of Lemma \ref{l4.1}.}\hspace{0.05in} A variant of the method in [Ma Scn 7.3] can be adapted to the present setting with a force term, to obtain the existence of a solution to (\ref{4.3}). \\
Denote by $\hat{\varphi }(\bar{\tau },\xi ,v)$, $\xi=(\xi_x,\xi_z) \in \Z^2 $ the Fourier transform of $\varphi $ with respect to space, and define $\hat{g} $ analogously. Then for $\xi\neq (0,0)$,
\begin{eqnarray*}
\frac{\partial \hat{\varphi }}{\partial \bar{\tau }}=  \frac{1}{\e }\widehat{{L}_J^*\varphi}-i\xi \cdot  v^\mu   \ \hat{\varphi}+\e G M^{-1}\frac{\partial 
(M\hat{\varphi}) }{\partial v_z} +\hat{g}- |v_z|r(-1)^{\xi_z}.
\end{eqnarray*}
Here  $v^\mu=(\mu v_x,v_z)$,  $r={\cal{F}}_x \varphi(\bar{\tau},\xi_x,\pi,v)$ for $v_z>0$,  $r={\cal{F}}_x\varphi(\bar{\tau},\xi_x,-\pi,v)$ for $v_z<0$, with ${\cal{F}}_x $ denoting Fourier transform with respect to the $x$-variable. Let $\beta $ be a truncation function belonging to $C^1(\R)$ with support in $(0,\infty] $, and such that $\beta (\bar{\tau })= 1$ for $\bar{\tau } >\tau_0 $ for some $\tau_0 >0 $. Let $\overline{\hat{\varphi }}= \hat{\varphi }\beta $. Then, for $(0,0)\neq \xi \in {Z}^2$,
\begin{eqnarray*}
\frac{\partial \overline{\hat{\varphi }}}{\partial \bar{\tau } }= \frac{1}{\e }\widehat{L_J^* \beta \varphi}+i\xi \cdot v^\mu\overline{\hat{\varphi }}+\e G M^{-1}\frac{\partial 
(M \bar{\hat{\varphi}})}{\partial v_z}+\hat{\varphi }\frac{\partial \beta }{\partial \bar{\tau }}+\hat{g}\beta -|v_z|r\beta(-1)^{\xi_z}.\end{eqnarray*}
Let $\cal{F}$ be the Fourier transform in $\bar{\tau }$ with Fourier variable $\sigma $. We put
$$\Phi = {\cal F}\overline{\hat{\varphi }},\quad \tilde{Z}= {\cal F} \left(\e ^{-1}\widehat{ L_J ^*\beta {\varphi }}+\e GM^{-1}\frac{\partial 
(M \bar{\hat{\varphi}}) }{\partial v_z}
+\hat{\varphi } \frac{\partial \beta }{\partial \bar{\tau }}+\hat{g} \beta-|v_z|r\beta (-1)^{\xi_z}\right),$$
$$ {Z}= {\cal F}\left (\e ^{-1}\widehat{ L_J^* \beta \varphi }+\hat{g} \beta-|v_z|r\beta(-1)^{\xi_z}\right),\  {Z'}= {\cal F} \left(\e ^{-1}\widehat{ L_J^* \beta \varphi }+\hat{g} \beta\right),\ \hat{U}= \left(i\sigma +i\xi \cdot v^{\mu }\right)^{-1}.$$
Let $\chi $ be the indicatrix function of the set $\displaystyle{
\{ v;\hspace*{0.05in}\mid \sigma+\xi \cdot v^\mu \mid <\alpha |\xi| \} ,
}$
for some positive $\alpha $ to be chosen later.  Similarly to Section 2, the  elements $\bar\psi _0,...,\bar\psi _4$ are an orthonormal   basis for the kernel of $L_{J}^*$. Let $\zeta_s(v)= (1+\mid v\mid )^s$. For $\xi \neq (0,0) $
\begin{eqnarray*}
\parallel P(\chi \Phi )\parallel&&\leq c\sum_{j=0}^4 \left| \int \chi \Phi (\sigma,\xi ,v)\bar{\psi}_jMdv\right|\   \parallel \bar{\psi}_j\parallel\\
&&\leq c\parallel \zeta_{-s} \chi\Phi \parallel\sum_{j=0}^4 \parallel \chi \zeta_s\bar{\psi}_j\parallel 
\leq c\sqrt{{\alpha }}\parallel \zeta_{-s}\chi \Phi \parallel.
\end{eqnarray*}
Use this estimate on the support of $\chi$ for 
 $
\alpha = \parallel \zeta_{-s}\Phi \parallel^{-1}\parallel \zeta_{-s}Z'\parallel.
$

As in Section 2, the previous estimate also holds with respect to $\text{\rm supp}\  \chi_1 $ where the indicatrix function $\chi _1$ is taken for $\alpha=\sqrt{\delta_1}$. We fix $\delta_1 $ so that $c\sqrt \delta_1 << 1$. Then the above estimate gives that the
hydrodynamic $P$-part of the right-hand side, $\parallel P( \chi_1\Phi )\parallel$, can be  absorbed by  $\parallel P( \chi_1 \Phi) \parallel $ in the left-hand side. The estimates hold in the same way when  $\chi_1$ is suitably smoothed around  $\sqrt{\delta_1} |\xi|$. For the remaining $(1-\chi )(1-\chi _1)\Phi=\chi^c\chi_1^c \Phi$ we shall use that
$\Phi = -\hat{U}\tilde{Z}$. Then
\begin{eqnarray*}
&&\parallel P\chi^c\chi_1^c\Phi \parallel ^2\leq c\Big( \parallel \zeta_s\chi^c\chi_1^c\hat{U}\parallel ^2 +
\parallel \zeta_{s+2}\chi^c\chi_1^c\hat{U}\parallel ^2\Big) \parallel \zeta _{-s}Z'\parallel ^2+\tilde\Theta\\
&&+\frac{C\parallel {\cal{F}} (\sqrt{|v_z|}\beta r) \parallel^2}{\delta_1|\xi|^2}\leq \frac{C}{{\mid \xi \mid ^2 \mid \alpha \mid }}\parallel \zeta_{-s}Z'\parallel ^2 +\frac{\parallel {\cal F} (\sqrt{|v_z|}\beta r)\parallel^2}{\delta_1|\xi|^2}+\tilde\Theta
\end{eqnarray*}
where
$$\tilde\Theta:=- 2\sum_{j=0}^4   \int \bar{\psi} _j \chi^c\chi_1^c \hat{U} {\cal{F}} \Big(\e GM^{-1}\frac{\partial 
(M \bar{\hat{\varphi}})}{\partial v_z}
+\hat{\varphi}\frac {\partial \beta}{\partial \bar {\tau}}\Big)Mdv \Big(\int \bar{\psi} _j\chi^c\chi_1^c({\cal{F}}\hat{\varphi} \beta -\hat{U} Z)M dv\Big)^*  .$$
We again replace $\alpha $ in the denominator by
 $\parallel \zeta _{-s}\Phi \parallel^{-1}\parallel \zeta _{-s}Z'\parallel$.  That gives
\begin{eqnarray*}
 \parallel P\Phi \parallel ^2\leq c(\parallel \zeta _{-s}\Phi \parallel\parallel \zeta _{-s}Z'\parallel +\frac{\parallel {\cal{F}}(\sqrt{|v_z|}\beta r)\parallel^2}{\delta_1|\xi|^2}+\sqrt{\delta_1}\parallel \zeta_{-s}(I-P)\Phi \parallel^2)+
\tilde\Theta. 
\end{eqnarray*}
Hence,
\begin{eqnarray*}
 \parallel P\Phi \parallel ^2 _H&&\leq C\Big( (\parallel P\Phi \parallel+\parallel \zeta _{-s}(I-P)\Phi \parallel )\parallel \zeta _{-s}Z'\parallel
 +\frac{\parallel {\cal{F}}\sqrt{|v_z|}\beta r\parallel^2}{\delta_1|\xi|^2}\\ &&+\sqrt{\delta_1}\parallel \zeta_{-s}(I-P)\Phi \parallel^2\Big )
+\tilde\Theta . 
\end{eqnarray*}
Consequently,
\begin{eqnarray*}
 \parallel P\Phi \parallel ^2&&\leq c\Big(  \parallel \zeta _{-s}Z'\parallel ^2+\frac{\parallel {\cal{F}}(\sqrt{|v_z|}\beta r)\parallel^2}{\delta_1|\xi|^2}
 +\parallel \zeta _{-s}(I-P)\Phi \parallel\parallel \zeta _{-s}Z'\parallel\\ && +\parallel\zeta_{-s}(I-P)\Phi\parallel^2 \Big) 
+\tilde\Theta .
\end{eqnarray*}
We next discuss the term $\tilde\Theta$.
 The first term in the first integral can be bounded by $\e $ times an integral of a product of $M$, $1+|\xi_z |$, a polynomial in $v $, $\mid {\cal{F}} \bar{\hat {\varphi}} \mid $ and $\hat U$ or $\hat U ^2$. So this factor is bounded by $\e C \parallel \Phi \parallel $. And so,
\begin{eqnarray*}
&&\parallel P\Phi \parallel^2\leq C\Big( \parallel \zeta _{-s}Z'\parallel^2+\frac{\parallel {\cal{F}}\sqrt{|v_z|}\beta r\parallel^2}{\delta_1|\xi|^2}+\parallel (I-P)\Phi \parallel^2\Big) \\
&&-2\sum_{j=0}^4  \int \bar{\psi }_j\chi^c\chi_1^c \hat{U}\Big({\cal{F}}\hat{\varphi}\frac {\partial \beta}{\partial \bar {\tau}}\Big)Mdv \Big( \int \bar{\psi}_j\chi^c\chi_1^c({\cal{F}}\hat{\varphi} \beta -\hat{U} Z)M dv\Big)^* .
\end{eqnarray*}
Therefore for $\xi\neq (0,0)$,

\begin{eqnarray*}
&&\int (P\Phi )^2(\sigma ,\xi ,v)Mdvd\sigma \hspace{5cm}\\
&&\leq C\Big( \frac{1}{\e ^2}\int d\sigma\parallel \zeta _{-s}(v){\cal F}\widehat{L_J^*\varphi} )(\sigma , \xi , \cdot ){\parallel}^2 
+\int d\sigma\Big(\parallel (I-P)\Phi (\sigma ,\xi ,\cdot )\parallel^2  \\
&&+\frac{\parallel {\cal{F}} \sqrt{|v_z|}\beta r\parallel^2}{\delta_1|\xi|^2}\Big)+\int \parallel \nu^{-\frac{1}{2}} \hat{g}\beta (\bar{\tau} ,\xi ,\cdot )\parallel^2d\bar{\tau} \Big) \\
&& - 2\sum_{j=0}^4 \int d\sigma \int \bar{\psi} _j\chi^c\chi_1^c \hat{U}({\cal{F}}\hat{\varphi}\frac {\partial \beta}{\partial \bar {\tau}})Mdv \Big( \int \bar{\psi} _j\chi ^c\chi _1^c({\cal{F}}\hat{\varphi} \beta -\hat{U} Z)M dv\Big)^*   .
\end{eqnarray*}
Sending $\tau_0$  to zero implies that
\begin{eqnarray}
&&\int _{0}^{\infty}d\bar{\tau} \int (P\hat{\varphi } )^2(\bar{\tau },\xi ,v)Mdv
\leq C\Big( \frac{1}{\e ^2}\int _{0}^{\infty}d\bar{\tau }
\Big( \parallel \zeta _{-s}(v)\widehat{L_J^*\varphi} )(\bar{\tau} , \xi , \cdot )\parallel^2\quad\label{parse}\\
&&+ \parallel (I-P)\hat{\varphi} (\bar{\tau} ,\xi ,\cdot )\parallel^2\Big)+\int _0^{\infty}d\bar{\tau }\int \nu^{-1} \hat{g}^2(\bar{\tau },\xi ,v)Mdv+\int_0^{\infty}d\bar{\tau }\frac{\parallel \sqrt{|v_z|}r\parallel^2}{\delta_1|\xi|^2}\Big) .\nonumber
\end{eqnarray}
Taking $\delta$ and $\e$ small enough and summing the previous inequality over all $0\neq\xi \in  \Z^2$, implies, by the Parseval inequality, that
\begin{eqnarray*}
&&\int _0^{\infty}\int (\widetilde{P\varphi })^2(\bar{\tau },x,z,v)Mdvdxdzd\bar{\tau }  \\
&&\leq c\Big( \frac{1}{\e ^2}\int _0^{\infty}\int \nu ((I-P){\varphi} )^2(\bar{\tau },x ,z,v)Mdvdxdzd\bar{\tau } \nonumber \\
&&+\int _0^{\infty}\int \nu ^{-1}{g}^2(\bar{\tau },x,z,v)Mdvdxdzd\bar{\tau }+\parallel \gamma ^{-}\varphi \parallel^2_{2\infty,2,\sim}+\eta\parallel\varphi\parallel^2_{2,2,2}\Big) \nonumber.
\end{eqnarray*}
As in Section 2, to use  an argument based on a variant of Green's formula, we 
multiply the equation (\ref{4.3})  by $2{\varphi} M \kappa$, and integrate over $[0,\bar{T}] \times [0,2\pi]^2 \times \R^3$, integrate by parts and obtain, by using the spectral inequality (\ref{spgap}) and the bounds $1\le\kappa(z)\le \text{\rm e}^{2\e G\pi}$,
\begin{eqnarray*}
&&\parallel \gamma ^- \varphi \parallel^2_{2\bar{T},2,\sim}+\parallel {\varphi} \parallel _{2\bar{T},2,2}^2+\frac{1}{\e }\parallel \nu ^{\frac{1}{2}}(I-P){\varphi} \parallel _{2 \bar{T},2,2}^2\nonumber \\
&&\leq c(\e \parallel \nu ^{-\frac{1}{2}}(I-P){g}\parallel _{2 \bar{T},2,2}^2+\eta _1\parallel P{\varphi} \parallel _{2\bar{T},2,2}^2+\frac{1}{\eta _1}\parallel Pg\parallel _{2\bar{T},2,2}^2).
\end{eqnarray*}
Inserting this into the previous inequality, the lemma follows. $\square$

\bigskip

We next decompose the operator $H$ in the remainder equation in accordance with the operator $L_J$: $H(\,\cdot\,)=\tilde{J}(q,P\,\cdot\,)+H_1(\,\cdot\,)$. We notice that ${H_1}(R)$ is of order zero in $\e$, and only depends on the nonhydrodynamic part $(I-P)R$.
As in Section 2, to solve the equation for $R$ we shall use an iteration procedure based on the  decomposition of $R$ in the sum $R_1+R_2$, where $R_1$ and $R_2$ are solutions of   two different problems. $R_1$ solves
\begin{eqnarray}
&&\frac{\partial R_1}{\partial t}+\frac{1}{\e }v^\mu\nabla\cdot R_1+\frac{1}{\e }v_z\cdot \frac{\partial R_1}{\partial z}-\frac{G}{M}\frac{\partial 
(M  R_1)}{\partial v_z}= \frac{1}{\e ^2}{L}_JR_1+\frac{1}{\e}H_1(R_1)+\frac{1}{\e }g,\quad\label{4.4}\\
&&R_1(0,x,z,v)= R_0(x,z,v),\nonumber \\
&&R_1(t,x,\mp\pi,v)=-\frac{1}{\e}\bar{\psi}(t,x,\mp\pi,v),\quad t>0,\hspace*{0.05in}v_z\gtrless0,\nonumber 
\end{eqnarray}
Here $R_1$ is periodic in $x$ of period $2\pi$, and $g$ is some given  function, $x$-periodic of period $2\pi$ with $\int Mg(\cdot,x,z,v)dxdzdv \equiv 0$. For the existence of the solution, see the discussion of (\ref{4.3}). The equation for $R_2$ will be introduced below in (\ref{4.8}).\\
The non-hydrodynamic part of ${R_1}$ is again estimated by Green's formula; multiply (\ref{4.4}) by $2{R}_1M\kappa$, integrate with respect to the variables $(\bar{\tau},x,z,v)$ over $[0,\bar{T}] \times [0,2\pi]^2 \times \R^3$, integrate by parts and use the spectral inequality for $L_J$ and the bounds $1\le\kappa(z)\le \text{\rm e}^{2\e G\pi}$, to obtain, for every $\eta _1>0$,
\begin{eqnarray}
\parallel \gamma ^{-}R_1\parallel^2_{2\bar{T},2, \sim}+ \parallel {R}_1(\bar{T})\parallel _{2,2}^2+\frac{1}{\e }\parallel \nu ^{\frac{1}{2}}(I-P_J){R}_1\parallel _{2\bar{T},2,2}^2
\leq c\Big( \parallel {R}_0\parallel _{2,2}^2\quad\label{4.5}\\
+\e \parallel \nu ^{-\frac{1}{2}}(I-P_J){g}\parallel _{2\bar{T},2,2}^2+\frac{\eta _1}{2 }\parallel P_J{R}_1\parallel _{2\bar{T},2,2}^2+\frac{1}{2\eta _1 }\parallel P_J{g}\parallel _{2\bar{T},2,2}^2+\frac{1}{\e ^2}\parallel {\bar{\psi}}\parallel^2_{2\bar{T},2,\sim}\Big)\nonumber.
\end{eqnarray}

An a priori bound for $P_J R_1 $ is obtained in the following lemma based on dual techniques involving the problem (\ref{4.3}). Consider first the problem (\ref{4.4}) without the term $H_1(R_1)$.\\
\begin{lemma}
\label{l4.2}
Set $h:=P_JR_1$. Then 
\begin{eqnarray*}
&&\parallel h \parallel ^2_{2,2,2} \leq  c( \parallel {R}_0\parallel ^2_{2,2}+ \parallel \nu^{-\frac{1}{2}}(I-P_J){g} \parallel ^2_{2,2,2}+\frac{1}{\e ^2}\parallel P_J{g} \parallel ^2_{2,2,2}+\frac{1}{\e^3}\parallel{\bar{\psi}}\parallel^2_{2,2,\sim}).
\end{eqnarray*}
\end{lemma} 
\noindent\underline{Proof of Lemma \ref{l4.2}.}\hspace*{0.05in}In the variables $(\bar{\tau },x,z,v)$, the function $R_1$ is $2\pi$-periodic in $x$ and solution to
\begin{eqnarray}
&&\frac{\partial R_1}{\partial \bar{\tau }}+\mu\ v_x\cdot \frac{\partial R_1}{\partial x} +v_z\cdot \frac{\partial R_1}{\partial z} -\e  G M^{-1}\frac{\partial 
(M  R_1)}{\partial v_z} = \frac{1}{\e }{L}_J  R_1 +g,\label{4.6}\\
&&R_1(0,x,z,v)= R_0(x,z,v),\hspace{1cm}\nonumber\\
&&R_1(\bar{\tau },x,\mp\pi,v)=-\frac{1}{\e}\bar{\psi}(\bar{\tau},x,\mp\pi,v),\quad \bar{\tau }>0,\hspace*{0.05in}v_z\gtrless 0,\hspace{1cm}\nonumber
\end{eqnarray}
Let $\varphi $ be a $2\pi$-periodic function in $x$, solution to
\begin{eqnarray*}
\frac{\partial \varphi }{\partial \bar{\tau }}+\mu \ v_x \frac{\partial \varphi}{\partial x} +v_z \frac{\partial \varphi}{\partial z} -\e  G M^{-1}\frac{\partial 
(M\varphi)}{\partial v_z} = \frac{1}{\e }{L_J^*}\varphi +h,
\end{eqnarray*}
with zero initial values and ingoing boundary values  at $z=-\pi,\pi$.  Multiply the equation for $\varphi $ by $\kappa MR_1$ and the one for $R$ by $\kappa M\varphi $ sum them and integrate by parts. Then, 
\begin{eqnarray*}
&&\int d\bar{\tau}dxdz\kappa(z)\frac{\partial }{\partial \bar{\tau }}({R}_1,{\varphi} )_{H}+ \int d\bar{\tau}dxdzdv\left( M\frac{\partial}{\partial z}(v_z\kappa R_1\varphi )-\e G\kappa \frac{\partial 
(MR_1\varphi)}{\partial v_z}\right)\\
&&=\int d\bar{\tau}dxdz{M}\kappa dv[ \frac{1}{\e }({L_J}((I-P_J)R_1)(I-P){\varphi} )+\frac{1}{\e }((I-P_J)(R_1)L_J^*(I-P){\varphi} )\\
&&+{g}{\varphi} +hP_J{R}_1].
\end{eqnarray*}
This gives 
\begin{eqnarray*}
\parallel h\parallel _{2\bar{\tau },2,2}^2&&\leq \frac{K_1}{2}\parallel {R}_1(\bar{\tau },\cdot ,\cdot )\parallel _{2,2}^2+\frac{1}{2K_1}\parallel {\varphi} (\bar{\tau },\cdot ,\cdot )\parallel _{2,2}^2\\
&&+\frac{K_1}{2}\parallel \gamma ^{-}R_1\parallel _{2\bar{\tau},2,\sim}^2+\frac{1}{2K_1}\parallel \gamma ^{-}\varphi \parallel _{2\bar{\tau},2,\sim}^2\\
&&+\frac{K_3}{\e }\parallel \nu ^{\frac{1}{2}}(I-P_{J}){R}_1\parallel _{2\bar{\tau},2,2}^2+\frac{1}{K_3\e }\parallel \nu ^{\frac{1}{2}}(I-P){\varphi} \parallel _{2\bar{\tau},2,2}^2 \\
&&+\frac{K_4}{2} \parallel \nu ^{-\frac{1}{2}}(I-P_{J}){g}\parallel _{2\bar{\tau},2,2}^2+ \frac{1}{2K_4}\parallel \nu ^{\frac{1}{2}}(I-P){\varphi} \parallel _{2\bar{\tau},2,2}^2\\
&&+\frac{\e^2}{2K_4}\parallel\nu^{\frac{1}{2}}\varphi\parallel^2_{2,2}+\frac{K_2}{2}\parallel P_{J}{g}\parallel _{2\bar{\tau},2,2}^2+\frac{1}{2 K_2}\parallel{ P\varphi} \parallel _{2\bar{\tau},2,2}^2+\frac{\e^2}{2K_2}\parallel\nu^{\frac{1}{2}}\varphi\parallel^2_{2,2},
\end{eqnarray*}
for any positive constants $K_j$, $j=1,...,4$. 
All the terms computed at time $\bar \tau$ on the l.h.s can be estimate using Lemma \ref{l4.1} and (\ref{4.5}), leading to 
\begin{eqnarray*}
&&\parallel h \parallel _{2,2,2}^2\leq c[(K_1 +K_3)\parallel {R}_0\parallel _{2,2}^2\\
&&+(\frac{K_1}{\e^2}+\frac{K_3}{\e ^2})\parallel{\bar{\psi}} \parallel_{2,2,\sim}^2+(\e K_1+K_4+K_3\e) \parallel \nu^{-\frac{1}{2}}(I-P_J){g} \parallel _{2,2,2}^2\\ 
&&+(\frac{1}{\e  K_1}+\frac{1}{\e ^2 K_2}+\frac{1 }{\e K_3}+\frac{1}{ K_4}+\frac{\eta_1}{\e^2 K_1} )\parallel h\parallel _{2 ,2,2}^2
+(\frac{K_1}{ \eta_1}+\frac{K_3}{\eta _1} +K_2)\parallel P_J{g}\parallel _{2,2,2}^2 \\
&&+\parallel P_J{R_1}\parallel ^2_{2 ,2,2} ({\eta_1K_1}+\eta_1 K_3 )+\eta (\frac{\e}{K_1}+\frac{\eta_1 }{K_1}+\frac{\e}{K_3}+\frac{\e^2}{K_4}+\frac{1}{K_2})\parallel<P_{J^*}\varphi>\parallel_{2,2}^2].
\end{eqnarray*}
We are left with the term  $<P_{J^*}\varphi>$. As in Section 2, we use an approach based on ordinary differential equations for the  Fourier transform  with respect to the time and $x$-variables.  Namely, the quantity $<\varphi>:=(2\pi)^{-1}\int \varphi(\cdot,x,\cdot)dx$ satisfies a $1$-d problem including a small perturbation of magnitude $\delta$ from the value at the bifurcation point. After a Fourier transform in time (Fourier variable $\sigma$) the case of $|\e\sigma|<\sigma_0$ with $0<\sigma_0$ sufficiently small, can be handled as in Lemma \ref{2.4}. For the remaining $\sigma$'s use the term $\-i\e\sigma \mathcal{F}_{\bar\tau}\mathcal{F}_x\beta \varphi(\sigma,\xi_x,z)$ to express the $\psi_0$-moment. With $c_2=(v_z^2,\psi_4)(v_z^2\bar{A},\psi_4)^{-1}$, project the equation along $v_z-c_2v_z\bar{A}$, and along $\psi_0$ and use the equation, leading to an expression for 
$$\displaystyle{
\mathcal{F}_{\bar\tau}\mathcal{F}_x(\e\frac{\partial}{\partial t}\beta(\varphi_{v_z}- c_2\varphi_{v_z\bar{A}})+\frac{\partial}{\partial z}(\beta\varphi_{0}+\vartheta_2))\quad \hbox{and}\  \ \mathcal{F}_{\bar\tau}\mathcal{F}_x(\e\frac{\partial}{\partial t}(\beta\varphi_{0}) +\frac{\partial}{\partial z}(\beta\varphi_{ v_z}))},$$
with $\vartheta_2$ a nonhydrodynamic moment of $\varphi$, thus to an expression for 
\begin{equation}
-i\e\sigma (|\mathcal{F}_{\bar\tau}\mathcal{F}_x(\beta \varphi_{0})|^2+|\mathcal{F}_{\bar\tau}\mathcal{F}_x(\beta \varphi_{v_z})|^2)+\frac{\partial}{\partial z}((\mathcal{F}_{\bar\tau}\mathcal{F}_x(\beta\varphi_{0}+\zeta_2))(\mathcal{F}_{\bar\tau}\mathcal{F}_x(\beta\varphi_{v_z}))^*).\nonumber
\end{equation} 
(For details cfr Lemma 5.5 below). Estimating the ingoing terms, this results in 
\begin{eqnarray*}
\parallel<P\varphi>\parallel_{2,2}&&\leq c\parallel<P\varphi>_x\parallel_{2,2,2}\leq c(\parallel<P\varphi>_x\parallel_{2,2,2} +\e\parallel \nu^{\frac{1}{2}}\varphi\parallel_{2,2,2})\\
&&\leq \frac{c}{\e}\parallel <h>_x\parallel_{2,2,2}+\eta\parallel \nu^{\frac{1}{2}}\varphi\parallel_{2,2,2}\leq \frac{c}{\e}\parallel h\parallel_{2,2,2}+\eta\parallel \nu^{\frac{1}{2}}\varphi\parallel_{2,2,2}.
\end{eqnarray*}
So choosing $\e <1$, then $K_1$ and $K_3$ (resp.$K_2$) of order $\e ^{-1}$ (resp. $\e ^{-2}$ ) and $\eta _1$ of order $\e $, leads to 
\begin{eqnarray*}
\parallel h\parallel _{2,2,2}^2&&\leq c\Big( \frac{1}{\e}  \parallel {R}_0\parallel _{2,2}^2+\parallel \nu^{-\frac{1}{2}}(I-P_J){g}\parallel _{2,2,2}^2+\frac{1}{\e ^2}\parallel P_J{g}\parallel _{2,2,2}^2\\
&&+\frac{1}{\e ^3}\parallel{\bar{\psi}}\parallel_{2,2,\sim}^2+\eta\parallel<P_JR_1>\parallel^2_{2,2,2}\Big).
\end{eqnarray*}
This ends the proof of Lemma \ref{l4.2}, when coming back to the $t$-variable. $\square$

\bigskip

We now give the final estimates for $R_1$.
\begin{lemma}
\label{l4.3}
The    solution $R_1$ to (\ref{4.4}) satisfies
\begin{eqnarray*}
\parallel \nu^{\frac{1}{2}}R_1\parallel _{2,2,2}&&\leq c\Big(\parallel R_0\parallel _{2,2}+\parallel \sqrt{\nu }(I-P_J)g\parallel _{2,2,2}+\frac{1}{\e }\parallel P_Jg\parallel _{2,2,2}
+
\frac{1}{\e^\frac{3}{2}}\parallel\bar{\psi}\parallel_{2,2,\sim}\Big) , \\
\parallel R_1\parallel _{\infty ,2,2}&&\leq c\Big( \parallel R_0\parallel _{2,2}+\parallel \sqrt{\nu}(I-P_J)g\parallel _{2,2,2}
+\frac{1}{\e }\parallel P_Jg\parallel _{2,2,2}+\frac{1}{\e^{\frac{3}{2}}}\parallel \bar{\psi}\parallel_{2,2,\sim}\Big) ,\\
\parallel \nu^{\frac{1}{2}}R_1\parallel _{\infty ,\infty ,2}
&&\leq c\Big( \e^{-1}\parallel R_0\parallel _{2,2}+\parallel R_0\parallel _{\infty ,2}+\frac{1}{{\e} }\parallel \nu ^{-\frac{1}{2}}(I-P_J)g\parallel _{2,2,2}\\
&&+\frac{1}{\e ^2 }\parallel P_J g \parallel _{2,2,2} 
+\e \parallel \nu^{-\frac{1}{2}}g\parallel_{\infty, \infty, 2}+\e^{-\frac{5}{2}}\parallel\bar{\psi}\parallel_{2,2,\sim}
+\frac{1}{\e}\parallel\bar{\psi}\parallel_{\infty,2,\sim}\Big) .
\end{eqnarray*}
\end{lemma}

\noindent\underline{Proof of Lemma \ref{l4.3}.}\hspace*{0.05in} The solution $R_1$ of (\ref{4.4}) without $H_1$-term satisfies
\begin{eqnarray*}
&&\frac{1}{\sqrt \e }\parallel \gamma ^{-}R_1 \parallel _{2,2\sim} + \sup _{t\geq 0}\parallel R_1(t)\parallel _{2,2}+\frac{1}{\e }\parallel \nu ^{\frac{1}{2}}(I-P_J)R_1\parallel _{2,2,2}\leq c\Big( \parallel R_0\parallel _{2,2}\\
&&+\frac{1}{\e^{\frac{3}{2}}}\parallel \bar{\psi}\parallel_{2,2,\sim}+\parallel \nu^{-\frac{1}{2}}(I-P_J)g\parallel _{2,2,2}+\frac{\eta }{\sqrt{\e }}\parallel P_JR_1\parallel _{2,2,2}+\frac{1}{\eta \sqrt{\e }}\parallel P_Jg\parallel _{2,2,2}\Big) ,
\end{eqnarray*}
for any $\eta >0$. Moreover, it follows from Lemma \ref{l4.2} that
\begin{eqnarray*}
\parallel P_JR_1\parallel _{2,2,2}&&\leq c\Big( \parallel R_0\parallel _{2,2}+\parallel \nu^{-\frac{1}{2}}(I-P_J)g\parallel _{2,2,2}+\frac{1}{\e }\parallel P_J g\parallel _{2,2,2}\\
&&+\frac{1}{\e\sqrt{\e}}\parallel \bar{\psi}\parallel_{2,2,\sim}\Big) .
\end{eqnarray*}
Choosing $\eta = \sqrt{\e }$ leads to the first two inequalities of Lemma \ref{l4.3}. Then, to get the $L^\infty$ estimates, one has to study the solution along the characteristics. This analysis is complicated by the presence of the force, but can be done along the lines in \cite{AEMN} and the result is
\begin{eqnarray*}
\parallel \nu^{\frac{1}{2}}R_1\parallel _{\infty ,\infty ,2}\leq c\Big( \frac{1}{\e }\parallel R_1\parallel _{\infty ,2,2}+\parallel R(0, \,\cdot\,)\parallel _{\infty ,2}+\e \parallel \nu ^{-\frac{1}{2}}g\parallel _{\infty ,\infty ,2}+\parallel \gamma^+R_1\parallel_{\infty ,2,\sim}\Big) ,
\end{eqnarray*}
which leads to the last inequality of Lemma \ref{l4.3}. Adding the term $\e^{-1}H_1(R_1)$ does not change these results. $\square$

\bigskip

The remaining part $R_2$ of $R$ satisfies the equation
\begin{eqnarray}
&&\e \frac{\partial R_2}{\partial t}+\mu\ v_x\frac{\partial R_2}{\partial x}+v_z\frac{\partial R_2}{\partial z} -\e G M^{-1}\frac {\partial (M R_2)}{\partial v_z}= \frac{1}{\e }{L_J}R_2+H_1(R_2),\label{4.8} \\
&&R_2(0,x,z,v)=0,\quad \quad\nonumber\\
 && R_{2} (t,x,\mp\pi,v)=  \frac{M_\mp(v)}{M(v)}\int _{w_z\lessgtr 0} \Big(R_1(t,x,\mp\pi,w)+R_2(t,x,\mp\pi,w) \nonumber \\
 &&\hspace*{1.09in}+\frac{1}{\e}\bar{\psi}(t,x,\mp\pi,w)\Big)|w_z|Mdw,    \quad t>0,\hspace*{0.03in}v_z\gtrless 0,\nonumber
\end{eqnarray}
Its analysis is more involved and requires a careful study of the Fourier transform of $R_2$.  As with the stationary case in Section 2, existence for the problem (\ref{4.8}) can be adapted from the corresponding study in [Ma], if one includes into that approach  the spectral estimate for $L_J$, and the characteristics due to the force term. 

 In (\ref{4.8}) the given indata part is  
\begin{eqnarray*}
f^-(t,x,\mp\pi,v)= \frac{M_\mp}{M}\int _{w_z\lessgtr 0} \Big(R_1(t,x,\mp\pi,w)+\frac{1}{\e}\bar{\psi}(t,x,\mp\pi,w)\Big)|w_z|Mdw, \quad    v_z\gtrless 0,\nonumber 
\end{eqnarray*}
By Green's formula for (\ref{4.8}), and noting that $H_1(R_2)$ only depends on $(I-P_0)R_2$, we get
\begin{equation}
\e \| R_2(t)\|^2_{2,2}+\parallel \gamma^-R_2 \parallel_{2t,2,\sim}^2 + \frac{c}{\e}\parallel \nu ^{\frac{1}{2}}(I-P_J)R_2 \parallel _{2t,2,2} ^2\label{green}\leq \parallel \gamma^+ R_2 \parallel_{2t,2,\sim} ^2\ . \end{equation}
This estimate is not yet final. In fact, compared with the analogous estimate for $\varphi$ and $R_1$, the boundary terms here are different, due to the diffusive boundary conditions for $R_2$. We follow the reasoning in the stationary case for (\ref{3.7.0}) - (\ref{3.7}). Taking into account the differences, like dependence on time, we get
 \begin{equation}
\label{4.9}
\e\|R_2\|^2_{2,2}(t)+\frac{c}{\e}\parallel \nu ^{\frac{1}{2}}(I-P_J)R_2 \parallel _{2t,2,2} ^2
\leq \frac{1}{\e\eta}\parallel f^-\parallel^2_{2t,2,\sim}+C\e\eta\parallel     P_JR_2\parallel^2_{2t,2,2},\quad
\end{equation}
\begin{equation}
\label{4.9.1}
\parallel \gamma^-R_2\parallel^2_{2t,2,\sim}\le \frac 1  {\e^2}\parallel f^-\parallel^2_{2t,2,\sim}+C\parallel P_JR_2\parallel^2_{2t,2,2}.
\end{equation}
The hydrodynamic estimates for $R_2$ are obtained similarly to the stationary case. We again start with the $1$-d ($x$-independent) case, with an inhomogeneous term $g_1$ which will take into account the $x$-dependence in  later proofs. Reduce the equation (\ref{4.8})  to a $1$-d problem for $\int dx R_2(t,x,z,v):=R_2(t,z,v)$, with an inhomogeneous term $g_1$ such that $g_{10}=\int_{\R^3}dv g_1=0$: 
\begin{equation}
\e \frac{\partial R_2}{\partial t}+v_z \frac{\partial R_2}{\partial z} - \e G M^{-1}\frac{\partial 
(M  R_2)}{\partial v_z} = \frac{1}{\e }{L}_J  R_2 +H_1(R_2)+g_1.
\label{1dt}
\end{equation}
Then, for the solution of this new problem 
the following lemma holds:
\begin{lemma}
\label{l4.4}
\begin{eqnarray*}
\parallel P_JR_2 \parallel^2_{2,2,2}\leq \frac{c_1}{\e^2}\parallel f^-\parallel_{2,2\sim}^2+c_2(\parallel P_JR_1\parallel^2_{2,2,2} +\parallel \nu ^{- \frac{1}{2}} g_1\parallel^2_{2,2,2}).
\end{eqnarray*}
\end{lemma}
\noindent \underline{Proof of Lemma \ref{l4.4}}.\hspace{0.05in} The equation for the Fourier transform, $\hat R_2$ of $R_2(t,z,v)$ with respect to the space variable $z$ is
\begin{equation}
\label{4.8.1}\e\frac{\partial}{\partial t}\hat R_2 +iv_z\xi_z\hat R_2-\e GM^{-1}\frac{\partial}{\partial{v_z}}(M\hat R_2)=\e^{-1}\widehat{L_J  R_2}+\widehat{ H_1( R_2)}-v_zr(-1)^{\xi_z}+\hat g_1,
\end{equation}
$r(v)$ now denoting the difference between the ingoing and outgoing boundary values,
\begin{equation}
r(v)= R_{2 }(t,\pi,v)- R_{2 }(t,-\pi,v).
\end{equation}

 Starting from the method of Lemma \ref{l4.1} with $L_J$ instead of its adjoint, and considering the ingoing boundary values as known, we reach (\ref{parse}) for $\xi_z\neq 0$ with obvious changes, 
 \begin{eqnarray*}
\int _{0}^{\infty}dt \int |P_{J0}\hat{R_2 } |^2(t ,\xi_z ,v)Mdv 
 \leq C\Big( \frac{1}{\e ^2}\int _{0}^{\infty}dt \big( \parallel \zeta _{-s}(v)\widehat{{L_J}{R_2}} (t , \xi_z , \cdot )\parallel^2\\
+ \parallel (I-P_{J0})\hat{R_2} (t ,\xi_z ,\cdot )\parallel^2\big)
+\int _0^{\infty}dt\int \xi_z^{-2}\nu^{-1} |\hat{g_1}|^2(t,\xi_z ,v)Mdv\quad \quad\\
+\int_0^{\infty}dt \big(\frac{\parallel \sqrt{|v_z|}r\parallel^2}{\delta_1|\xi_z|^2}+\e^2\parallel\hat{R}_2\parallel^2\big)\Big) .\nonumber
\end{eqnarray*}  
Here, similarly to Section 2, we have fixed $z= z_0$, $t= t_0$ in the basis elements $\bar{\psi }_1$, ..., $\bar{\psi }_4$, writing $P_{J0}$ for the corresponding kernel projection. At the end we replace $P_{J0}$ with $P_J$, since $P_J-P_{J0}= O(\e )$. 
The sum $R=R_1+R_2$  satisfies by hypothesis $\langle \hat{R}(0,\,\cdot\,)\rangle=\hat{R}_{0}(0)=0$ for $t=0$. The coefficients in the asymptotic expansion can be chosen so that, moreover, $\hat{R}_{0}(t,0)=0$ for $t>0$. That gives an estimate for $\hat{R}_{20}(0)$ in terms of  $\hat R_{10}$. 

For an estimate of the $r$-term, for $z=\pi$ we use  (\ref{4.9.1}) giving an estimate of the outflow at $\pi$ in terms of known quantities and $ P_{J0}{R_2 } $. Then, for $\xi_z\ge \bar\xi_z$,  and $\bar\xi_z$ large enough, the latter appears multiplied by a small factor and can be absorbed by the total sum on the left-hand side, taken over  $\xi_z\ge \bar\xi_z$. Hence, for $\xi$ large the bound stated in Lemma \ref{l4.4} is proved.

For the remaining bounded number of $\xi$'s, each hydrodynamic mode is estimated separately. We obtain estimates for all the hydrodynamic moments in terms of $\hat R_{20}$, which will be the last one to be estimated.

For the $v_z$-moments multiply (\ref{4.8.1}) by $v_z$ (resp. one), integrate with respect to $Mdv$, and multiply with $\hat{R}_{20}$ (resp. $\hat{R}_{2v_z}$). Use the notation $[f]_-=f(\xi_z)-(-1)^{\xi_z}f(0)$. Combining the results, it implies  that  for each fixed $\xi_z$
\begin{eqnarray*}
&\displaystyle{\e \frac{\partial}{\partial t} \left(\hat{R}_{20}[\hat R_{2v_z}]_-^*\right)} &=i\xi_z \hat{R}_{2v_z}[\hat R_{2v_z}]_-^*
+i\xi \hat{R}^*_{2v_z^2}\hat{R}_{20 }
+\e G \int  \Big\{\hat{R}_{20}(\xi_z)v_z\frac{\partial}{\partial v_z}( M\hat{R}^*_2)\Big\}dv\\&&
+[\hat R_{2v_z}]_-^*\int M\Big\{v_zr(-1)^{\xi_z}+\hat g_1\Big\}dv. 
\end{eqnarray*}
We integrate over  $t\in[0,\bar t]$, to get an estimate for the  term $i\xi_z |\hat{R}_{2v_z}(\xi_z)|^2$ on the right-hand side. The time derivative,   then, produces a term $\e\hat{R}_{20}(\xi_z)[\hat R_{2v_z}]_-^*$ computed at time $\bar t$ which is bounded as  $C\e( |\hat{R}_{20}(\xi_z)|^2 +|\hat{R}_{2v_z}(\xi_z)|^2+ |\hat{R}_{2v_z}(0)|^2)\le \e\| PR_2\|_{2,2}^2 (\bar t)$. The latter is  estimated   by using  Green's formula for  (\ref{1dt}), getting
$$
\e\| PR_2\|_{2,2}^2\le \frac{c}{\eta }(\parallel f^-\parallel^2_{2\bar t,2,\sim}+\parallel \nu ^{- \frac{1}{2}} g_1\parallel^2_{2\bar t,2,2})+\eta \parallel     P_JR_2\parallel^2_{2\bar t,2,2} .
$$
The other terms can be easily estimated, but for  the one containing
the boundary $r$-term, which is equal to $\hat{R}_{2v_z}(\xi_z)(R_{2v_z}(t,\pi)-R_{2v_z}(t,-\pi))$. This term can be estimated by
$$\int \big( \eta |\hat{R}_{2v_z}|^2 + \frac{1}{\eta} \parallel f^-\parallel_{2\sim}^2\big) dt.$$
In conclusion, we have that for $\xi_z \neq 0$ 
\begin{eqnarray}
\label{4.10}
&&\int |\hat{R}_{2v_z}(\xi_z)|^2dt\\&&\leq C\int dt\Big(\mid \hat{R}_{20}(\xi_z) \mid^2+\mid \hat{R}_{2v_z}(0) \mid^2+\eta\parallel PR_2\parallel_{2,2}^2+ \frac{1}{\eta} \parallel f^-\parallel_{2\sim}^2+\parallel \nu ^{- \frac{1}{2}} g_1\parallel_{2,2}^2\Big).\nonumber
 \end{eqnarray}
We  then compute $\hat{R}_{2v_z}(t,0)$ for $\xi_z=0$. Multiply (\ref{1dt})  by $M$, integrate over  velocity  and  over $z'\in[-\pi,z]$, followed by an integration over $[-\pi,\pi]$. We get, after multiplication by $\hat R_{2v_z}(t, 0)$,
\begin{eqnarray}\label{eq4.31}
&&\e\hat R_{2v_z}(t, 0)\frac{\partial \overline{R_{20}}}{\partial t}+ (\hat R_{2v_z}(t, 0))^2\nonumber\\&&=\hat R_{2v_z}(t, 0)\int_{-\pi}^\pi dz\int_{-\pi}^zdz'dvMg_1+2\pi \hat R_{2v_z}(t, 0)\int dv v_z MR_2(t,v,-\pi),
\end{eqnarray}
where $\overline{R_{20}}:=\int_{-\pi}^\pi dz\int_{-\pi}^zdz'dv MR_2$.
Multiply (\ref{4.8.1}) by $Mv_z$, integrate over velocity and multiply by  $\overline{R_{20}}$,
\begin{eqnarray*}
\e \overline{R_{20}}\frac{\partial}{\partial t} \hat{R}_{2v_z}=\overline{R_{20}}\Big(r_{v_z^2}+\int  v_z \hat g_1Mdv\Big)+\e G \overline{R_{20}}\int v_z \frac{\partial}{\partial v_z}M\hat{R}_2(0,v)dv.
\end{eqnarray*}
Summing the last two equations,
\begin{eqnarray}
&&\e\frac{\partial }{\partial t}(\overline{R_{20}} \hat{R}_{2v_z}) +  \hat{R}_{2v_z}^2= \overline{R_{20}}\Big(r_{v_z^2}+\int Mv_z\hat g_1dv\Big)+  \e G \overline{R_{20}}\int v_z \frac{\partial}{\partial v_z}M\hat{R}_2(0,v)dv\nonumber\\
&&+\hat{R}_{2v_z}\int_{-\pi}^\pi dz\int_{-\pi}^zdz'dvMg_1+\hat{R}_{2v_z}2\pi \int dv v_z MR_2(t,v,-\pi)
\end{eqnarray}
We use this relation to bound the time integral of $ \hat{R}_{2v_z}^2$. We integrate over time, and use inequality  (\ref{4.9.1})  to control the boundary term $r_{v_z^2}$. The result is
$$\int |\hat{R}_{2v_z}(t,0)|^2dt\le C\Big(\frac 1\eta\int dt \|R_{20}\|^2_{2}+\eta\parallel P_JR_2\parallel_{2,2}^2 +\parallel \nu ^{- \frac{1}{2}} g_1\parallel_{2,2}^2 +\parallel f^-\parallel_{2\sim}^2\Big) .$$
By Parseval identity, $\|R_{20}\|^2_{2}=\sum_{\xi_z\ne 0} |\hat R_{20}(\xi_z)|^2 +|\hat R_{20}(0)|^2$. The last term is equal to $|\hat R_{10}(0)|^2$ because $\int dz dv R(t,z,v)=0$. The sum for $\xi_z>\bar\xi_z$ was estimated above. In conclusion,
\begin{eqnarray}
\nonumber &&\int |\hat{R}_{2v_z}(0)|^2dt\leq C\Big( \sum_{0<\xi_z\le\bar\xi_z } |\hat R_{20}(\xi_z)|^2+\parallel P_JR_{1}\parallel_{2,2,2}^2+\eta\parallel PR_2\parallel_{2,2,2}^2+\parallel \nu^{\frac{1}{2}} g_1\parallel_{2,2,2}^2\\
&&\qquad\qquad\qquad+ \parallel\nu^{\frac{1}{2}}(I-P)R_2\parallel_{2,2,2}^2 
+\frac{1}{\e^2}\parallel\nu^{\frac{1}{2}}(I-P_J)R_2\parallel_{2,2,2}^2+\parallel f^-\parallel_{2,2\sim}^2\Big).\label{4.11}
\hspace{1cm}
\end{eqnarray}

For the control of $v_x$-moments, we use $\hat{R}_{2v_x}(\xi)=\hat{R}_{2v_z^2v_x}(\xi)-\hat{R}^\perp_{2 v_xv_z^2}(\xi)$ (recall that $\int v_z^2v_x^2 Mdv=1$). Multiply (\ref{4.8.1}) by $v_xv_z$ (resp. $v_xv_z^2$), integrate with respect to $Mdv$, and multiply with $\hat{R}_{2v_xv_z^2}$ (resp. $\hat{R}_{2v_xv_z}$). Adding the results implies for each $\xi_z$ that                   
\begin{eqnarray*}
&&\e \frac{\partial}{\partial t} (\hat{R}_{2v_xv_z}\hat{R}^*_{2v_xv_z^2}) =i\xi_z |\hat{R}_{2v_xv_z^2}|^2+i\xi_z \hat{R}^*_{2v_xv_z^3}\hat{R}_{2v_xv_z }+\e G \int  \Big(\hat{R}^*_{2v_xv_z^2}v_xv_z\frac{\partial M\hat{R}_2}{\partial v_z}\\
&&+\hat{R}_{2v_xv_z}v_xv_z^2\frac{\partial M\hat{R}^*_2}{\partial v_z}\Big)dv
+\hat{R}^*_{2v_xv_z^2}\int Mv_xv_z\Big(\frac{1}{\e}\widehat{L_JR}_2+\widehat{H_1(R_2)}+v_zr(-1)^{\xi_z}+\hat{g}_1\Big)dv\\ 
&&+\hat{R}_{2v_xv_z}\int Mv_xv_z^2\Big(\frac{1}{\e}\widehat{L_JR}_2+\widehat{H_1(R_2)}+ v_zr(-1)^{\xi_z}+\hat{g}_1\Big)^*dv.
\end{eqnarray*} 
We do not estimate directly  the terms involving higher moments of the boundary term $r$. To  remove these terms we  subtract the same expression for $\xi_z=0$ multiplied by $(-1)^{\xi_z}$, getting
\begin{eqnarray*}
&&\e \frac{\partial}{\partial t}  ([\hat{R}_{2v_xv_z}]_-[\hat{R}_{2v_xv_z^2}]_-^*)
=i\xi_z [\hat{R}_{2v_xv_z^2}]_-^*\hat{R}_{2v_xv_z^2}(\xi_z)+i\xi_z \hat{R}^*_{2v_xv_z^3}[\hat{R}_{2v_xv_z}]_-\\
&&+\e G \int  \left \{[\hat{R}_{2v_xv_z^2}]_-^*v_xv_z\frac{\partial}{\partial v_z}(M[\hat{R}_2]_-)])
+[\hat{R}_{2v_xv_z}]_-v_xv_z^2\frac{\partial}{\partial v_z}( M[\hat{R}_2]_-)^*\right\}dv\\
&&+[\hat{R}_{2v_xv_z^2}]_-^*\int Mv_xv_z\Big(\frac{1}{\e}[\widehat{L_JR}_2]_-
+[\widehat{H_1(R_2)}]_-+ [\hat{g}_1]_-\Big)dv\\
&&+[\hat{R}_{2v_xv_z}]_-\int Mv_xv_z^2\Big(\frac{1}{\e}[\widehat{L_JR}_2]_-]+[\widehat{H_1(R_2)}]_-+ [\hat{g}_1]_-\Big)^*dv.
\end{eqnarray*} 
Integrate with respect to $t$ to obtain for $\xi\neq 0$   
\begin{eqnarray*}
\int |\hat{R}_{2v_x}(\xi_z)|^2dt&&\leq C\int \Big(\mid \hat{R}_{2v_x}(0) \mid^2+\eta\parallel PR_2\parallel_{2,2}^2+ \parallel\nu^{\frac{1}{2}}(I-P)R_2\parallel_{2,2}^2 \nonumber\\
&&+\parallel \nu^{-\frac{1}{2}}g_1\parallel _{2,2}^2+\frac{1}{\e^2}\parallel\nu^{\frac{1}{2}}(I-P_J)R_2\parallel_{2,2}^2\Big)dt.
\hspace{1cm}
\end{eqnarray*}
We have used that only $(I-P)R_2$ contributes to $\hat{R}_{2v_xv_z}$ and $\hat{R}_{2v_xv^3_z}$.

Now we discuss the estimate of $\hat R_{2v_x}$ for  $\xi_z=0$. To eliminate the outgoing boundary terms, we multiply (\ref{1dt}) by $Mv_xv_z$ and consider first the equation we get by taking the integral $\int_{-\pi}^{\pi}dz \int_{-\pi}^z dz'\int_{\R^3} dv$ and then the one we get by taking the integral $2\pi\int_{-\pi}^{\pi}dz\int_{v_z<0}dv$. By taking the difference of the two equations we get for the l.h.s. (but for the force terms)
\begin{eqnarray*}
\e\frac{\partial}{\partial t}\Big(\int dv \int_{-\pi}^{\pi}dz \int_{-\pi}^zdq  v_xv_zMR_2(t,q,v)-2\pi\int_{v_z<0}v_xv_zM\hat{R}_2(t,0,v)dv\Big)
+\hat R_{2v_xv_z^2}(t,0)
\end{eqnarray*}
 The remaining ingoing boundary terms are zero. Now, notice that $\int_{v_z<0}v_xv_zM\hat{R}_2(t,0,v)=\zeta\hat{R}_{2v_x}(t,0)+\hat{R}_{2\perp}(t,0)$, 
where $\zeta=\int_{v_z<0}v_x^2v_zMdv$ and $\hat{R}_{2\perp}(t,0)=\int_{v_z<0}v_xv_zM\hat{R}^\perp_2(t,0,v)dv$ depends on the nonhydrodynamic part of $\hat R_2$. We multiply by $\hat{R}_{2v_x}(t,0)$ and, by using the equation for  it,  we get  in the l.h.s (but for the force terms)
\begin{equation}
\e\frac{\partial}{\partial t}\Big(\mathcal {D}\hat{R}_{2v_x}(t,0)-2\pi\e\zeta\hat{R}_{2v_x}^2(t,0)\Big)+ \hat{R}_{2v_x}\hat{R}_{2v_xv^2_z}+\mathcal {D}\int dv v_xv_z(\gamma ^{-}R_2(\pi)-\gamma ^{-}R_2(-\pi))
\nonumber
\end{equation}
where $\mathcal {D}:\int \int_{-\pi}^{\pi} \int_{-\pi}^z v_xv_zM R_2(t,q,v)dqdzdv -2\pi \hat{R}_{2\perp}(t,0)$
  is nonhydrodynamic. We use again $\hat{R}_{2v_x}(\xi)=\hat{R}_{2v_xv_z^2}(\xi)-\hat{R}^\perp_{2 v_xv_z^2}(\xi)$ and estimate $\gamma ^{-}R_2$ by (\ref{4.9.1}). This gives
\begin{eqnarray*}
\int |\hat{R}_{2v_x}(0)|^2dt&&\leq C\int dt \Big(\eta\parallel PR_2\parallel_{2,2}^2 +\parallel\nu^{\frac{1}{2}}(I-P)R_2\parallel_{2,2}^2 \\&&
+\frac{1}{\e^2}\parallel\nu^{\frac{1}{2}}(I-P_J)R_2\parallel_{2,2}^2+\parallel f^-\parallel _{2,\sim }^2+\parallel \nu ^{- \frac{1}{2}} g_{1}\parallel^2_{2,2}\Big).
\end{eqnarray*}
The $v_y$-moments are analogous, hence for all $\xi_z$
\begin{eqnarray}
\label{4.12}
\int |\hat{R}_{2v_x}(\xi_z)|^2dt+\int |\hat{R}_{2v_y}(\xi_z)|^2dt&&\leq c\int (\eta\parallel PR_2\parallel_{2,2}^2+ \parallel f^-\parallel _{2,\sim }^2+\parallel\nu^{\frac{1}{2}}(I-P)R_2\parallel_{2,2}^2 \nonumber\\
&&+\frac{1}{\e^2}\parallel\nu^{\frac{1}{2}}(I-P_J)R_2\parallel_{2,2}^2+\parallel \nu ^{- \frac{1}{2}} g_{1}\parallel^2_{2,2})dt.
\end{eqnarray}

Consider next the $\psi_4$-moment for $\xi_z \neq 0$. Multiply (\ref{4.8.1}) by $v_z\bar{A}$ (resp. $v_z^2\bar{A}$), and integrate with respect to $Mdv$. Similarly to the proof of Lemma \ref{2.6}, this gives
\begin{eqnarray*}
\e \frac{\partial}{\partial t}(v_z\bar{A},\hat{R}_2(\xi_z))&&+(-1)^{\xi_z}r_{v_z^2\bar{A}}+ i\xi_z (v_z^2\bar{A},\hat{R}_2({\xi_z}))+\e G
\int\frac{\partial}{\partial v_z}(v_z\bar{A}M)\hat{R}_2(\xi_z)dv\\
&&=\frac{1}{\e}(v_z\bar{A},\widehat{L_JR_2}(\xi_z))+(v_z\bar{A},\widehat{H_1(R_2)}(\xi_z))+(v_z\bar{A},\hat{g}_1),
\end{eqnarray*}
\begin{eqnarray*}
\e\frac{\partial}{\partial t} (v_z^2\bar{A},\hat{R}_2(\xi_z))&&+(-1)^{\xi_z}r_{v_z^3\bar{A}}+ i\xi_z (v_z^3\bar{A},\hat{R}_2({\xi_z}))+\e G
\int\frac{\partial}{\partial v_z}(v_z^2\bar{A}M)\hat{R}_2(\xi_z)dv\\&&\frac{1}{\e}(v_z^2\bar{A},\widehat{L_JR_2}(\xi_z))+(v_z^2\bar{A},\widehat{H_1(R_2)}(\xi_z))+(v_z^2\bar{A},\hat{g}_1).\ 
\end{eqnarray*}
Similarly to the $v_x$-case,  we manipulate the equations to remove the boundary terms,  leading to, 
\begin{eqnarray*}
&&\e \frac{\partial }{\partial t}([\hat{R}_{2v_z\bar{A}}]_-[\hat{R}_{2v_z^2\bar{A}}]_-^*)i\xi [\hat{R}_{2v_z^2\bar{A}}]_-^*\hat{R}_{2v_z^2\bar{A}}(\xi_z)-i\xi_z \hat{R}^*_{2v_z^3\bar{A}}(\xi_z)[\hat{R}_{2v_z\bar{A}}]_-\\
&&+\e G \int  dv \Big([\hat{R}_{2v_z^2\bar{A}}]_-^*\ v_z\bar{A}\frac{\partial }{\partial v_z}(M[\hat{R}_2]_-)+[\hat{R}_{2v_z\bar{A}}]_-\ v_z^2\bar{A}\frac{\partial}{\partial v_z}( M[\hat{R}_2]_-)\Big)
\\
&&+[\hat{R}_{2v_z^2\bar{A}}]_-^*\int dv Mv_z\bar{A}\Big(\frac{1}{\e}[\widehat{L_JR}_2]_-
+[\widehat{H_1(R_2)}]_-+[\hat{g}_1]_-
\Big)
\\&&+[\hat{R}_{2v_z\bar{A}}]_-\int dv Mv_z^2\bar{A}\Big(\frac{1}{\e}[\widehat{L_JR}_2]_-
+[\widehat{H_1(R_2)}]_-+[\hat{g}_1]_-\Big)^*.
\end{eqnarray*} 
It follows that for $\xi_z\neq 0$
\begin{eqnarray}
\label{4.13}
&&\int dt\mid \hat{R}_{24}(\xi)\mid^2\leq C\int dt \Big(\mid \hat{R}_{24}(0) \mid^2 +\eta\mid \hat{R}_{20}(\xi_z) \mid^2+\eta\mid \hat{R}_{2v_z}(0) \mid^2+\eta\parallel P{R}_{2} \parallel_{2,2}^2\nonumber\\
&&+ \parallel(I-P)R_2\parallel_{2,2}^2
+\frac{1}{\e^2}\parallel\nu^{\frac{1}{2}}(I-P_J)R_2\parallel_{2,2}^2+\parallel f^-\parallel_{2,\sim}^2+\parallel \nu ^{- \frac{1}{2}} g_{1}\parallel^2_{2,2}\Big).\end{eqnarray}  
For $\xi_z=0$ multiplying (\ref{1dt}) with $v_z\bar{A}$ and arguing similarly to the proof of  (\ref{4.12}), gives
\begin{eqnarray}
\label{4.14}
\int |\hat{R}_{24}(0)|^2dt\leq c\int dt \Big(\eta\parallel PR_2\parallel_{2,2}^2+ \parallel\nu^{\frac{1}{2}}(I-P)R_2\parallel_{2,2}^2 \nonumber\\
+\frac{1}{\e^2}\parallel\nu^{\frac{1}{2}}(I-P_J)R_2\parallel_{2,2}^2+\parallel f^-\parallel^2_{2,\sim}+\parallel \nu ^{- \frac{1}{2}} g_{1}\parallel^2_{2,2}\Big).
\hspace{1cm}
\end{eqnarray}
The only moments still to be estimated are  the $\psi_0$-moments for $\xi_z\neq 0$. With $c_2=(v_z^2,\psi_4)(v_z^2\bar{A},\psi_4)^{-1}$ and $c_3=(v_z^4,1)(v_z^6,1)^{-1}$, proceed similarly to the corresponding $\psi_4$-case discussed earlier, but start from $v_z-c_2v_z\bar{A}$ instead of $v_z\bar{A}$, and $v_z^2-c_3v_z^4$ instead of $v_z^2\bar{A}$. That gives
\begin{eqnarray}
\label{4.15}
\int dt\mid \hat{R}_{20}^2(\xi_z)\mid&&\leq C\int dt \Big(\parallel P_JR_1\parallel^2_{2,2}+\eta\parallel P{R}_{2} \parallel_{2,2}^2+\parallel \nu ^{- \frac{1}{2}} g_{1}\parallel^2_{2,2}\nonumber\\
&&+ \parallel(I-P)R_2\parallel_{2,2}^2
+\frac{1}{\e^2}\parallel\nu^{\frac{1}{2}}(I-P_J)R_2\parallel_{2,2}^2+\parallel f^-\parallel^2_{2,\sim}\Big).\quad  \quad
\end{eqnarray}  
The lemma is proved by collecting  the estimates above. $\square$

\bigskip

By using this $1$-d analysis, it follows in the $2$-d case that
\begin{lemma}
\label{l4.5}
 The  solution $R_2$ of (\ref{4.8}) satisfies
\begin{eqnarray*}
\parallel P_JR_2 \parallel^2_{2,2,2}\leq c(\frac{1}{\e^2}\parallel f^-\parallel_{2,2\sim}^2+\parallel P_JR_1\parallel^2_{2,2,2}).
\end{eqnarray*}
\end{lemma}
\noindent\underline{Proof of Lemma \ref{l4.5} }\hspace*{.05in} We can apply Lemma \ref{l4.4} to  $\hat R_2(0,\xi_z,v)=\int dx R_2(x,z,v)$ and, taking into account that $g_1$ is of order $\delta$, get a bound for the Fourier components $P_J\hat{R}_2(0,\xi_z)$, for $\delta$ small.    As discussed at the beginning of the proof of Lemma \ref{l4.4},  the components with $\xi$ large are under control, since the $r$-terms are small after division by $|\xi|^2$. The remaining components in the  case $\xi_x\ne 0$, $\xi_x,\xi_z$ finite, are estimated by analyzing the equations for the moments of $R_2$ and applying in a suitable way Lemma  \ref{l4.4}. The proof follows closely the one of Lemma  \ref{l4.4}, so we will not give all details but only point out the differences.

Consider first the $v_x$-moment for $\xi_z=0$. Multiply the (spatial) Fourier version of (\ref{4.8}) by $Mdv$ (resp. $v_xMdv$) and integrate. Combining the results 
\begin{eqnarray*}
&&\e \frac{\partial}{\partial t} (\hat{R}_{20}(\xi_x,0)\hat{R}_{2v_x}^*(\xi_x,0)) =i\mu\xi_x \hat{R}_{2v_x}(\xi_x,0)\hat{R}^*_{2v_x}(\xi_x,0)
+i\mu\xi_x \hat{R}^*_{2v_x^2}(\xi_x,0)\hat{R}_{20 }(\xi_x,0)
\\&&+\hat{R}^*_{2v_x}(\xi_x,0)\int dv M
v_zr 
+\hat{R}_{20}(\xi_x,0)\int dv Mv_xv_zr^*.
\end{eqnarray*}
This gives
\begin{eqnarray}
\label{4.16}
\int |\hat{R}_{2v_x}|^2(\xi_x,0)dt&&\leq C\int dt \Big(|\hat{R}_{20}|^2(\xi_x,0)+\eta\parallel PR_2\parallel_{2,2}^2+ \parallel\nu^{\frac{1}{2}}(I-P)R_2\parallel_{2,2}^2 \nonumber\\
&&+\frac{1}{\e^2}\parallel\nu^{\frac{1}{2}}(I-P_J)R_2\parallel_{2,2}^2+\parallel f^-\parallel^2_{2\sim}\Big).
\hspace{1cm}
\end{eqnarray}
To bound the $v_z$-moments for $\xi_z\neq 0$, we use a variant of the proof in Lemma \ref{l4.4}.
Multiply   the Fourier version of (\ref{4.8}) by $((v_x^2,v_z^2\bar{B})-v_z^2\bar{B})Mdv$ and integrate. That removes the hydrodynamic $\xi_x$-term. To remove the $v_zr$-term we also subtract the same expression for $\xi_z=0$ multiplied with $(-1)^{\xi_z}$. We use the notation $[f]^x:=f(\xi)-(-1)^{\xi_z}f(\xi_x,0)$. Proceeding as in Lemma \ref{l4.4} gives
\begin{eqnarray*}
&&\e \frac{\partial}{\partial t} ([\hat{R}_{20}(v_x^2,v_z^2\bar{B})-\hat{R}_{v_z^2\bar{B}})]^x
[\hat{R}^*_{2v_z}]^x)=i\xi_z (\hat{R}_{2v_z}(v_x^2,v_z^2\bar{B})
-\hat{R}_{2v_z^3\bar{B}})[\hat{R}^*_{2v_z}]^x\\&&
+i\mu \xi _x \Big( [\hat{R}_{v_x}(v_x^2,v_z^2\bar{B})-\hat{R}_{v_xv_z^2\bar{B}}]^x
+[\hat{R}^*_{2v_xv_z}]^x\Big) [\hat{R}^*_{2v_z}]^x+\xi_z\hat{R}^*_{2v_z^2}(\xi))[\hat{R}_{20}(v_x^2,v_z^2\bar{B})
-\hat{R}_{v_z^2\bar{B}}]^x\\
&&-\e G [\hat{R}^*_{2v_z}]^x \int  
v_z^2\bar{B}\frac{\partial}{\partial v_z} ( M[\hat{R}_2]^x)dv
+\e G [\hat{R}_{20}(\xi)-\hat{R}_{v_z^2\bar{B}}]^x\int 
 v_z\frac{\partial}{\partial v_z}( M[\hat{R}^*_2]^xdv)\\
&&+[\hat{R}^*_2]^x\int M\Big(\frac{1}{\e}v_z^2\bar{B}[\widehat{L_JR}_2]^x
-v_z^2\bar{B}[\widehat{H_1(R_2)}]^x\Big) dv.
\end{eqnarray*}
It follows after integration with respect to $t$ that for $\xi_z \neq 0$
\begin{eqnarray}
\label{4.17}
\int |\hat{R}_{2v_z}|^2(\xi)dt&&\leq C\int dt \Big(|\hat{R}_{2v_z}|^2(\xi_x,0)+|\hat{R}_{20}|^2(\xi)+\eta\parallel PR_2\parallel_{2,2}^2\\&&+ \parallel\nu^{\frac{1}{2}}(I-P)R_2\parallel_{2,2}^2 
+\frac{1}{\e^2}\parallel\nu^{\frac{1}{2}}(I-P_J)R_2\parallel_{2,2}^2\Big).\nonumber
\end{eqnarray}
The moment $\hat{R}_{2v_x}(\xi_x,0)$ is under control by (\ref{4.16}), and so the previous approach for the $v_x$-moment when $\xi_z\neq 0$, gives
\begin{eqnarray}
\label{4.18}
\int |\hat{R}_{2v_x}|^2(\xi)dt&&\leq C\int dt  \Big(|\hat{R}_{20}|^2(\xi)+|\hat{R}_{24}|^2(\xi)+\eta\parallel PR_2\parallel_{2,2}^2\nonumber \\&&+ \parallel\nu^{\frac{1}{2}}(I-P)R_2\parallel_{2,2}^2 
+\frac{1}{\e^2}\parallel\nu^{\frac{1}{2}}(I-P_J)R_2\parallel_{2,2}^2+\parallel f^-\parallel^2_{2\sim}\Big).\quad  \quad
\end{eqnarray}
The $\hat{R}_{2v_y}(\xi)$-moment for $\xi_z\neq 0$  is similarly treated, starting from $$\e\frac{\partial}{\partial t}\Big\{ (\hat{R}_{2v_yv_z}(\xi)-(-1)^{\xi_z}\hat{R}_{2v_yv_z}(\xi_x,0))(\hat{R}_{2v_yv_z^2}(\xi)-\hat{R}_{2v_yv_z^2}(\xi_x,0)\Big\}.$$ Inserting the corresponding right-hand sides and estimating the upcoming moments, gives for $\xi_z\neq 0$
\begin{eqnarray}
\label{4.19}
\int |\hat{R}_{2v_y}(\xi)|^2dt&&\leq C\int dt  \Big(\mid \hat{R}_{2v_y}(0) \mid^2+\eta\parallel PR_2\parallel_{2,2}^2+ \parallel\nu^{\frac{1}{2}}(I-P)R_2\parallel_{2,2}^2 \nonumber\\
&&+\frac{1}{\e^2}\parallel\nu^{\frac{1}{2}}(I-P_J)R_2\parallel_{2,2}^2\Big).
\hspace{1cm}
\end{eqnarray}
For the $\hat{R}_{2v_z}(\xi_x,0)$-moment,
use the procedure of Lemma \ref{l4.4} with $i\mu\xi_x\mathcal{F}_x R_{2v_x}(\xi_x,\cdot)$ added as an inhomogeneous term. Thus
\begin{eqnarray*}
\e\frac{\partial}{\partial t} \int_{-\pi}^{\pi}dz\int_{-\pi}^zdq\mathcal{F}_xR_{20}(\xi_x,q)&&=i\mu\xi_x\int_{-\pi}^{\pi}dz\int_{-\pi}^z dq\mathcal{F}_x R_{2v_x}(\xi_x,q)+\hat{R}_{2v_z}(\xi_x,0)\\
&&-2\pi\int_{v_z>0}Mv_z\mathcal{F}_xf^-(\xi_x,-\pi)dv.
\end{eqnarray*}
The equation for  $\hat{R}_{2v_z}$ is
\begin{eqnarray*}
\e\frac{\partial}{\partial t} \hat{R}_{2v_z}=i\mu\xi_x\hat{R}_{2v_xv_z}(\xi_x,0)+r_{v_z^2}(\xi_x)+\e G \int v_z \frac{\partial}{\partial v_z}M\hat{R}_2(\xi_x,0,v)dv.
\end{eqnarray*}
The resulting terms are of the same type we get   before, except for 
$$\int_{-\pi}^{\pi}dz\int_{-\pi}^z dq\mathcal{F}_x R_{2v_x}(\xi_x,q) \qquad \hbox{ and}  \qquad \int_{-\pi}^{\pi}dz\int_{-\pi}^zdq \mathcal{F}_x R_{20}(\xi_x,q)\ .$$ The former can be controlled using the observation 
$\parallel \mathcal{F}_xR_{2v_x}(\xi_x,.)\parallel_2^2
=\sum_{\xi_z}|\hat{R}_{2v_x}(\xi_x,\xi_z)|^2 $, together with the estimates (\ref{4.16}), (\ref{4.18}) for the right-hand side. We conclude:
\begin{eqnarray}
\label{4.20}
&&\int |\hat{R}_{2v_z}|^2(\xi_x,0)dt\leq C\int dt \Big(\parallel {R}_{20}\parallel_{2,2}^2+\eta\parallel PR_2\parallel_{2,2}^2\nonumber \\
&&+ \parallel\nu^{\frac{1}{2}}(I-P)R_2\parallel_{2,2}^2 
+\frac{1}{\e^2}\parallel\nu^{\frac{1}{2}}(I-P_J)R_2\parallel_{2,2}^2+\parallel f^-\parallel^2_{2\sim}\Big).\quad  \quad
\end{eqnarray}
For the moment $\hat{R}_{2v_y}(\xi_x,0)$ we follow the corresponding procedure of Lemma \ref{l4.4}. The new terms which result from the $x$-derivative, $\hat{R}_{2v_xv_y}(\xi_x,0)$ and $\hat{R}_{2v_xv_yv_z}(\xi_x,0)$, are nonhydro\-dynamic, and so the estimate  (\ref{4.12}) 
also holds for this moment.

The $\psi_4$-moments for $\xi_z\neq0$ are treated as in the $1$-d case. The extra terms resulting from the $x$-derivative, contain at least one nonhydrodynamic factor. The resulting inequality is again (\ref{4.13}).\\ 
The moment $\hat{R}_{24}(\xi_x,0)$ is obtained as in the $1$-d case, but here with an additional term $\int|\hat{R}_{20}|^2(\xi_x,0)dt $ to the right in the final estimate (\ref{4.14}). Thus
\begin{eqnarray}
\label{4.21}
\int |\hat{R}_{24}|^2(\xi_x,0,t)dt&&\leq C\int dt\Big(|\hat{R}_{20}|^2(\xi_x,0,t)+\eta\parallel PR_2\parallel_{2,2}^2+ \parallel\nu^{\frac{1}{2}}(I-P)R_2\parallel_{2,2}^2 \nonumber\\
&&+\frac{1}{\e^2}\parallel\nu^{\frac{1}{2}}(I-P_J)R_2\parallel_{2,2}^2+\e^2\parallel f^-\parallel^2_{2\sim}\Big).
\hspace{1cm}
\end{eqnarray}
We now discuss the moment $\hat{R}_{20}(\xi_x,0)$  for $\xi_x\neq0$.
For $\e|\sigma|>\sigma_1$ and $\sigma_1$ sufficiently large, consider the equation (\ref{4.8}) written in Fourier variables for the time and $x$-dependence. Introduce also the cutoff function $\beta$ as in Lemma \ref{l4.1}. Use the term $\-i\e\sigma \mathcal{F}_t\mathcal{F}_x\beta R_2(\sigma,\xi_x,z, v)$ to express the $\psi_0$-moment. For this, project the equation along $v_z-c_2v_z\bar{A}$, and along $c_3\psi_0+v_x^2\bar{B}$ with $c_3=-(v_x^2,v_x^2\bar{B})>0$ to remove a $\mathcal{F}_t\mathcal{F}_x\beta R_{2v_x}$-moment. That leads to an expression for 
\begin{equation}
\mathcal{F}_t\mathcal{F}_x\beta (-i\e\sigma(R_{2v_z}- c_2R_{2v_z\bar{A}})-i\mu\xi_x\zeta_1+\frac{\partial}{\partial z}(R_{20}+\zeta_2)),\nonumber
\end{equation} 
and for
\begin{equation}
\mathcal{F}_t\mathcal{F}_x\beta \big(-i\e\sigma(c_3R_{20}+R_{2v_x^2\bar{B}}\big)-i\mu\xi_x\zeta_3 +\frac{\partial}{\partial z}\big(\frac{3}{2}c_3R_{2 v_z}+\zeta_4)\big), \nonumber
\end{equation}
with $\zeta_j$, $j=1,...,4$ certain nonhydrodynamic moments of $R_2$. Thus we get an expression for 
\begin{equation*}
-i\e\sigma c_3\Big(\frac{3}{2}|\mathcal{F}_t\mathcal{F}_x\beta R_{20}|^2+|\mathcal{F}_t\mathcal{F}_x\beta R_{2v_z}|^2\Big)+\frac{\partial}{\partial z}\Big((\mathcal{F}_t\mathcal{F}_x\beta(R_{20}+\zeta_2))(\frac{3}{2}c_3\mathcal{F}_t\mathcal{F}_x\beta(R_{2v_z}+\zeta_4))^*\Big).
\end{equation*} 
After division by $\sigma$ and integration, the boundary term is multiplied by the small coefficient $\sigma_1^{-1}$. This can now be estimated separately at $-\pi$ and at $\pi$ using Green's formula (\ref{4.9.1}).  It follows that 
\begin{eqnarray}
\label{4.22}
&&\parallel(1-\chi_{\sigma_1})\mathcal{F}_t\mathcal{F}_xR_{20}(\,\cdot\,,\xi_x,\,\cdot\,)\parallel^2_{2,2}\leq C\Big(
\sigma_1^{-1}(\parallel \mathcal{F}_xR_2(\,\cdot\,,\xi_x,\,\cdot\,)\parallel^2_{2,2,2}\\
&&+\frac{1}{\e^2}\parallel(I-P_j)\mathcal{F}_xR_2(\,\cdot\,,\xi_x,\,\cdot\,)\parallel^2_{2,2,2})+ \parallel(I-P)\mathcal{F}_xR_2(\,\cdot\,,\xi_x,\,\cdot\,)\parallel^2_{2,2,2}\Big).\nonumber
\end{eqnarray}

The case of $\e\sigma$ small requires a different argument. 
Consider equation (\ref{4.8}) and its Fourier transform in $t$, $x$ and $z$. We denote the total Fourier transform of a function $h$, ${\cal F}_t{\cal F}_x{\cal F}_zh$ by $\hat h$, and by $\hat h{^z}$ the Fourier transform $\mathcal{F}_t\mathcal{F}_xh$. With $\beta $ defined in Lemma \ref{4.1} put $\overline{\hat R_2}:=\widehat{\beta R_2}$ and $\overline{\hat R_2}{^z}:=\widehat{\beta R_2}{^z}$. We have
\begin{eqnarray}&&
\e i\sigma\overline{\hat R_2}+\mu v_xi\xi_x\overline{\hat R_2}+iv_z\xi_z\overline{\hat R_2}+v_z \overline{\hat r}(-1)^{\xi_z}=\nonumber\\
&&\e M^{-1}G\partial_{v_z}(M\overline{\hat R_2})+
\e^{-1}\widehat{L_J \beta R_2}+\widehat{\beta H_1( R_2)}+\e \widehat{R_2\partial_t\beta}:= {\cal N}.\nonumber 
\\\label{noz}
&&\e i\sigma\overline{\hat R_2}{^z}+\mu v_xi\xi_x\overline{\hat R_2}{^z}+v_z\partial_z\overline{\hat R_2}{^z}=\\
&&\e M^{-1}G\partial_{v_z}(M\overline{\hat R_2})
+\e^{-1}\widehat{ L_J  \beta R_2}{^z} +\widehat{\beta H_1( R_2)}{^z}+\e \widehat{ R_2\partial_t\beta}{^z}:={\cal N}^z\nonumber.
\end{eqnarray}
We notice that the right-hand sides contain only terms that can be estimated by contributions either involving the nonhydrodynamic part or the hydrodynamic one multiplied by a small factor.

For  $\xi_z=0$ we have
\begin{equation}\e i\sigma\overline{\hat R_2}+\mu v_xi\xi_x\overline{\hat R_2}+v_z \overline{\hat r}= {\cal N}(\sigma,\xi_x,0,v).\label{siz}\end{equation}
We take the integral $\int_{-\pi}^\pi dz\int_{-\pi}^zdq$ of   (\ref{noz}) 
\begin{equation}
-\int_{-\pi}^\pi dz\int_{-\pi}^zdq  \big[ \e \sigma \overline{\hat R_2}{^z}+\mu\xi_xv_x\overline{\hat R_2}{^z}\big]+ iv_z\overline{\hat R_2}(\sigma,\xi_x,0)-2\pi iv_zr(-\pi)=\int_{-\pi}^\pi dz\int_{-\pi}^zdq{\cal N}^z\label{noz1}.
\end{equation}
Let $W(|v|)$ be a smooth function such that
\begin{eqnarray}
\int_0^\infty \r^3(W'(\r))^2 M(\r)d\r< \infty  ,\quad \int_0^\infty \r^5W(\r)M(\r)d\r=1,\quad\int_0^\infty \r^6W(\r)M(\r)d\r=0,\nonumber\\
\int_0^\infty \r^7W(\r)M(\r)d\r=3,\quad\int_0^\infty \r^8W(\r)M(\r)d\r=0 \label{propF}.
\end{eqnarray}

Then, multiply (\ref{noz1}) by $v_xv_zMW(|v|)$ and integrate over $v$. The first term does not contribute to the hydrodynamic part of $\overline{\hat R_2^z}$. The contribution to the hydrodynamic part from the second and third terms are respectively
$$\int_{-\pi}^\pi dz\int_{-\pi}^zdq(\psi_3,\overline{\hat R_2}{^z})\mu\xi_x\int dv v^2_xv^2_zM(|v|)W(|v|),\quad 2\pi i(\psi_1,\overline{\hat R_2})\int dv v^2_xv^2_zM(|v|)W(|v|).$$
Let us use the polar coordinates to compute the $v$-integrals 
$$\int dv v^2_xv^2_zM(|v|)W(|v|)=\int_{S_2} d\omega \o^2_x\o^2_z\int_0^\infty d\r \r^6M(\r)W(\r)=0.$$
The vanishing of the integral is due to the second condition in  (\ref{propF}). And so on the left-hand side there are only boundary and nonhydrodynamic terms. We put the latter  on the right-hand side, denoted  now by ${\cal N}^z_1$,
$$\int _{v_z<0}dv  v_xv_z^2M(|v|)W(|v|)\gamma ^{-}\overline{\hat R_2}(-\pi)={\cal N}^z_1.$$
The $v$-integral of the ingoing part of $r$ times $v_xv_z^2M(|v|)W(|v|)$ is zero because of the boundary conditions. In this way we have reached a control of the boundary term $\gamma ^{-}\overline{\hat R_2 }(-\pi)$. Now we reproduce this term by 
 multiplying (\ref{siz}) by $v_xv_zMW$ and integrating over $v_z<0$,
$$i\e\sigma \overline{\hat R}_{2v_x}(\sigma,\xi_x,0)c_1 +i\mu\xi_x\bar{\hat R}_{20}(\sigma,\xi_x,0)c_1 +\int _{v_z<0}dv  v_xv_z^2M(|v|)W(|v|) \gamma ^{-}\bar{\hat R}_2(-\pi)={\cal N}_1,$$
where ${\cal N}_1$ incorporates ${\cal N}$ and  all the other nonhydrodynamic terms and
$$c_1=\int_{S_2^-} d\omega {\o^2_x\o_z}\int_0^\infty d\r \r^5M(\r)W(\r)=\int_{S_2^-} d\omega {\o^2_x\o_z},$$
where $S_2^-={\{\o\in S_2, \o_z<0\}}$, because of the first condition in (\ref{propF}). 
We have used the second condition in (\ref{propF}) to cancel the remaining hydrodynamic terms and as before the ingoing part does not contribute. The conclusion is
\begin{equation}
c_1\big[\e\sigma \overline{\hat R}_{2v_x}(\sigma,\xi_x,0)+\mu\xi_x\bar{\hat R}_{20}(\sigma,\xi_x,0)\big ]={\cal N}_1-{\cal N}^z_1.
\label{con}
\end{equation}
We will get now  a second equation involving  $\overline{\hat R}_{2v_x}$ and $\overline{\hat R}_{20}$ with a similar procedure. The two equations together will give us  the wanted estimate for these terms.

Multiply (\ref{noz1}) by $v^2_xv_zW(|v|)$ and integrate over $v$. The first  and second term do  not contribute to the hydrodynamic part of $\hat R_2$. The contribution to the hydrodynamic part of the   third term is
$$ i(\psi_4+\psi_0,\hat R_2)\int dv v^2_xv^2_z(\psi_4+\psi_0)M(|v|)W(|v|).$$
The $v$-integrals in polar coordinates vanish because of the second and fourth conditions in  (\ref{propF}).
 The boundary integral is then given as before in terms of a r.h.s., denoted by  ${\cal N}^z_2$, involving non-hydrodynamic terms and ${\cal N}^z$,
 \begin{equation}i\int _{v_z<0}dv  v^2_xv_z^2M(|v|)W(|v|) \bar{\hat R}_2 (-\pi)={\cal N}^z_2.
\label{BT}
\end{equation}
The  ingoing part vanishes this time  because  of the second  condition in  (\ref{propF}). 
Now multiply (\ref{siz}) by $v^2_xv_zMW$  and integrate over $v_z<0$,
$$ \mu\xi_x\overline{\hat R}_{2v_x}(\sigma,\xi_x,0)c_2 +\e\sigma\overline{\hat R}_{20}(\sigma,\xi_x,0)c_1 +\int _{v_z<0} v_x^2v_z^2M(|v|)W(|v|) \bar{\hat R}_2(-\pi)dv={\cal N}_2,$$
where 
$$ c_2=\int_{S_2^-} d\omega \o^4_x\o_z\int_0^\infty d\r\r^7M(\r)W(\r)=3\int_{S_2^-} d\omega \o^4_x\o_z$$
because of the first and third conditions in (\ref{propF}). 
This time  the ingoing part 
$$i\int _{v_z<0}dv  v^2_xv_z^2M(|v|)W(|v|) r_{in}(\pi)$$ gives a contribute of order $\e$ due to the presence of the Maxwellian $M_+$ in the boundary conditions. We put 
this term and all the other non-hydrodynamic terms  in the r.h.s. term denoted by ${\cal N}_2$. By using  (\ref{BT}) we finally get
\begin{equation}
c_2\e\sigma \overline{\hat R}_{2v_x}(\sigma,\xi_x,0)+c_1\mu\xi_x\overline{\hat R}_{20}(\sigma,\xi_x,0)={\cal N}_2-{\cal N}^z_2.
\label{con1}
\end{equation}
Equations  (\ref{con}) and (\ref{con1}) give a system of two equations that allows to express $\hat R_{2v_x}$ and $\hat R_{20}$ in terms of quantities under control provided that  $c_1\ne c_2$. This  can be easily checked by direct computation.

As an end result, it holds for $\e \sigma\leq\sigma_1$ that
\begin{eqnarray}
\label{4.30}
\parallel\chi_{\sigma_1}\hat{R}_{20}(\cdot,\xi_x,0)\parallel_2^2 \leq c\Big(\frac{1}{\e^2} \parallel(I-P_J)R_2\parallel^2_{2,2,2}+\parallel(I-P)R_2\parallel ^2_{2,2,2}+\eta\parallel R_2\parallel^2_{2,2,2} \Big).\quad \quad   
\end{eqnarray}

The same estimate also holds for the $\psi_0$-moment when $\xi_z\neq0$. The proof is simpler. The boundary term is  this time removed by subtracting the same equation for $\xi_z=0$ (times $(-1)^{\xi_z}$) and then multiplying first by $v_xv_zMW$  and then by $v^2_xv_zMW$. This gives two equations
$$ \e\sigma c_1[\bar{\hat R}_{2v_x}]_- +c_1\mu\xi_x[\bar{\hat R}_{20}]_-=B,\quad \e\sigma c_2[\bar{\hat R}_{2v_x}]_- +c_1\mu\xi_x[\bar{\hat R}_{20}]_-=B_1,$$
where $\displaystyle{[f]_-=f(\sigma,\xi_x,\xi_z,v)-(-1)^{\xi_z}f(\sigma,\xi_x,0,v)}$.
\vskip.1cm

The lemma follows by collecting the previous estimates. $\square$

\bigskip

The above study of $R_2$ leads to
\begin{lemma}
\label{l4.6}
Any solution $R_2$ to the problem (\ref{4.8}) satisfies the a priori estimates
\begin{eqnarray*}
\parallel \nu^{\frac{1}{2}}(I-P_J)R_2\parallel_{2,2,2}^2&&\leq c\Big(\e  \parallel R_0\parallel_{2,2}^2\\ 
&&+\e  \parallel \nu^{-\frac{1}{2}}(I-P_J)g\parallel_{2,2,2}^2+\frac{1}{\e}\parallel P_J g\parallel_{2,2,2}^2+\frac{1}{\e^2} \parallel \bar{\psi} \parallel_{2,2,\sim}^2\Big) ,\\
\parallel P_J R_2\parallel_{2,2,2} ^2&&\leq c\Big(\frac{1}{\e} (\parallel R_0\parallel_{2,2}^2+\parallel\nu^{-\frac{1}{2}}(I-P_J)g\parallel_{2,2,2}^2)+\frac{1}{\e^3}\parallel P_J g\parallel_{2,2,2}^2\\
&&+\frac{1}{\e^4 }\parallel \bar{\psi} \parallel_{2,2,\sim}^2\Big),\\
\parallel \nu^{\frac{1}{2}}R_2\parallel ^2_{\infty,\infty,2}&&\leq c(\frac{1}{\e^2}\parallel R_2\parallel^2_{\infty,2,2}+\parallel \gamma ^-R_1\parallel_{\infty,2,\sim}^2+\frac{1}{\e^2} \parallel \bar{\psi} \parallel^2_{\infty,2, \sim})\\
&&\leq c\Big(\frac{1}{\e^3}\parallel R_0\parallel ^2_{2,2}+ \frac{1}{\e^3}\parallel \nu^{-\frac{1}{2}}(I-P_J)g\parallel ^2_{2,2,2} +\frac {1}{\e ^5}\parallel P_Jg\parallel ^2_{2,2,2}\\
&&+\frac{1}{\e^6}\parallel \bar{\psi}\parallel^2_{2,2,\sim}+\parallel R_0\parallel^2_{\infty,2}+\e^2\parallel\nu^{-\frac{1}{2}} g\parallel ^2_{\infty,\infty,2}+\frac{1}{\e^2}\parallel\bar{\psi}\parallel^2_{\infty,2,\sim}\Big).
\end{eqnarray*}
\end{lemma}
\bigskip

\begin{theorem}
\label{l4.7}
There exists a solution $R$ to the rest term problem (\ref{4.1}) such that 
\begin{equation}
\label{4.31}
\int _{0}^{+\infty }\int _{[-\pi,\pi] }\int _{[-\pi,\pi] }\int _{\R^3} |R(t,x,z,v)|^2M(v)dtdxdzdv< c \e ^7.
\end{equation}
\end{theorem}
\noindent\underline{Proof of Theorem \ref{l4.7}}\hspace*{.05in} Take the asymptotic expansion of fifth order in $\e$. We shall prove that $R$ can be obtained as the limit of an approximating sequence, and that  $R$ satisfies (\ref{4.31}).
Since $R$ is a solution to the initial boundary value problem for the rescaled rest term, (\ref{4.31}) in turn implies the $L_M^2$-convergence to zero of $R(t)$, when time tends to infinity.\\
Let the approximating sequence $\{R^n\}$ be defined by $R^0= 0$, and
\begin{eqnarray*}
&&\frac{\partial R^{n+1}}{\partial t}+\frac{1}{\e }v^\mu\cdot \nabla R^{n+1} -GM^{-1} \frac{\partial(MF^{n+1})}{\partial v_z}=\frac{1}{\e ^2}{L}_J R^{n+1}+\frac{1}{\e }H_1(R^{n+1})
+\frac{1}{\e }{J}(R^{n},R^n)+A,\\
&&R^{n+1}(0,x,z,v)= R_0(x,z,v),\\
&&R^{n+1}(t,x,\mp\pi,v)= \frac{M_\mp}{M}\int _{w_z\lessgtr0} (R^{n+1}(t,x,\mp\pi,w)+\frac{1}{\e}\bar{\psi}(t,x,\mp\pi,w))|w_z|Mdw \\
&&\hskip3.1cm-\frac{1}{\e}\bar{\psi}(t,x,\mp\pi,v),\quad
  x\in[-\pi,\pi],\quad t>0,\hspace*{0.03in}v_z\gtrless0.
\end{eqnarray*} 
Here the initial value $R_0$ is of $\e$-order four, $A$ has been chosen so that  $\int\int\int A(\cdot,x,z,v)Mdxdzdv\equiv 0$,  $(I-P_J)g=\e (I-P_J) A$ is of order four, and $P_Jg=\e P_JA$ is of order five.\\
The function $R^1$ is solution to  
\begin{eqnarray*}
&&\frac{\partial R^{1}}{\partial t}+\frac{1}{\e }v^\mu\cdot \nabla   R^{1} - GM^{-1}\frac{\partial(M R^1)}{\partial v_z}= \frac{1}{\e ^2}L_JR^{1}
+\frac{1}{\e }H_1(R^{1})
+A,\\
&&R^{1}(0,x,z,v)= R_0(x,z,v),\\
&&R^{1}(t,x,\mp\pi,v)= (M^{-1}M_\mp)(v)\int_{w_z\lessgtr0}(R^{1}(t,x,\mp\pi,w)
+\frac{1}{\e}\bar{\psi}(t,x,\mp\pi,w))w_zMdw \\
&&\hskip3.1cm- \frac{1}{\e}\bar{\psi}(t,x,\mp\pi,v),\quad
 x\in[-\pi,\pi],\quad  t>0,\quad v_z\gtrless 0.
\end{eqnarray*}
Split $R^1$ into two parts $R_1$ and $R_2$, solutions of (\ref{4.4}) and (\ref{4.8}), respectively, with $g=\e A$. Then using the corresponding a priori estimates, Lemma \ref{l4.3} and Lemma \ref{l4.6} together with the exponential decrease of $\bar{\psi}$, and the $\e$-orders of $R_0$ and  $A$, we get for some constant $c_1$. 
\begin{eqnarray*}
\parallel \nu^{\frac{1}{2}}R^1\parallel _{\infty ,\infty ,2}\leq c_1\e^{\frac{5}{2}},\quad \parallel \nu^{\frac{1}{2}}R^1\parallel _{2,2,2}\leq c_1 \e^{\frac{7}{2}} ,
\end{eqnarray*}
By induction  
\begin{eqnarray*}
\parallel \nu^{\frac{1}{2}} R^j\parallel _{\infty ,\infty ,2}&&\leq 2c_1 {\e}^{\frac{5}{2}} ,\quad j\leq n+1, \\
\parallel \nu^{\frac{1}{2}}(R^{n+1}-R^n)\parallel _{2,2,2}&&\leq c_2{\e}^2\parallel\nu^{\frac{1}{2}}( R^n-R^{n-1})\parallel _{2,2,2},\quad n\geq 0,
\end{eqnarray*}
for some constant $c_2$. Namely, if this holds up to $n^{th}$ order, then 
\begin{eqnarray*}
&&\frac{\partial }{\partial t}(R^{n+2}-R^{n+1})+\frac{1}{\e }v^\mu\cdot \nabla(R^{n+2}-R^{n+1})- GM^{-1}\frac{\partial}{\partial v_z}(M (R^{n+2}-R^{n+1}))\\
&&= \frac{1}{\e ^2}L_J(R^{n+2}-R^{n+1})+\frac{1}{\e }H_1(R^{n+2}-R^{n+1})
+\frac{1}{\e }G^{n+1},\\
&&(R^{n+2}-R^{n+1})(0,x,z,v)= 0,\\
&&(R^{n+2}-R^{n+1})(t,x,\mp\pi,v)=M^{-1}M_\mp\int _{w_z\lessgtr 0} (R^{n+2}-R^{n+1})(t,x,\mp\pi,w)|w_z|Mdw\\
\\&& \hskip5cm x\in[-\pi,\pi], \quad t>0,\hspace*{0.03in}v_z\gtrless 0.
\end{eqnarray*}
Here
\begin{eqnarray*}
G^{n+1}= (I-P)G^{n+1}= \tilde{J}(R^{n+1}+R^n,R^{n+1}-R^n).
\end{eqnarray*}
It follows that
\begin{eqnarray*}
&&\parallel \nu^{\frac{1}{2}}( R^{n+2}-R^{n+1})\parallel _{2,2,2}\leq {c}\e^{-\frac{1}{2}}\parallel \nu^{-\frac{1}{2}}G^{n+1}\parallel _{2,2,2}\\
&&\qquad\qquad\leq {c}\e^{-\frac{1}{2}}\Big( \parallel \nu^{\frac{1}{2}} R^{n+1}\parallel _{\infty ,\infty ,2}+\parallel \nu^{\frac{1}{2}} R^{n}\parallel _{\infty ,\infty ,2}\Big) \parallel \nu^{\frac{1}{2}}( R^{n+1}-R^{n})\parallel _{2,2,2}\\
&&\qquad\qquad\leq c_2{\e}^2\parallel \nu^{\frac{1}{2}}( R^{n+1}-R^{n})\parallel _{2,2,2}.
\end{eqnarray*}
Consequently,
\begin{eqnarray*}
\parallel \nu^{\frac{1}{2}}R^{n+2}\parallel _{2,2,2}&&\leq \parallel \nu^{\frac{1}{2}}(R^{n+2}-R^{n+1})\parallel _{2,2,2}+...+\parallel \nu^{\frac{1}{2}}(R^{2}-R^{1})\parallel _{2,2,2}+\parallel \nu ^{\frac{1}{2}}R^1\parallel _{2,2,2}\\
&&\leq 2c_1\e^{\frac{7}{2}}  ,
\end{eqnarray*}
for $\e $ small enough. Similarly $\parallel R^{n+2} \parallel _{\infty,\infty,2}\leq 2c_1 {\e}^{\frac{5}{2}}$. In particular $\{R^n\}$ is a Cauchy sequence in $L_M^{2}([0,+\infty)\times \Omega \times \R^3)$. The existence of a solution $R$ to (\ref{4.1}) follows, and the estimate (\ref{4.31}) holds. This means that there is a sequence of Lebesgue points in time,
 $\{t_j\}_{j=1}^{\infty}$ with $t_j$ tending to infinity with $j$, where the
$\|\,\cdot\,\|_{2,2}$- norm of the solution ${R}$ tends
 to zero. But the $\|R(t,\,\cdot\,)\|_{2,2}$ for fixed $t\geq t_j$ is
 uniformly bounded by the norm at $t_j$ plus some tail integrals from
$t_j$
to $\infty$, hence tends to zero when time tends to infinity.
$\square$ \\
This completes the study of the $R$-term and the stability theorem follows.

\bigskip

\noindent {\bf Acknowledgments.}  R.M. and R. E.  wish to thank the  Centre de Math\'ematiques et Informatique, 
Universit\'e de Provence,  Marseille, and the Institut Henri Poincare - Centre Emile Borel, where part of this work has been done, for the kind hospitality and support. They would also like to thank A. Mielke for useful discussions. The work of R. M. and R. E.  was supported  in part by MIUR, INDAM-GNFM, INFM.

\end{document}